\pdfoutput=1
\documentclass[11pt,a4paper]{article}
\usepackage{jheppub}
\usepackage{graphicx}
\usepackage{subfigure}
\usepackage{lineno}
\usepackage{array}
\usepackage{rotating}
\usepackage{afterpage}

\usepackage{preprintcover}  
\PreprintCoverPaperTitle{Measurements of fiducial and differential cross sections for Higgs boson production in the diphoton decay channel at ${\bf \sqrt{s}=8}$~TeV with  ATLAS}  
\PreprintIdNumber{CERN-PH-EP-2014-148}  
\PreprintCoverAbstract{Measurements of fiducial and differential cross sections are presented for Higgs boson production in proton--proton collisions at a centre-of-mass energy of $\sqrt{s}=8$~TeV. The analysis is performed in the $H \rightarrow \gamma\gamma$ decay channel using 20.3~fb$^{-1}$ of data recorded by the ATLAS experiment at the CERN Large Hadron Collider. The signal is extracted using a fit to the diphoton invariant mass spectrum assuming that the width of the resonance is much smaller than the experimental resolution. The signal yields are corrected for the effects of detector inefficiency and resolution. The  $pp\rightarrow H \rightarrow \gamma\gamma$ fiducial cross section is measured to be 
$43.2 \, \pm  9.4 \, ({\rm stat.})  \,  {}^{+3.2}_{-2.9}  \, ({\rm syst.}) \, \pm  1.2 \, ({\rm lumi})$~fb
for a Higgs boson of mass 125.4~GeV decaying to two isolated photons that have transverse momentum greater than 35\% and 25\% of the diphoton invariant mass 
and each with absolute pseudorapidity less than 2.37. 
Four additional fiducial cross sections and two cross-section limits are presented in phase space regions that test the theoretical modelling of different Higgs boson production mechanisms, or are sensitive to physics beyond the Standard Model. Differential cross sections are also presented, as a function of variables related to the diphoton kinematics and the jet activity produced in the Higgs boson events. The observed spectra are statistically limited but broadly in line with the theoretical expectations. }  
\PreprintJournalName{JHEP}  

\def\mh{\ensuremath{m_H}}
\def\pt{\ensuremath{p_{\rm T}^{}}}

\def\mgg{\ensuremath{m_{\gamma\gamma}}}
\def\ptgg{\ensuremath{p^{\gamma\gamma}_{\rm T}}}
\def\pttgg{\ensuremath{p^{\gamma\gamma}_{\rm Tt}}}

\def\met{\ensuremath{E_{\rm T}^{\rm miss}}}

\def\ptjl{\ensuremath{p_{\rm T}^{ j_1}}}
\def\ptjsl{\ensuremath{p_{\rm T}^{ j_2}}}
\def\ptjssl{\ensuremath{p_{\rm T}^{ j_3}}}
\def\htj{\ensuremath{H_{\rm T}}}
\def\yjl{\ensuremath{|y_{j_1}|}}
\def\yjsl{\ensuremath{|y_{j_2}|}}
\def\ptggjj{\ensuremath{p^{}_{\rm T, \gamma\gamma{}jj}}}
\def\ygg{\ensuremath{|y_{\gamma\gamma}|}}
\def\mjj{\ensuremath{m_{jj}}}
\def\deltayjj{\ensuremath{|\Delta{y_{jj}|}}}
\def\deltaygg{\ensuremath{|\Delta{y_{\gamma\gamma}|}}}
\def\dphiggjj{\ensuremath{|\Delta\phi_{\gamma\gamma,jj}|}}
\def\dphijj{\ensuremath{|\Delta\phi_{jj}|}} 
\def\njet{\ensuremath{N_{\rm jets}}}
\def\costhetastar{\ensuremath{\rm | cos \, \theta^{*}|}}
\def\taujet{\ensuremath{\tau_{1}}}
\def\sumtaujet{\ensuremath{\sum_i \tau_{i}}}

\def\antikt{anti-\ensuremath{k_{t}}}

\def\geant{{\sc Geant 4}}
\def\ctten{{\sc CT10}}

\def\sherpa{{\sc Sherpa}} 
 
\def\herwig{{\sc Herwig}} 
\def\jimmy{{\sc Jimmy}} 
 
\def\hres{{\sc Hres}} 
\def\minlo{{\sc Minlo}} 
\def\minlohj{{\sc Minlo~HJ}} 
\def\minlohjj{{\sc Minlo~HJJ}} 
\def\powheg{{\sc Powheg}} 
\def\powhegpyt{{\sc Powheg-Pythia}} 
\def\powhegher{{\sc Powheg-Herwig}} 
\def\powhegbox{{\sc Powheg Box}} 
\def\pythia{{\sc Pythia8}} 
\def\pythiasix{{\sc Pythia6}}

\def\c{\ensuremath{\cal{C}}} 

\DeclareGraphicsExtensions{.pdf}

\title{Measurements of fiducial and differential cross sections for Higgs boson production in the diphoton decay channel at ${\bf \sqrt{s}=8}$~TeV with  ATLAS}

\author{ATLAS Collaboration}
\abstract{
Measurements of fiducial and differential cross sections are presented for Higgs boson production in proton--proton collisions at a centre-of-mass energy of $\sqrt{s}=8$~TeV. The analysis is performed in the $H \rightarrow \gamma\gamma$ decay channel using 20.3~fb$^{-1}$ of data recorded by the ATLAS experiment at the CERN Large Hadron Collider. The signal is extracted using a fit to the diphoton invariant mass spectrum assuming that the width of the resonance is much smaller than the experimental resolution. The signal yields are corrected for the effects of detector inefficiency and resolution. The  $pp\rightarrow H \rightarrow \gamma\gamma$ fiducial cross section is measured to be 
$43.2 \, \pm  9.4 \, ({\rm stat.})  \,  {}^{+3.2}_{-2.9}  \, ({\rm syst.}) \, \pm  1.2 \, ({\rm lumi})$~fb
for a Higgs boson of mass 125.4~GeV decaying to two isolated photons that have transverse momentum greater than 35\% and 25\% of the diphoton invariant mass 
and each with absolute pseudorapidity less than 2.37. 
Four additional fiducial cross sections and two cross-section limits are presented in phase space regions that test the theoretical modelling of different Higgs boson production mechanisms, or are sensitive to physics beyond the Standard Model. Differential cross sections are also presented, as a function of variables related to the diphoton kinematics and the jet activity produced in the Higgs boson events. The observed spectra are statistically limited but broadly in line with the theoretical expectations. 
}

\begin{document}
\maketitle

\section{Introduction}
\label{sec:intro}

In July 2012, the ATLAS and CMS Collaborations announced the observation of a new particle~\cite{Aad:2012tfa,Chatrchyan:2012ufa} in the search for the Standard Model Higgs boson \cite{Englert:1964et,Higgs:1964ia,Higgs:1964pj,Guralnik:1964eu,Higgs:1966ev,Kibble:1967sv}. With an increasing dataset, the emphasis has now shifted to determining the properties of the new particle and testing the consistency of the Standard Model against the data. The mass of the particle has been measured to be $\mh = 125.36 \pm 0.41$~GeV and $\mh = 124.70 \pm 0.34$~GeV by ATLAS~\cite{atlasmass} and CMS~\cite{Khachatryan:2014ira}, respectively. The spin, charge conjugation and parity  of the particle have been probed by examining the angular distributions of the decay products in the $H \rightarrow \gamma\gamma$, $H \rightarrow ZZ$ and $H\rightarrow WW$ decay channels, with the data favouring a CP-even spin-zero particle~\cite{Aad:2013xqa,Chatrchyan:2013mxa,Chatrchyan:2013iaa}. Finally, the strengths of the couplings between the new particle and the gauge bosons and fermions have been explored for a number of benchmark models, using a global fit to the signal yields obtained in different decay channels~\cite{Aad:2013wqa,Chatrchyan:2012ufa,Chatrchyan:2013mxa,Chatrchyan:2013iaa,Chatrchyan:2014vua}. In all cases, the results are consistent with those expected for a Standard Model Higgs boson. 

In this paper, measurements of fiducial and differential cross sections of $pp \rightarrow H \rightarrow \gamma\gamma$ are presented, using 20.3~fb$^{-1}$ of proton--proton collision data at a centre-of-mass energy of $\sqrt{s}=8$~TeV, which was recorded by the ATLAS experiment at the CERN Large Hadron Collider (LHC). 
The investigation of these observables is an 
alternative approach to studying the properties of the Higgs boson and allows a diverse range of physical phenomena to be probed, such as the theoretical modelling of different Higgs boson production mechanisms and physics beyond the Standard Model. Furthermore, the cross sections are designed to be as model independent as possible to allow comparison to any current or future theoretical prediction. For each fiducial region (or bin of a differential distribution), the signal yield is extracted using a fit to the corresponding diphoton invariant mass spectrum. The cross sections are determined  by correcting these yields for detector inefficiency and resolution, and by accounting for the integrated luminosity of the dataset.

The  $pp \rightarrow H\rightarrow \gamma \gamma$ cross section is measured in a fiducial region defined by two isolated photons that have absolute pseudorapidity\footnote{ATLAS uses a right-handed coordinate system with its origin at the nominal interaction point (IP) in the centre of the detector and the $z$-axis along the beam pipe. The $x$-axis points from the IP to the centre of the LHC ring, and the $y$-axis points upward. Cylindrical coordinates $(r,\phi)$ are used in the transverse plane, $\phi$ being the azimuthal angle around the beam pipe. The pseudorapidity is defined in terms of the polar angle $\theta$ as $\eta=-\ln\tan(\theta/2)$.} in the interval $|\eta|<2.37$, with the leading (subleading) photon satisfying $ \pt / \mgg > $~0.35 (0.25), where $\pt$ is the transverse momentum of the photon and \mgg\ is the diphoton invariant mass.\footnote{For a Higgs boson of mass 125.4~GeV and narrow (approximately zero) width, the transverse momentum selection criteria correspond to $\pt>43.9$~GeV and $\pt>31.4$~GeV for the leading and subleading photon, respectively.} These `baseline' diphoton selection criteria  are made for all cross sections presented in this article. Four additional cross sections and two cross-section limits are presented in fiducial regions that allow the theoretical modelling of specific Higgs boson production mechanisms to be studied. Three fiducial regions are defined for events that contain at least one jet, at least two jets, or at least three jets with $\pt>30$~GeV and absolute rapidity $|y|<4.4$. A single-lepton region selects events that contain an electron or muon with $\pt>15$~GeV and $|\eta|<2.47$, enhancing the contribution from Higgs bosons produced in association with a vector boson ($VH$ production). Similarly, a fiducial region is defined for events that have large missing transverse momentum, with magnitude $\met>80$~GeV, which is sensitive to $VH$ production and possible contributions from Higgs bosons produced in association with dark matter particles. Finally, the cross section is measured for events that contain at least two jets that have large dijet invariant mass, $\mjj>400$~GeV, large rapidity separation, $\deltayjj > 2.8$, and  diphoton-dijet systems that are back-to-back in azimuthal angle, $\dphiggjj > 2.6$. This region enhances the contribution from Higgs boson production via vector-boson fusion (VBF)~\cite{Cahn:1986zv}. The details of the photon, lepton, jet and missing transverse momentum selection are documented in sections~\ref{sec:evsel} and \ref{sec:fits} for detector-level and particle-level objects, respectively.

The differential cross sections are measured in the baseline fiducial region for four categories of kinematic variables. 
\begin{enumerate}
\item \underline{Higgs boson kinematics}: The transverse momentum, \ptgg, and absolute rapidity, \ygg, of the diphoton system. Inclusive Higgs boson production is dominated by gluon fusion for which the transverse momentum of the Higgs boson is largely balanced by the emission of soft gluons and quarks. Measuring \ptgg\ therefore probes the perturbative-QCD modelling of this production mechanism. The rapidity distribution of the Higgs boson is also sensitive to the modelling of the gluon fusion production mechanism, as well as the parton distribution functions (PDFs) of the colliding protons.
\item \underline{Jet activity}: The jet multiplicity, \njet, the transverse momentum and absolute rapidity of the leading jet, \ptjl\ and \yjl, the transverse momentum of the subleading jet, \ptjsl, and the scalar sum of jet transverse momenta, \htj. The jet variables are sensitive to the theoretical modelling and relative contributions of the different Higgs boson production mechanisms. In the Standard Model, events with zero or one jet are dominated by gluon fusion and the transverse momentum and rapidity of the leading jet probe the theoretical modelling of hard quark and gluon radiation in this process. The  contribution from the
VBF and $VH$ processes becomes more important for two-jet events. The small contribution from top--antitop production in association with the Higgs boson ($t\bar{t}H$) becomes increasingly relevant at the highest jet multiplicities and for large \htj.
\item \underline{Spin--CP sensitive variables}: The cosine of the angle between the beam axis and the photons in the Collins--Soper frame~\cite{Collins:1984kg} of the Higgs boson, \costhetastar, and the azimuthal angle between the two leading jets, \dphijj, in events containing two or more jets. The \costhetastar\ variable can be used to study the spin of the Higgs boson. The \dphijj\ variable is sensitive to the charge conjugation and parity properties of the Higgs boson's interactions with gluons and weak bosons in the gluon fusion and VBF production channels, respectively~\cite{Plehn:2001nj,Klamke:2007cu,Andersen:2010zx,Dolan:2014upa}.
\item \underline{VBF-sensitive variables} for events containing two or more jets: The dijet rapidity separation, \deltayjj, and the azimuthal angle between the dijet and diphoton systems, \dphiggjj. The distribution of these variables are sensitive to the differences between the gluon fusion and VBF production mechanisms. In vector-boson fusion, the $t$-channel exchange of a $W$ boson typically results in two high transverse momentum jets that are well separated in rapidity.   Furthermore, quark/gluon radiation in the rapidity interval between the two jets is suppressed in the VBF process when compared to the gluon fusion process, because there is no colour flow between the two jets. The \dphiggjj\ distribution for VBF is therefore steeper and more closely peaked at $\dphiggjj=\pi$ than for gluon fusion. 
\end{enumerate}
\section{The ATLAS detector}
\label{sec:atlas}

The ATLAS detector is described in detail elsewhere \cite{cite:atlas}. Charged-particle tracks and interaction vertices are reconstructed using information from the pixel detector, silicon microstrip detector and the transition radiation tracker, which are collectively referred to as the inner detector. The inner detector has full azimuthal coverage over the pseudorapidity interval $|\eta|<2.5,$ and is immersed in a 2\,T axial field to allow charged-particle transverse momentum reconstruction. 
The energies of photons and electrons are measured in the electromagnetic (EM) liquid-argon sampling calorimeter, which is split into barrel and end-cap regions that cover $|\eta|<1.475$ and  $1.375<|\eta|<3.2$, respectively. For $|\eta|<2.5$, the EM calorimeter is divided into three layers longitudinal in shower depth. The first layer, referred to as the strip layer,
has a fine segmentation in the regions $|\eta| < 1.4$ and $1.5 < |\eta| < 2.4$ to facilitate the separation of
photons from neutral hadrons and to allow shower directions to be measured, while most
of the energy is deposited in the second layer. In the range of $|\eta| < 1.8$ a presampler layer allows for the correction of energy losses upstream of the calorimeter. The energies of jets are measured in the EM and hadronic calorimeters. The hadronic calorimeter is divided into three sub-regions; the barrel region ($|\eta|<1.7$) consists of an active scintillator tiles and steel absorbers, whereas the end-cap ($1.5<|\eta|<3.2$) and forward ($3.1<|\eta|<4.9$) regions are based on liquid-argon technology. The muon spectrometer comprises separate trigger and precision tracking chambers, with the latter providing muon reconstruction over the region $|\eta|<2.7$. The spectrometer is immersed in the magnetic field provided by three air-core toroids, deflection in which allows the muon momenta to be determined.

Events are retained for analysis using a three-level trigger system~\cite{Aad:2012xs}, which identifies events consistent with predefined topologies of interest. The Level-1 trigger algorithms are implemented in hardware, using coarse detector information to reduce the event rate to less than 75~kHz. The Level-2 and Event Filter run software-based trigger algorithms that use the full granularity of the detector to refine the event selection, reducing the final rate of events to below 400~Hz.

\section{Object and event selection}
\label{sec:evsel}

The measurements are performed using proton--proton collision data recorded between April and December 2012 at $\sqrt{s}=8$~TeV. This dataset corresponds to an integrated luminosity of 20.3~fb$^{-1}$. Candidate $H \rightarrow \gamma \gamma$ events were retained for analysis using a diphoton trigger, which selected events that contained two electromagnetic clusters with transverse energy greater than 35~GeV and 25~GeV and  shower shapes that matched the expectations for EM showers initiated by photons. This diphoton trigger is more than 99\% efficient for events passing the final analysis selection. Events are also required to have at least one reconstructed collision vertex, defined by at least three inner detector tracks with $\pt > 400$~MeV. The inelastic collisions that occur in addition to the hard interaction produce mainly low transverse momentum particles that form the so-called `pileup' background. The events are also required to be in a data-taking period in which the detector was fully operational. 

Photon candidates are reconstructed from clusters of energy deposited in the electromagnetic calorimeter. They are required to have $\pt>25$~GeV and $|\eta|<2.37$, but excluding the transition regions between the barrel and end-cap calorimeters, $1.37<|\eta|<1.56$. Unconverted and converted photon candidates are both used in the analysis. Unconverted photon candidates are defined as clusters without any matching track in the inner detector. Converted photon candidates are identified by matching the clusters with one or two inner detector tracks that originate from a conversion vertex in the inner detector. The photon reconstruction efficiency is  approximately 96\%, averaged over the transverse momentum and pseudorapidity expected for photons originating from the decay of a Higgs boson with a mass of 125~GeV. The converted and unconverted photon energies are corrected for energy losses in the material preceding the calorimeter, as well as shower leakage outside of the clusters, using a combination of simulation-based and data-driven correction factors~\cite{egammanew}. 
All photons are required to satisfy `loose' identification criteria~\cite{cite:photonid}, which are based on the shower shapes in the second layer of the electromagnetic calorimeter and the energy deposition in the hadronic calorimeter. The loose identification criteria are also applied to photon candidates reconstructed in the trigger.

The two highest transverse momentum photons are identified as the decay products of a Higgs boson candidate. The invariant mass of the diphoton pair is required to lie in the range $105 \leq \mgg < 160$~GeV and the leading (subleading) transverse momentum photon must satisfy $ \pt / \mgg >$~0.35~(0.25). These photons are also required to satisfy `tight' selection criteria~\cite{cite:photonid}, which place additional requirements relative to  the `loose' ones and have been reoptimised for the pileup conditions in 2012 data. The efficiency of the photon identification criteria ranges between 85\% and 95\%, depending on the photon transverse momentum and pseudorapidity. To further reduce the misidentification of jets, the photons are required to be isolated in both the inner detector and the calorimeter. The scalar summed transverse momentum of inner detector tracks that have $\pt>1$~GeV, originate from the primary vertex (see below) and lie within a cone of size $\Delta R = [(\Delta\eta)^2 + (\Delta\phi)^2]^{1/2}=0.2$ about the photon direction, is required to be less than 2.6~GeV. Tracks matched to a converted photon are excluded from the isolation definition. The isolation energy in the calorimeter is defined by summing the transverse energy of positive-energy topological clusters\footnote{Topological clusters are three-dimensional clusters of variable size, built by associating calorimeter cells on the basis of the signal-to-noise ratio~\cite{cite:clustering}.} reconstructed in the electromagnetic and hadronic calorimeters within $\Delta R<0.4$ from the photon candidate, excluding the region of size $0.125\times0.175$ in $\eta\times\phi$ around the barycentre of the photon cluster. This isolation energy is corrected for leakage of the photon energy outside of the excluded region, as well as contamination from pileup interactions~\cite{Cacciari:2007fd,Aad:2010sp}, and is required to be less than 6~GeV. The photon isolation efficiency is approximately 95\% per photon.

Once the Higgs boson candidate has been identified, the primary interaction vertex is identified using the photon direction determined from calorimeter pointing information\footnote{The direction of the photon candidates can be measured using the longitudinal segmentation of the EM calorimeter.} as input parameters to a multivariate algorithm~\cite{Aad:2013wqa}, which also accounts for the summed transverse momenta of tracks with $\pt>400$~MeV associated with each interaction vertex, the difference in azimuth between the direction of the vector sum of the tracks momenta and the diphoton system, and the track information from converted photons. The photon direction, and hence the photon momentum, is defined with respect to this primary vertex.

Electrons are reconstructed from clusters of energy in the electromagnetic calorimeter matched to inner detector tracks. They are required to have $\pt > 15$~GeV and $|\eta|< 2.47$. All electrons are required to satisfy the `medium' identification criteria~\cite{,Aad:2011mk,Aad:2014fxa}, which have been reoptimised for the pileup conditions in 2012 data~\cite{cite:electronid}. The electrons are also required to be isolated in both the inner detector and the calorimeter.  The summed transverse momenta of tracks within $\Delta R<0.2$ of the electron direction is required to be less than 15\% of the electron transverse energy. Similarly, the transverse energy deposited in calorimeter cells within $\Delta R<0.4$ of the electron direction is  required to be less than 20\% of the electron transverse energy, after excluding the transverse energy due to the electron and correcting for the expected pileup contribution. Electrons that overlap with the selected photons ($\Delta R <0.4$) are removed from the analysis. 

Muons are identified as inner detector tracks that are matched and combined with track segments from the muon spectrometer~\cite{Aad:2014zya, cite:muon-eff}. They are required to have $\pt > 15$~GeV and $|\eta|<2.47$. Track quality requirements are imposed in order to suppress backgrounds, and impact parameter requirements reduce the impact of muons from pileup interactions. 
The muons are required to be isolated in both the inner detector and the calorimeter, using the same isolation criteria that are applied to the electron candidates. Muons that overlap with the selected photons ($\Delta R <0.4$) are removed from the analysis. 

Jets are reconstructed using the \antikt\ algorithm~\cite{Cacciari:2008gp} with a radius parameter of 0.4. The inputs to the algorithm are three-dimensional topological clusters. The jets are corrected for soft energy deposits originating from pileup~\cite{jet-pile} and then calibrated using a combination of simulation-based and data-driven correction factors that correct for calorimeter non-compensation and inactive regions of the calorimeter~\cite{Aad:2011he,Aad:2014bia}. Jets are required to have $\pt>30$~GeV and $|y|<4.4$. Jets that do not originate from the primary vertex are identified using the jet vertex fraction (JVF). Tracks are ascribed to a jet using ghost-association~\cite{Cacciari:2008gn} and the JVF is defined as the scalar summed transverse momentum of tracks from the primary interaction vertex divided by the summed transverse momentum of tracks from all vertices. Jets with $\pt<50$~GeV and $|\eta|<2.4$ are required to have $\mathrm{JVF}>0.25$. Jets are also required to be separated from photons ($\Delta R>0.4$) and electrons ($\Delta R > 0.2$).

Missing transverse momentum is calculated using an algorithm that performs the vectorial sum of all transverse energies associated with the reconstructed physics objects (such as photons, electrons, muons and jets) as well as individually calibrated calorimeter topological  clusters and inner detector tracks that are not associated with any reconstructed physics object. A full description of this algorithm can be found elsewhere~\cite{Aad:2012re}.

\section{Monte Carlo simulation}
\label{sec:mc}

Simulated samples are used to determine the shapes of the diphoton mass spectra for signal and background processes, and to correct the data for detector inefficiency and resolution. Monte Carlo event generators are used to produce events at the particle level for signal and background processes. The signal events are passed through a \geant~\cite{geant-1,geant-2} simulation of the ATLAS detector~\cite{Aad:2010ah} and reconstructed using the same analysis chain as used for the data. Pileup is included in the simulation by adding inelastic proton--proton collisions, such that the average number of interactions per bunch crossing reproduces that observed in the data. The inelastic proton--proton collisions were produced using \pythia~\cite{Sjostrand:2007gs} with the A2 set of parameters~\cite{ATLASUE2} that are tuned to data. The average number of interactions per bunch crossing, $\langle \mu  \rangle$,  is typically in the range $10 < \langle \mu  \rangle< 35$ for 2012 data.

Higgs boson production via gluon fusion is simulated at next-to-leading-order (NLO) accuracy in QCD using the \powhegbox~\cite{Nason:2004rx,Frixione:2007vw,Alioli:2010xd,Alioli:2008tz}, with the \ctten\ parton distribution function (PDF)~\cite{Lai:2010vv}. The mass and width of the Higgs boson is chosen to be $\mh=125$~GeV and $\Gamma_H = 4.07$~MeV, respectively. The parton-level events produced by the \powhegbox\ are passed to \pythia\ to provide parton showering, hadronisation and multiple parton interactions (MPI), using the AU2 tune for the underlying event~\cite{ATLASUE2}. This sample, referred to as \powhegpyt{}, is used as the default sample for Higgs boson production via gluon fusion. Additional gluon fusion samples are produced to assess the impact of generator modelling when correcting the data for detector effects. One such sample is produced by passing the parton-level events produced by the \powhegbox\ through \herwig\ \cite{Corcella:2000bw,Corcella:2002jc} and \jimmy~\cite{Butterworth:1996zw} (tune AUET2 \cite{ATLASUE}), which assesses the modelling of the parton shower, hadronisation and MPI. A sample of $H+1\,{\rm jet}$ events is produced at NLO accuracy in QCD using the \powhegbox, with the \minlo\ feature~\cite{Hamilton:2012np} applied to include $H+0\,{\rm jet}$ events at NLO accuracy and interfaced to \pythia\ to produce the fully hadronic final state. This sample is referred to as \minlohj. Similarly, a sample of $H+2\,{\rm jet}$ events is produced at NLO accuracy (referred to as \minlohjj), with the $H+0/1\,{\rm jet}$ events included with up to leading-order (LO) accuracy. A final gluon fusion sample is produced using \sherpa~1.4.3~\cite{Gleisberg:2008ta}, which produces $H+n\,{\rm jet}$ events ($n=0,1,2,3,4$) at LO accuracy in QCD and uses the  CKKW method~\cite{Catani:2001cc} to combine the various final-state topologies and match to a parton shower. The \sherpa\ sample is produced using the authors'  default tune for underlying event and  the \ctten\ PDF. All gluon fusion samples are normalised such that they reproduce the total cross section predicted by a next-to-next-to-leading-order plus next-to-next-to-leading-logarithm (NNLO+NNLL) QCD calculation with NLO electroweak corrections applied~\cite{Heinemeyer:2013tqa, 
Djouadi:1991tka,Dawson:1990zj,Spira:1995rr, 
Kramer:1996iq, Chetyrkin:1997un,
Harlander:2002wh,Anastasiou:2002yz,Ravindran:2003um,
Catani:2003zt,
Aglietti:2004nj, Actis:2008ug,
Anastasiou:2008tj, deFlorian:2009hc,
Moch:2005ky, Laenen:2005uz, Idilbi:2005ni, Ravindran:2005vv,
Anastasiou:2011pi}.

Higgs boson production via vector-boson fusion is generated at parton level to NLO accuracy in QCD using the \powhegbox\ \cite{Nason:2009ai} with the \ctten\ PDF. 
The parton-level events  are passed to \pythia\ to provide parton showering, hadronisation and MPI, using the AU2 tune for the underlying event. The VBF sample is normalised to an approximate-NNLO QCD cross section with NLO electroweak corrections applied~\cite{Heinemeyer:2013tqa,Ciccolini:2007jr,Ciccolini:2007ec,Figy:2003nv,Arnold:2008rz,Bolzoni:2010xr,Figy:2010ct}. 
 Higgs boson production in association with a vector boson ($ZH$, $WH$) or a top--antitop pair ($t\bar{t}H$) are produced at leading-order accuracy using \pythia\ with the CTEQ6L1 PDF and the 4C tune for underlying event~\cite{ATLASUE}. The $ZH$ and $WH$ samples are normalised to cross sections calculated at NNLO in QCD with NLO electroweak corrections \cite{Heinemeyer:2013tqa,Han:1991ia,Brein:2003wg,Ciccolini:2003jy}. The $t\bar{t}H$ sample is normalised to a cross-section calculation accurate to NLO in QCD~\cite{Heinemeyer:2013tqa,Beenakker:2001rj,Beenakker:2002nc,Dawson:2002tg,Dawson:2003zu}.

Samples of prompt diphoton ($\gamma \gamma$) and photon+jet ($\gamma j$) events are simulated with up to three additional partons in the final state using the \sherpa\ event generator, with the \ctten\ PDF and the 
authors' default tune for underlying event activity. Samples of dijet ($jj$) background events are simulated with \pythia. These samples are used to determine the form of the functions used to model the background diphoton invariant mass spectrum when extracting the signal, as discussed in the following section. The large sample size for these background processes prevents the use of the full ATLAS detector simulation and a simplified detector model is used to account for the photon and jet energy resolutions as well as the photon reconstruction, identification and mistag efficiencies~\cite{Aad:2012tfa,Aad:2013wqa}. 

\section{Extraction of signal yield and correction for detector effects}
\label{sec:fits}

The signal is extracted using the approach adopted in previous ATLAS measurements of $H \rightarrow \gamma \gamma$~\cite{Aad:2012tfa,Aad:2013wqa}. An unbinned maximum likelihood fit is performed on the \mgg\ spectrum in each fiducial region or bin of a differential distribution. The likelihood function, $\mathcal{L}$, is given by
\begin{equation}
  \label{eq:emlmaster}
  \mathcal{L}(\mgg, \nu^{\rm sig},  \nu^{\rm bkg}, \mh )  = \prod_{i} \, \left\{  \frac{{\rm e}^{-\nu_i}}{n_i!} \, \prod_{j}^{n_i} \left[  \nu_i^{\rm sig} \, \mathcal{S}_i( \mgg^j; \mh) +  \nu_i^{\rm bkg} \, \mathcal{B}_i(\mgg^j) \right]   \right\} \times\, \prod_{k} G_k
\end{equation}
where $i$ labels the categories (bins) being simultaneously fitted, $\nu_i^{\rm sig}$ is the fitted number of signal events, $\nu_i^{\rm bkg}$ is the fitted number of background events, $\nu_i= \nu_i^{\rm sig} + \nu_i^{\rm bkg}$ is the mean value of the underlying Poisson distribution for the $n_i$ events, $\mgg^j$ is the diphoton invariant mass for event $j$, $\mathcal{S}_i( \mgg^j; m_H)$ and $\mathcal{B}_i(\mgg^j)$ are the signal and background probability distribution functions, and the $G_k$ are normal or log-normal constraints incorporating  
uncertainties on the photon energy scale and resolution, as well as the uncertainty in the fitted peak position from the chosen background parameterisation. Other uncertainties that do not affect the shape of the diphoton mass spectrum are not included in the fit and are dealt with as part of the correction for detector effects. The fitted number of signal events is not constrained to be positive.

The signal probability distribution function is modelled as the sum of a Crystal Ball function and a Gaussian function and the fit is performed after fixing the Higgs boson mass to be $\mh = 125.4$~GeV \cite{atlasmass}. The Gaussian and Crystal Ball functions are required to have the same peak position and the parameters of the model that define the shape of the signal distribution are determined using simulated samples. The background probability distribution is modelled as the exponential of a first-, second- or third-order polynomial. The form of the background function is chosen, in each fiducial region or bin of a distribution, to minimise the bias observed in the extracted yield \cite{Aad:2012tfa,Aad:2013wqa} when fitting a background-only distribution constructed from the $\gamma \gamma$, $\gamma j$ and $jj$ simulated samples, after normalising the samples using data-driven scale factors determined in designated control regions. The control regions are defined by reversing the isolation and tight identification criteria for each photon and the relative composition of each background process is determined as a function of the jet multiplicity.
 
All events selected in the baseline fiducial region are included in the signal extraction for each of the observables, with any uncategorised events placed into an additional bin and included in the fit.  For example, events containing zero or one jets are included in this additional bin when fitting the \mjj\ distribution. The use of all events in each fit helps to constrain the systematic uncertainties from the photon energy scale and resolution.

\begin{figure}[t]
  \begin{center}
    \subfigure[] {
      \includegraphics[width=0.47\textwidth]{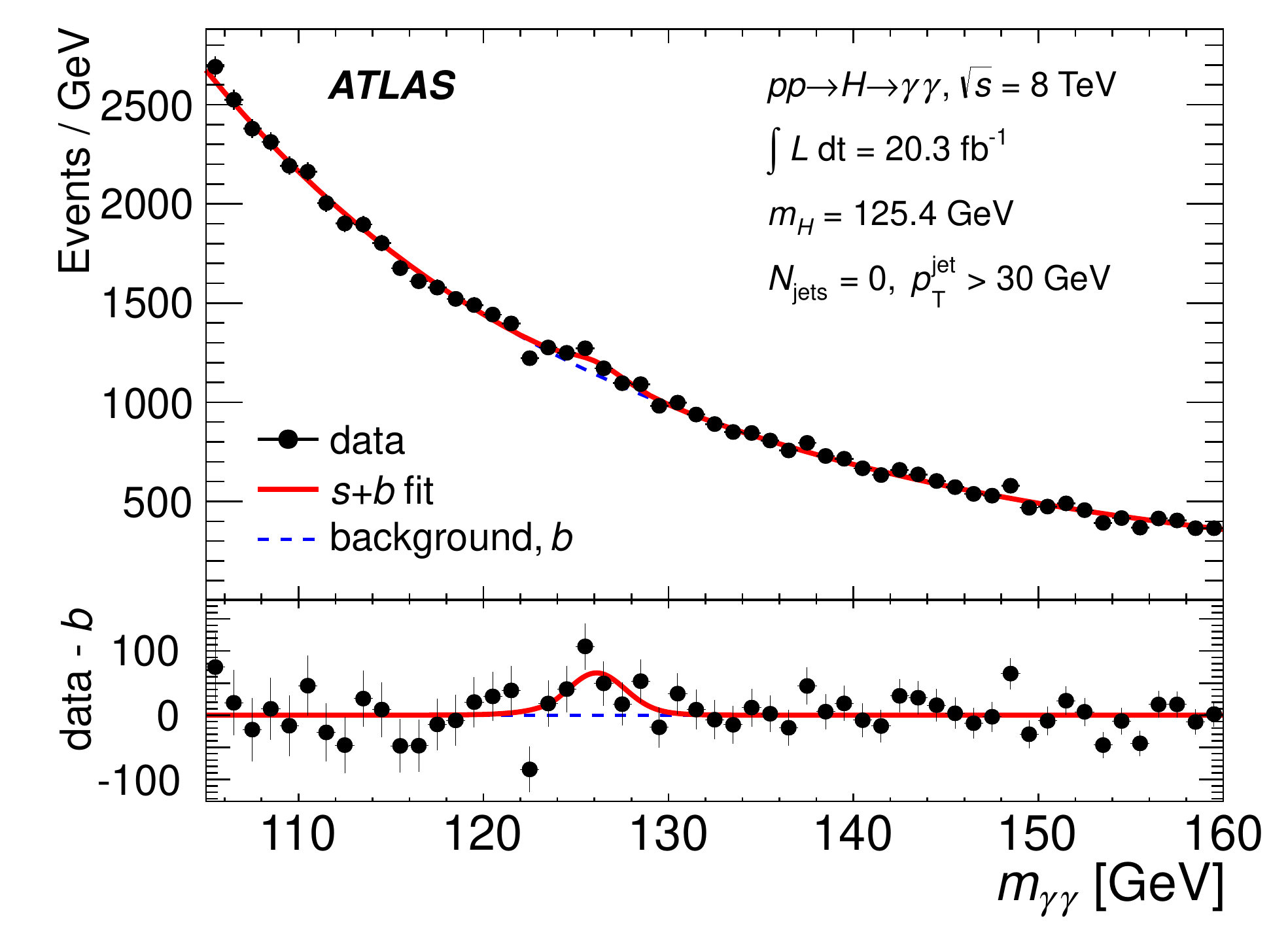}\quad
    }
    \subfigure[] {
      \includegraphics[width=0.47\textwidth]{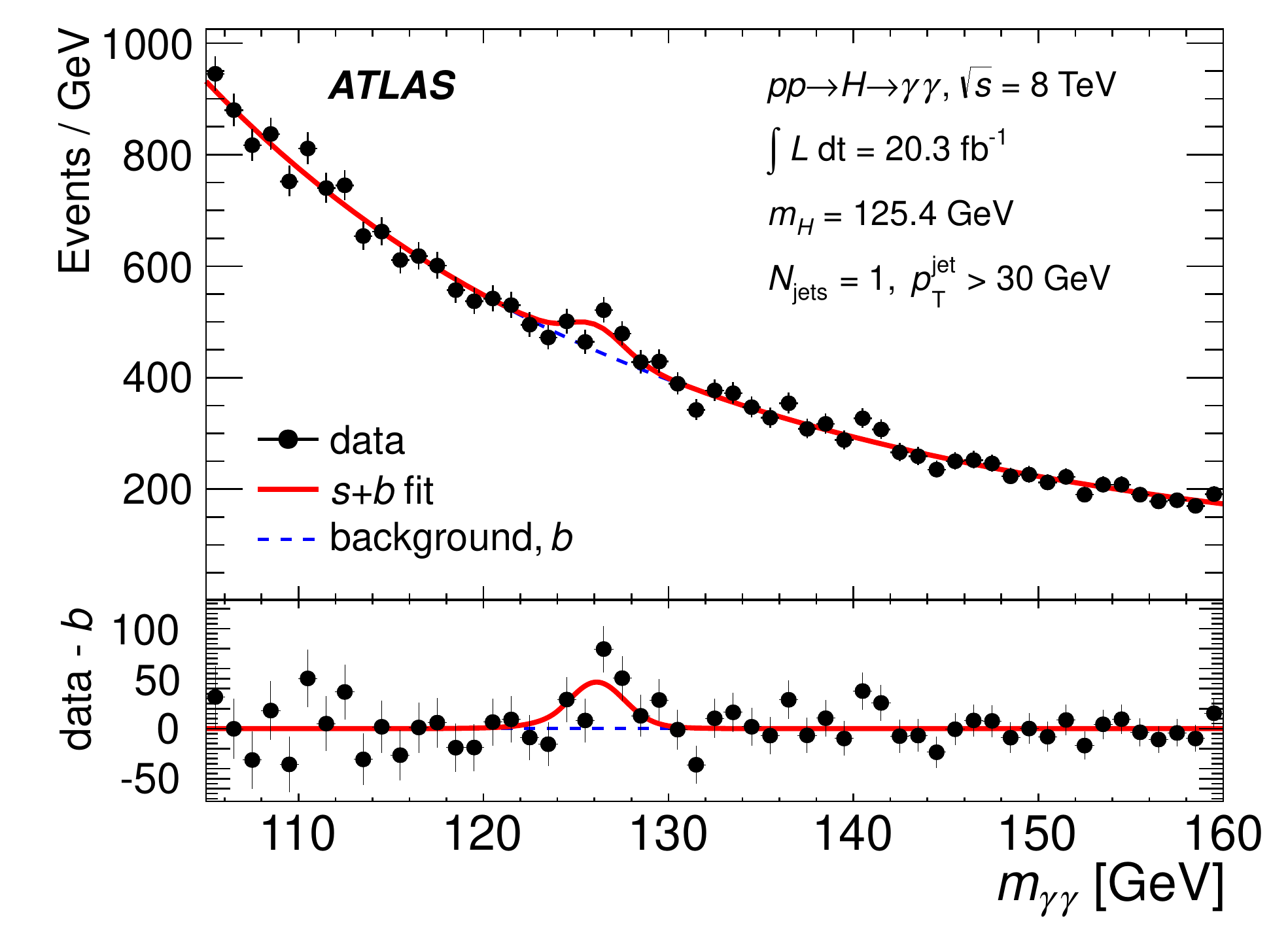}
    }
    \subfigure[] {
      \includegraphics[width=0.47\textwidth]{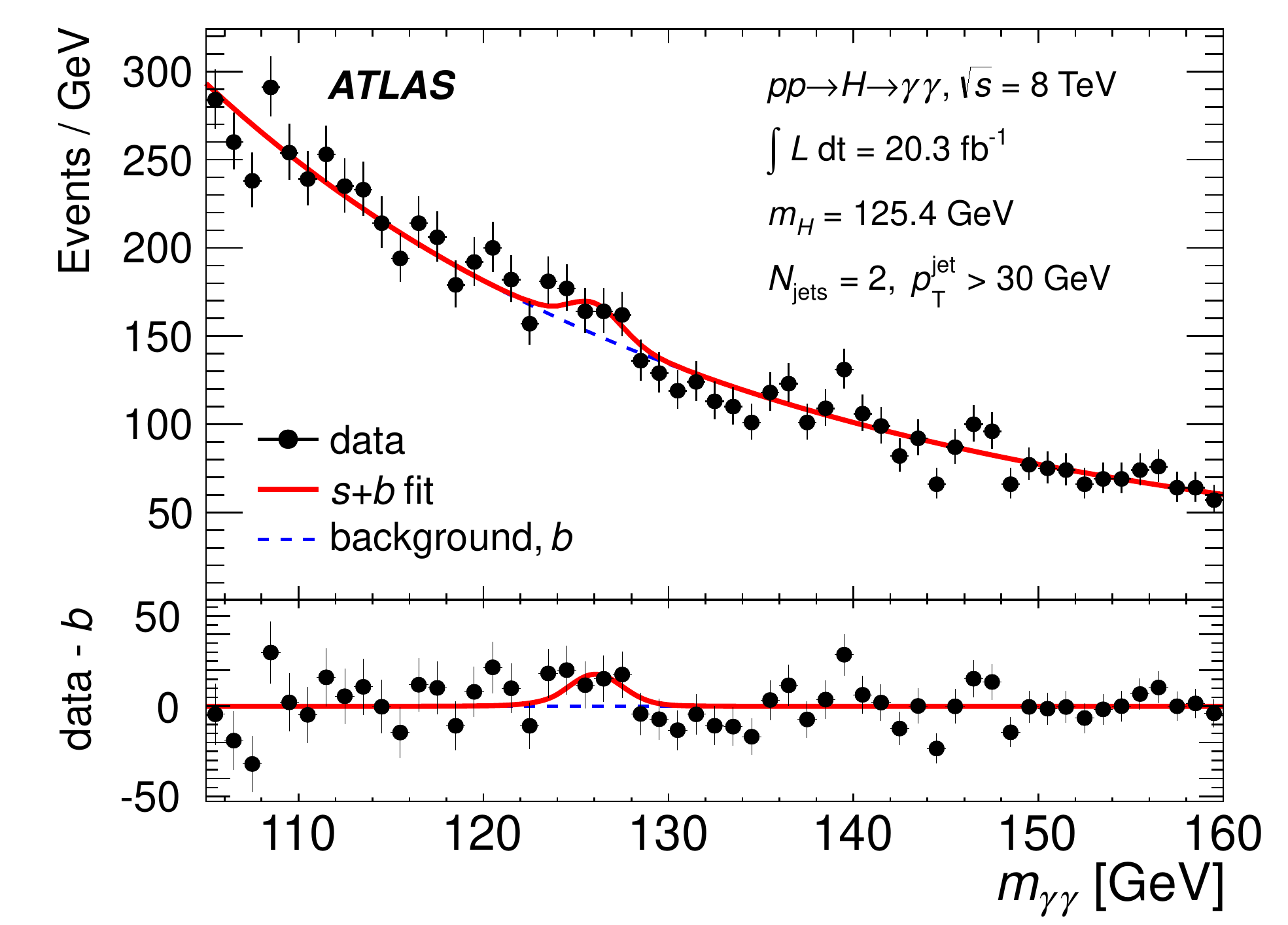}\quad
    }
    \subfigure[] {
      \includegraphics[width=0.47\textwidth]{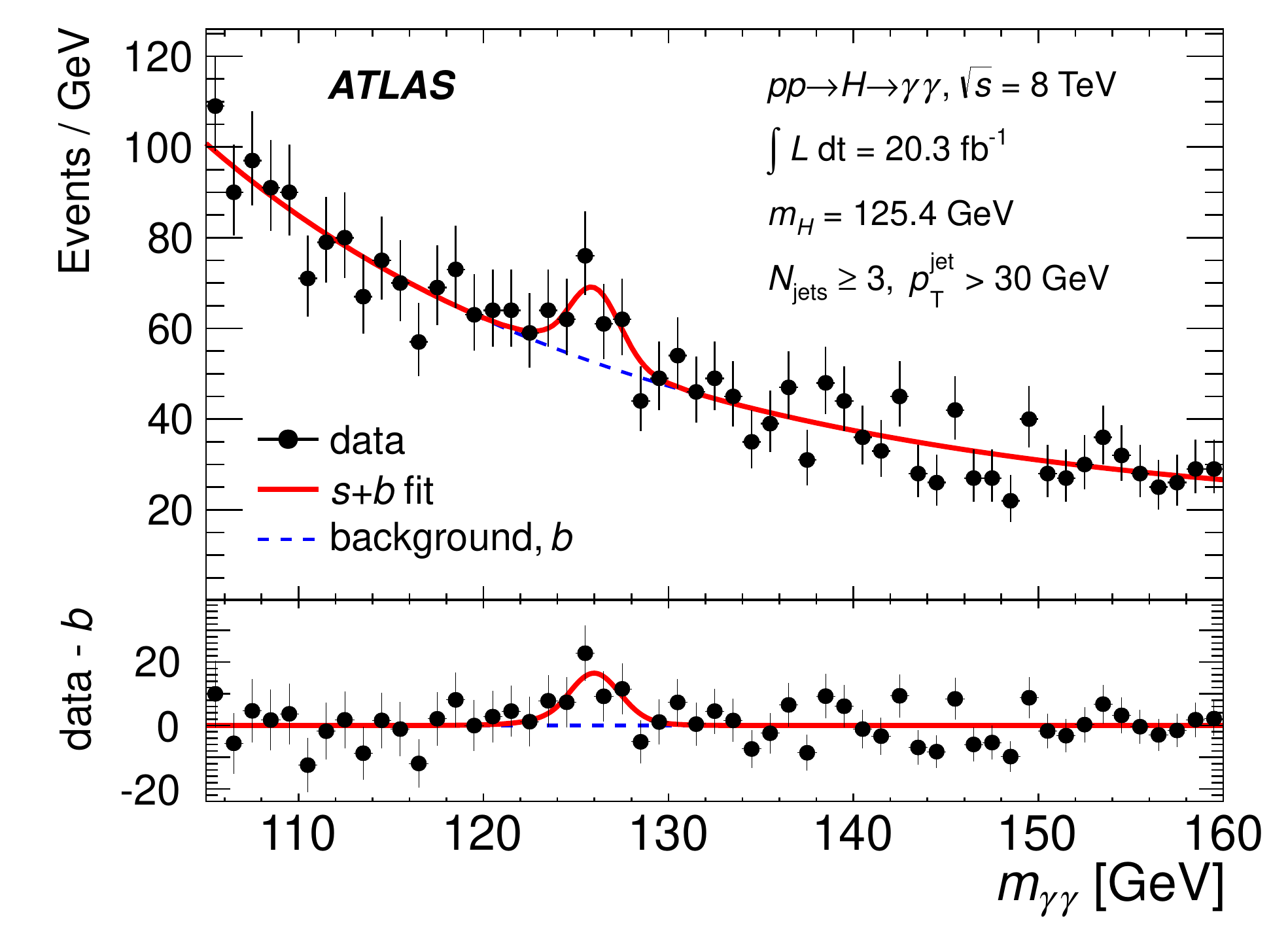}
    }
    \caption[]{The diphoton invariant mass spectrum for four bins of jet multiplicity as described in the legend. The curves show the results of the single simultaneous fit to data for all multiplicity bins, where the Higgs boson mass is fixed to be $\mh=125.4$~GeV. The red line is the combined signal and background probability distribution functions, and the dashed line shows the background-only probability distribution function. The difference of the two curves is the extracted signal yield. The bottom inset displays the residuals of the data with respect to the fitted background component.} 
    \label{fig:fit_njet}
  \end{center}
\end{figure}

Figure \ref{fig:fit_njet} shows the result of the signal-plus-background fit to the diphoton invariant mass reconstructed in different jet multiplicity bins. The difference in the extracted signal yield between fixing the Higgs boson mass and allowing it to float in the fit is 3.2\% in the baseline fiducial region, with the largest effect being 16\% for $\njet=1$. These differences are smaller than statistical uncertainties in the fit itself for all the results presented in this paper. The total number of selected diphoton events in each fiducial region, the extracted signal yields and the expected yields from simulation are presented in table~\ref{tab:yields}. 

\begin{table}
\centering
  \begin{tabular}{| l | c | c | c | }
  \hline
  Fiducial region &  $N_{\rm data}$ & $N_{\rm MC}^{\rm sig}$ & $\nu_i^{\rm sig}$  \\
  \hline
Baseline & 94627 & $403 \pm 45$ & $570 \pm 130$ \\            
$N_{\rm jets} \geq 1$ & 34293 & $178 \, {}^{+31}_{-26}$& $308 \pm 79$  \\
$N_{\rm jets} \geq 2$ & 10699 & $63 \pm 11$& $141 \pm 43$  \\
$N_{\rm jets} \geq 3$ & 2840 & $17 \pm 4$& $64 \pm 22$  \\
VBF-enhanced & 334 &$13 \pm 2$ & $24 \pm 9$  \\
$N_{\rm leptons} \geq 1$ & 168 & $3.5 \pm 0.4$ & $-3 \pm 5$  \\
$\met > 80$~GeV & 154 & $2.6 \pm 0.4$ & $-2 \pm 4$  \\       
 
  \hline
  \end{tabular}
  \caption{The total number of events selected in data in each fiducial region, $N_{\rm data}$, the expected signal yield obtained from the simulation samples discussed in section~\ref{sec:mc}, $N_{\rm MC}^{\rm sig}$, and the fitted yield obtained from data, $\nu_i^{\rm sig}$. The uncertainty on the fitted yield is the total uncertainty on the signal extraction, including the statistical and systematic uncertainties. The uncertainty on the expected yields include both the theoretical and experimental systematic uncertainties.}
  \label{tab:yields}
\end{table}

The cross section, $\sigma_i$, in a given fiducial region (or bin of a differential distribution) is defined by
\begin{equation}
\label{eq:corr}
\sigma_i = \frac{ \nu_i^{\rm sig}}{ c_i \int L \, \mathrm{d}t},
\end{equation}
where $ \int L\, \mathrm{d}t$ is the integrated luminosity of the dataset and $c_i$ is a correction factor that accounts for the difference in the event yield at detector level and particle level that arises from detector inefficiencies and resolutions. The correction factors are determined using the simulated Higgs boson event samples discussed in section \ref{sec:mc}.

The particle-level prediction is defined using particles that have mean lifetimes that satisfy $c\tau>10$~mm. The selection criteria applied to the particles are chosen to be very similar to the criteria applied at detector level to ensure minimal model dependence in the final measurement. The two highest transverse momentum photons with $|\eta|<2.37$ that do not originate from the decay of a hadron are required to satisfy  $ \pt / \mgg > 0.35$ and  $ \pt / \mgg > 0.25$, respectively. Furthermore, the summed transverse momentum of other particles (excluding muons and neutrinos) within a cone of $\Delta R= 0.4$ centred on the photon direction is required to be less than 14~GeV.\footnote{The particle-level criterion is determined using the simulated Higgs boson event samples, by comparing the calorimeter isolation energy to the particle-level isolation on an event-by-event basis. An isolation energy of 14~GeV at particle-level isolation is found to produce a mean calorimeter isolation energy of 6~GeV. The difference between the values is due to the low response of the calorimeters to soft-energy deposits. An additional charged-particle isolation (to replicate the track isolation at detector level) is found to not be necessary. After applying the isolation criterium, the two photons are found to originate from the decay of the Higgs boson for more than 99.99\% of the selected events.}  
Leptons are required to have $\pt>15$~GeV, $|\eta|<2.47$ and not to originate from the decay of a hadron. The lepton four momentum is defined as the combination of an electron (or muon) and all nearby photons with $\Delta R< 0.1$ that do not originate from the decay of a hadron. 
Jets are reconstructed from all particles with $c\tau>10$~mm, excluding muons and neutrinos, using the \antikt\ algorithm with a radius parameter of 0.4. Jets are required to have $\pt>30$~GeV, $|y|<4.4$ and be well separated from photons ($\Delta R > 0.4$) and electrons ($\Delta R> 0.2$). The missing transverse momentum is defined as the vector sum of neutrino transverse momenta. 

The correction factor (equation (\ref{eq:corr})) is 0.66 in the baseline fiducial region and the deviation from unity is mostly due to the effect of photon reconstruction and identification efficiency, including an extrapolation over the small region in pseudorapidity excluded from the photon reconstruction. The correction factor also accounts for migrations into and out of the fiducial volume caused by the finite photon energy resolution.\footnote{The correction factor also removes a small fraction of events (0.3\%) that originate from $H\rightarrow f\bar{f}\gamma$ decays that satisfy the diphoton analysis selection, where $f\bar{f}$ refers to a quark--antiquark or lepton--antilepton pair. No correction is applied to the data for interference between signal and background. Such interference effects are known to have a 1\% effect for events that satisfy the baseline selection, although the effects are known to have kinematic dependence.} The correction factor in the VBF-enhanced fiducial region is 0.71, which additionally corrects for migration into the fiducial volume at reconstruction level due to the jet selection requirements and the finite jet energy resolution.

The binning of the differential variables is determined using two criteria. First, the purity of all bins is required to be larger than 60\%, where the purity of a given bin is defined using simulation as the fraction of events at detector level that occupy the same bin at particle level. Second, the value of $s/\sqrt{b}$ in each bin is required to be larger than 1.5, where $s$ is the expected number of signal events in a diphoton mass window of $\pm4$~GeV about the Higgs boson mass and $b$ is the corresponding number of background events estimated from the data by linearly extrapolating the number of events observed outside of that window. In the rare case of the fit to data producing a negative yield in a differential distribution, the affected bin is merged with a neighbouring bin in order to ensure a positive yield (only one such case occurs).

\section{Systematic uncertainties}
\label{sec:syst}

The systematic uncertainties can be grouped according to whether they impact  the extraction of the signal yield, the correction factor, or the luminosity, which collectively define the cross-section measurement as given in equation (\ref{eq:corr}).

The impact of the photon energy scale and resolution uncertainties, as well as the impact of the background modelling on the fitted peak position, are included in the fit as nuisance parameters as discussed in section~\ref{sec:fits}. The uncertainty on the photon energy resolution and scale has been determined using $Z \rightarrow e^+ e^-$ events~\cite{egammanew}. The uncertainty due to the background modelling on the fitted peak position is estimated through fitting signal and background simulated samples with the chosen signal and background function.  
The impact of these systematic uncertainties on the extracted signal yield is studied by constructing an `Asimov dataset'~\cite{Cowan:2010js}, which is the expected diphoton invariant mass spectrum constructed from the final form of the background and signal probability distribution functions after fitting to the data. This Asimov dataset is fit twice, once allowing the nuisance parameters to float and once with the nuisance parameters fixed to their profiled values. The systematic uncertainty on the extracted yield due to the fit procedure is defined by subtracting, in quadrature, the uncertainty on the signal yield obtained with fixed nuisance parameters from the uncertainty on the signal yield obtained with floated nuisance parameters. The systematic uncertainty is $\pm$6.2\% in the baseline fiducial region. This uncertainty is added in quadrature to the uncertainty on the fitted yields due to the background modelling, which is determined by fitting background-only ($\gamma\gamma$, $\gamma j$, $jj$) simulated samples with the chosen background function and estimated to be 2.0\% in the  baseline fiducial region. 

The luminosity of the 2012 dataset is derived, following the same methodology as that detailed in ref.~\cite{Aad:2013ucp}, from a preliminary calibration of the luminosity scale determined from beam-separation scans performed in November 2012. The uncertainty in the integrated luminosity affects all fiducial and differential cross sections and is estimated to be 2.8\%.

The remaining systematic uncertainties are associated with the experimental and theoretical modelling of the simulated Higgs boson samples that are used to calculate the correction for detector effects (equation (\ref{eq:corr})). Uncertainties in the trigger efficiency, the photon energy scale and resolution, the photon identification efficiency and the photon isolation also affect all the differential and fiducial cross sections by changing the number of detector-level events and, therefore, the detector correction factors. The photon energy scale and resolution cause migrations into and out of the fiducial region and are estimated by shifting and smearing the photon energies by the known uncertainties and recalculating the correction factor. The effect on the measured cross section is typically less than 0.1\% for the photon energy scale and resolution. The uncertainty in the photon identification and trigger efficiencies have been determined from data~\cite{cite:photonid, Aad:2013wqa}. The impact of each uncertainty is estimated by applying event-level weights for each photon that cover the differences observed between data and simulation. The uncertainty on the cross section measured in the baseline fiducial region is 1.0\% and 0.5\% for the photon identification and trigger efficiencies respectively. The uncertainty in the photon isolation is dependent on the level of hadronic activity in the event, with a 1\% impact for events that satisfy the baseline selection and a 4\% impact for events containing three or more jets.

Distributions or fiducial regions that are sensitive to jet activity in the event are affected by uncertainties in the jet energy scale, jet energy resolution, jet vertex fraction efficiency and the modelling of jets originating from pileup interactions. The uncertainties associated with the jet energy scale and resolution are estimated by shifting or smearing the reconstructed jet energies by an amount commensurate with the uncertainties derived from the transverse momentum balance in $\gamma$--jet, $Z$--jet, dijet and multijet topologies \cite{Aad:2011he,jes2011,Aad:2012ag}. The difference in the cross section arising from  the systematically shifted and nominal correction factors is taken to be the systematic uncertainty. The effect of the jet energy scale and resolution depends on the variable, being 4\% for $\njet = 0$ and rising to 14\% for $\njet=3$, for example. The uncertainty associated with the jet vertex fraction selection is estimated by shifting the required fraction by $\pm0.03$, which encompasses the differences between the JVF distributions in simulation and data, and recalculating the correction factor. The uncertainty due to JVF modelling is less than 0.4\% for all jet multiplicities. The uncertainty associated with the modelling of pileup jets is estimated by removing a fraction of the jets originating from pileup interactions and recalculating the correction factor. The fraction is estimated by comparing the data to simulation in pileup-enriched control regions of $Z+{\rm jets}$ events \cite{Aad:2014dta,Aad:2012mfa}. The uncertainty due to pileup jet modelling is 0.7\% for $\njet=0$, rising to 3.3\% for events containing three or more jets. 

The systematic uncertainties on the lepton reconstruction, identification and isolation efficiencies, as well as the lepton momentum scale and resolution, have been determined using $Z$ bosons reconstructed in data \cite{cite:muon-eff,Aad:2014zya,Aad:2011mk}. The lepton-based uncertainties only have a non-negligible impact on the cross-section limit extracted for events containing one or more leptons.

Uncertainties in the correction factor due to theoretical modelling are estimated in three ways. First, the uncertainty in the gluon fusion modelling is taken to be the envelope of correction factors obtained by replacing the default \powhegpyt\ sample with alternative fully simulated samples, which include  the \powhegher, \minlohj, \minlohjj\ and \sherpa\ samples discussed in section \ref{sec:mc} as well as a \powhegpyt\ sample with MPI turned off. The inclusion of the \powhegpyt\ sample generated without MPI provides a conservative estimate of the impact of double parton scattering in phase space regions containing two or more jets. Second, the effect of increased or decreased contributions from the VBF and $VH$ production mechanisms is estimated by changing the relevant cross sections by a factor of 0.5 and 2.0 and recalculating the correction factors. This variation is consistent with the current uncertainty on the VBF and $VH$ signal strengths measured by the ATLAS Collaboration~\cite{Aad:2013wqa}. Similarly, the possible impact of an increased or decreased contribution from $t\bar{t}H$ events is estimated by increasing the cross section by a factor of five, or removing the contribution entirely, which  is consistent with the current limit on $t\bar{t}H$ production measured by the CMS Collaboration~\cite{Chatrchyan:2013yea}. Finally, the simulation events are reweighted to reproduce the \ptgg\ and  \ygg\ distributions observed in the data and the correction factors are recalculated. The gluon fusion modelling and signal composition uncertainties are added in quadrature and the total theoretical modelling uncertainty is then taken to be the envelope of that uncertainty and the uncertainty derived from the data-driven reweighting. The total theoretical modelling uncertainty on the cross section is ${}^{+3.3}_{-1.0}$\% for the baseline fiducial region, but can be as large as ${}^{+6.3}_{-4.9}$\% for events containing three or more jets.

The impact on the measured cross section due to destructive interference between Higgs boson production via gluon fusion and the $gg \rightarrow \gamma\gamma$ background was assessed by reweighting the \powhegpyt\ gluon fusion simulation on an event-by-event basis to include the expected interference contribution~\cite{Dixon:2003yb} and rederiving the detector correction factors. The applied weights  are dependent on the photon pseudorapidity values and are valid at low Higgs boson transverse momentum. Although the interference typically reduces the cross section by 1\%  (depending on Higgs boson kinematics), the impact on the correction factors is less than 0.1\% in all regions. 

\begin{figure}[t]
  \begin{center}
    \subfigure[] {
      \includegraphics[width=0.47\textwidth]{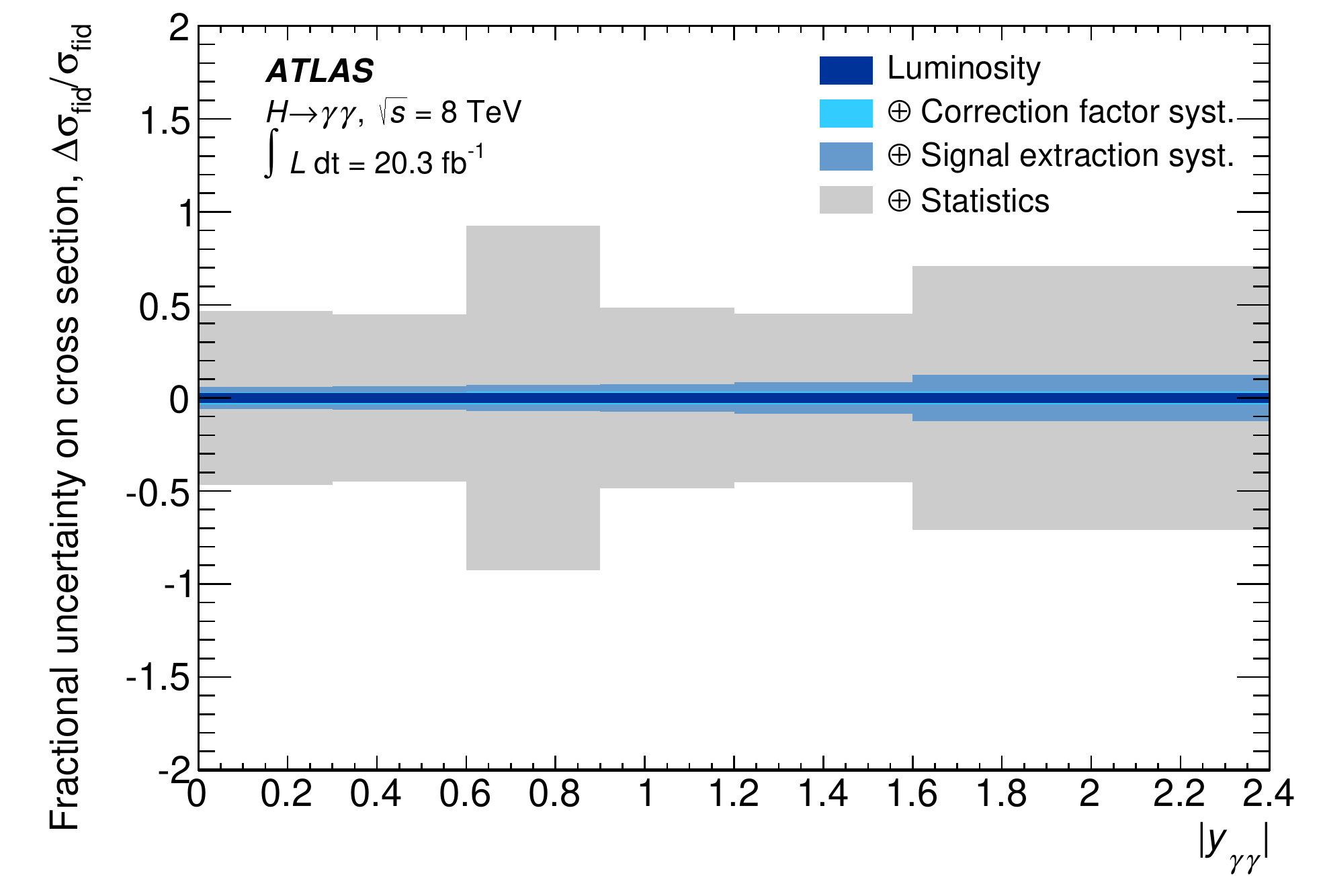}\quad
    }
    \subfigure[] {
      \includegraphics[width=0.47\textwidth]{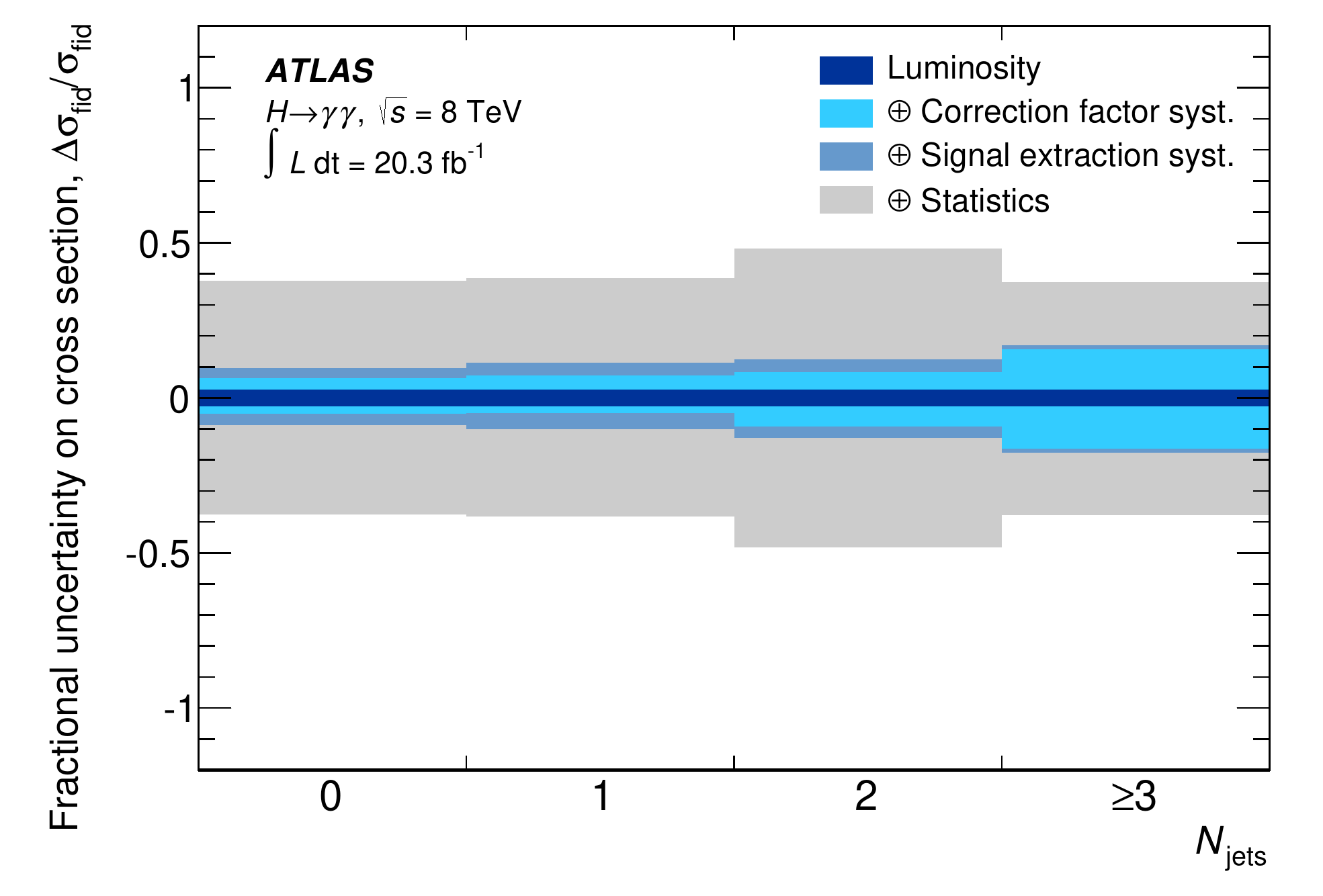}
    }
    \caption[]{The effect of systematic uncertainties associated with the signal extraction, the correction for detector effects (experimental and theoretical modelling) and the luminosity on the differential cross section as a function of (a) \ygg\ and (b) \njet. The statistical uncertainty associated with the signal extraction is also shown as a grey band.} 
    \label{fig:syst-breakdown}
  \end{center}
\end{figure}

The total systematic uncertainty is obtained from the sum in quadrature of the individual systematic uncertainties. A summary of the uncertainties on the measured fiducial cross sections are shown in table \ref{tab:syst-summary}.  Similarly, a breakdown of the systematic uncertainties on the differential cross sections as a function of  \ygg\ and \njet\ is shown in figure \ref{fig:syst-breakdown}. The variations on the fractional uncertainties derive from fluctuations of the yield, rather than from variations in the absolute size of the uncertainties. The dominant uncertainty is that of the signal extraction, which is primarily statistical in origin, although the jet energy scale and resolution uncertainties become increasingly important for high jet multiplicities and in the VBF-enhanced phase space.

\begin{table}
\centering
  \begin{tabular}{| l | c | c | c | c | c | }
  \hline
  Source  & \multicolumn{5}{c |}{Uncertainty on fiducial cross section (\%)} \\
  & Baseline & $N_{\rm jets} \geq 1$ & $N_{\rm jets} \geq 2$ & $N_{\rm jets} \geq 3$& VBF-\\ 
  & & & & & enhanced \\
  \hline     
  Signal extraction (stat.) & $\pm$22 & $\pm$25 & $\pm$30& $\pm$33 & $\pm$34 \\
  Signal extraction (syst.) & $\pm$6.5 & $\pm$7.4 & $\pm$7.1& $\pm$6.5 & $\pm$9.0 \\
  Photon efficiency & $\pm 1.5$ & $\pm 2.1$& $\pm 3.1$& $\pm 4.2$& $\pm 2.3$ \\
  Jet energy scale/resolution & - & ${}^{+6.2}_{-5.8}$& ${}^{+11}_{-10}$& ${}^{+15}_{-13}$& ${}^{+12}_{-11}$ \\
  JVF/pileup-jet  & - & $\pm$1.3& $\pm$2.2& $\pm$3.3& $\pm$0.5 \\ 
  Theoretical modelling & ${}^{+3.3}_{-1.0}$ &  ${}^{+5.0}_{-2.6}$&  $\pm 4.1$&  ${}^{+6.3}_{-4.9}$& ${}^{+2.2}_{-3.2}$ \\
  Luminosity     & $\pm$2.8 & $\pm$2.8 & $\pm$2.8 & $\pm$2.8 & $\pm$2.8 \\                             
  \hline
  \end{tabular}
  \caption{Uncertainties, expressed as percentages, on the cross sections measured in the baseline, $N_{\rm jets} \geq 1$, $N_{\rm jets} \geq 2$, $N_{\rm jets} \geq 3$ and  VBF-enhanced fiducial regions. The signal extraction systematic uncertainty contains the  effect of the photon energy scale and resolution, the impact of the background modelling on the signal yield and the uncertainty in the fitted peak position from the chosen background parameterisation.}
  \label{tab:syst-summary}
\end{table}

\section{Limit setting in the absence of a signal}
\label{sec:limit}

The extracted signal yields in the single-lepton and high-\met\ fiducial regions are consistent with zero and the data are used to place limits on the fiducial cross section in these regions. For each measurement the data are split into two categories, one of which contains those events that satisfy the baseline selection and are in the specified fiducial region and one that contains those events that are not. The diphoton spectrum in both categories are simultaneously fitted using the likelihood function given in equation (\ref{eq:emlmaster}), including systematic uncertainties on the photon energy scale and resolution as nuisance parameters. The agreement between the data and the expected yield for a hypothesised input cross section is quantified by the test statistic, $q$, defined as
\begin{equation}
  q  = \left\{
 \begin{array}{ccc}
 -2 \, \ln \, \tfrac{L(\mu^{\rm sig})}{L(\hat{\mu}^{\rm sig})}  & & 0 <  \hat{\mu}^{\rm sig} \le \mu^{\rm sig}  \\
 0 & &  \mu^{\rm sig} < \hat{\mu}^{\rm sig} \\
 \end{array}   \right. ,
\end{equation}
where $\hat{\mu}^{\rm sig} \ge 0 $ is the fitted cross section and $\mu^{\rm sig}$ is a given input cross section. The observed value of the test statistic, $q_{\rm obs}$, is determined from the ratio of the likelihood obtained by fixing the number of signal events to that predicted for a given cross section, to the likelihood obtained by allowing the number of signal events to float in the fit. An ensemble of pseudo-experiments is used  to determine the agreement between the data and a given input cross section and the background hypothesis. In each pseudo-experiment, a value of $q$ is calculated after selecting the predicted signal yield at random from a Poisson distribution with its mean determined by the input cross section. Systematic uncertainties associated with migrations into and out of the fiducial region are included by assuming that the uncertainties are Gaussian distributed. The 95\% confidence limit on the cross section is determined following the $CL_s$ prescription~\cite{Read:2002hq}, defined as the input cross section for which the fraction of pseudo-experiments that produce a value of $q$ that is smaller than $q_{\rm obs}$ is $0.95(1-p_{\rm b}) + p_{\rm b}$, taking into account the penalisation of the background hypothesis probability,~$p_{\rm b}$.

The Standard Model predicted cross section is $0.27\pm0.02$~fb and $0.14\pm0.01$~fb in the single-lepton and high-\met\ fiducial regions, respectively, estimated using the MC event generators presented in section~\ref{sec:mc}. The expected cross-section limit at 95\% confidence level in the single-lepton fiducial region is 1.23~fb, with a 68\% probability interval of [0.82,1.79]~fb. The expected cross-section limit at 95\% confidence level in the high-\met\ fiducial region is 1.06~fb, with a 68\% probability interval of [0.76,1.58]~fb. The systematic uncertainties degrade the limits by less than 5\% in total.

The fiducial cross-section limits are presented at particle level and are therefore sensitive to the modelling of underlying kinematic distributions, as a change in the shape of a distribution could change the amount of migrations into and out of the fiducial region. In practice, the presented limits are quite stable unless there is a sharply peaked (spiked) contribution from new physics at the boundary of the fiducial region. For example, a sharply peaked distribution at $\met \sim 80$~GeV results in the quoted limit corresponding to 90\% confidence level instead of 95\% confidence level. No such effect is observed for broad \met\ distributions or sharply peaked distributions away from boundary of the fiducial region.

\section{Theoretical predictions}
\label{sec:theory}

The most accurate theoretical predictions for Higgs boson production via gluon fusion in the baseline fiducial region are calculated at the parton level. The LHC Higgs cross section working group recommends using a calculation for the cross section of Higgs boson production via gluon fusion that is accurate to NNLO+NNLL in QCD and incorporates NLO electroweak corrections~\cite{Heinemeyer:2013tqa}. This is the prediction used by default in Higgs boson analyses at the LHC and is referred to as LHC-XS in the following discussion. More recently, a calculation of the cross section for Higgs boson production via gluon fusion was performed using soft and collinear effective theory~\cite{Stewart:2013faa}. This prediction, referred to as STWZ, is also accurate to NNLO+NNLL, but performs a different type of resummation and does not include any electroweak corrections. Both the LHC-XS and STWZ predictions are provided with uncertainties associated with renormalisation, factorisation and resummation scale variation, as well as an uncertainty from PDF variation. These predictions are corrected to the particle level to allow comparison to data, using diphoton acceptance, photon isolation and non-perturbative correction factors. 
The diphoton acceptance and photon isolation correction factors account for the decay of the Higgs boson to two isolated photons in the geometrical acceptance of the detector. They  are determined using \powheg+\pythia\ events with associated uncertainties from PDF and renormalisation/factorisation scale variations. The non-perturbative correction factors account for the impact of hadronisation and underlying event activity. They are defined as the ratio of cross sections produced with and without hadronisation and underlying event. The default non-perturbative correction factor is taken to be the centre of the envelope of correction factors obtained from multiple event generators and/or event generator tunes, with the uncertainty taken to be half of the envelope. The variations in non-perturbative correction factors were obtained using the AU2 (\pythia\ \cite{ATLASUE2}), UE-EE-4-LO (\herwig++ \cite{Gieseke:2003hm,Bellm:2013lba}) and AUET2B-LO, AUET2B-CTEQ6L1, AMBT2B-LO and AMBT-CTEQ6L1 (\pythiasix, \cite{ATLASUE}) tunes. The diphoton acceptance, photon isolation and non-perturbative correction factors are documented in appendix \ref{app:corr}. The $H\rightarrow \gamma\gamma$ branching ratio is taken to be $0.228\pm0.011$\%~\cite{Heinemeyer:2013tqa}. The total uncertainty on the theoretical predictions is taken to be the sum in quadrature  of the scale, PDF, branching ratio, diphoton acceptance, photon isolation and non-perturbative uncertainties.

For the differential distributions that probe the kinematics of the diphoton system, the \hres\ 2.2 calculation~\cite{deFlorian:2012mx,Grazzini:2013mca} is used to provide the prediction for Higgs boson production via gluon fusion. \hres\ is accurate to NNLO+NNLL in QCD but does not contain any electroweak corrections. 
The uncertainty associated with missing higher orders in the calculation is derived from the envelope of cross-section predictions obtained by simultaneously varying the renormalisation, factorisation and resummation scales by a factor of 0.5 or 2.0 (all combinations of scales are considered when forming the envelope, except those for which the renormalisation and factorisation scales differ by a factor of four). 
The uncertainty in the theoretical prediction from the choice of parton distribution function is estimated by (i) varying the \ctten\ eigenvectors and (ii) using the central values and uncertainties of two other PDF sets, {\sc MSTW2008nlo} \cite{Martin:2009iq} and {\sc NNPDF2.3} \cite{Ball:2012cx}. For each PDF set, the uncertainty on the cross section is calculated using the recommended procedure from each collaboration, with the \ctten\ results scaled to reflect 68\% probability, and the overall uncertainty is derived from the envelope of the individual uncertainties from each PDF set. The \hres\ calculation contains the decay products of the Higgs boson and is scaled to reproduce the default branching ratio of 0.228\%. The prediction is also corrected to the particle level to account for the small effect of photon isolation, using the photon isolation and non-perturbative correction factors determined independently for each bin of the differential distribution. The total uncertainty on the theoretical predictions is taken to be the sum in quadrature  of the scale, PDF, branching ratio, photon isolation and non-perturbative uncertainties.

For events containing one or more jets, a parton-level cross section has been calculated for Higgs boson production via gluon fusion using soft--collinear effective theory, by combining NNLO+NNLL zero-jet and NLO+NLL one-jet cross sections~\cite{Boughezal:2013oha} (referred to as BLPTW). A prediction for this fiducial cross section is also obtained at the parton level using the NNLO+NNLL prediction for the zero-jet efficiency provided by JetVHeto~\cite{Banfi:2012jm}. The BLPTW calculation also provides a prediction for the cross section for events containing two or more jets, which is accurate to approximate-NLO plus NLL in QCD.  The BLPTW and JetVHeto predictions are provided with uncertainties from renormalisation scale, factorisation scale, resummation scale and PDF variation. The parton level cross sections are corrected to the particle level to allow comparison to data, using diphoton acceptance, photon isolation and non-perturbative correction factors and accounting for the Higgs boson branching ratio to two photons. The total uncertainty on these predictions is taken to be the sum in quadrature  of the scale, PDF, branching ratio, diphoton acceptance, photon isolation and non-perturbative uncertainties.

The  cross section for Higgs boson production via gluon fusion in association with at least one jet (or at least two jets) can be calculated at NLO+LL accuracy in QCD using \minlohj\ (or \minlohjj). The uncertainties on each prediction associated with missing higher orders in the calculation is derived from the envelope of cross-section predictions obtained by simultaneously varying the renormalisation and factorisation scales by a factor of 0.5 or 2.0 (all combinations of scales are considered when forming the envelope, except those for which the renormalisation and factorisation scales differ by a factor of four). The uncertainty from the choice of parton distribution function is estimated in the same way as for \hres, taking the envelope of variations obtained using the \ctten\ eigenvectors and the central values and uncertainties of {\sc MSTW2008nlo} and {\sc NNPDF2.3}. The small uncertainties associated with non-pertubative modelling are included for both predictions, and are estimated in the same way as for the non-perturbative correction factors discussed above. \minlohj\ is also used for differential  distributions containing one or more jets and \minlohjj\ is used for differential distributions containing two or more jets. 

The contributions to the Standard Model predictions from VBF, $VH$ and $t\bar{t}H$ production are determined using the particle-level prediction obtained from the \powhegpyt\ and \pythia\ event generators, with the samples normalised to state-of-the-art theoretical calculations as discussed in section~\ref{sec:mc}. The uncertainty from scale and PDF variations on the VBF, $VH$ and $t\bar{t}H$ contributions are taken from these calculations, with an additional shape-dependent scale uncertainty derived for the VBF component by simultaneously varying the renormalisation and factorisation scale in the event generator by factors of 0.5 and~2.0.

\section{Fiducial cross section measurements and limits}
\label{sec:results_fid}

\begin{figure}[t]
  \begin{center}
      \includegraphics[width=0.9\textwidth]{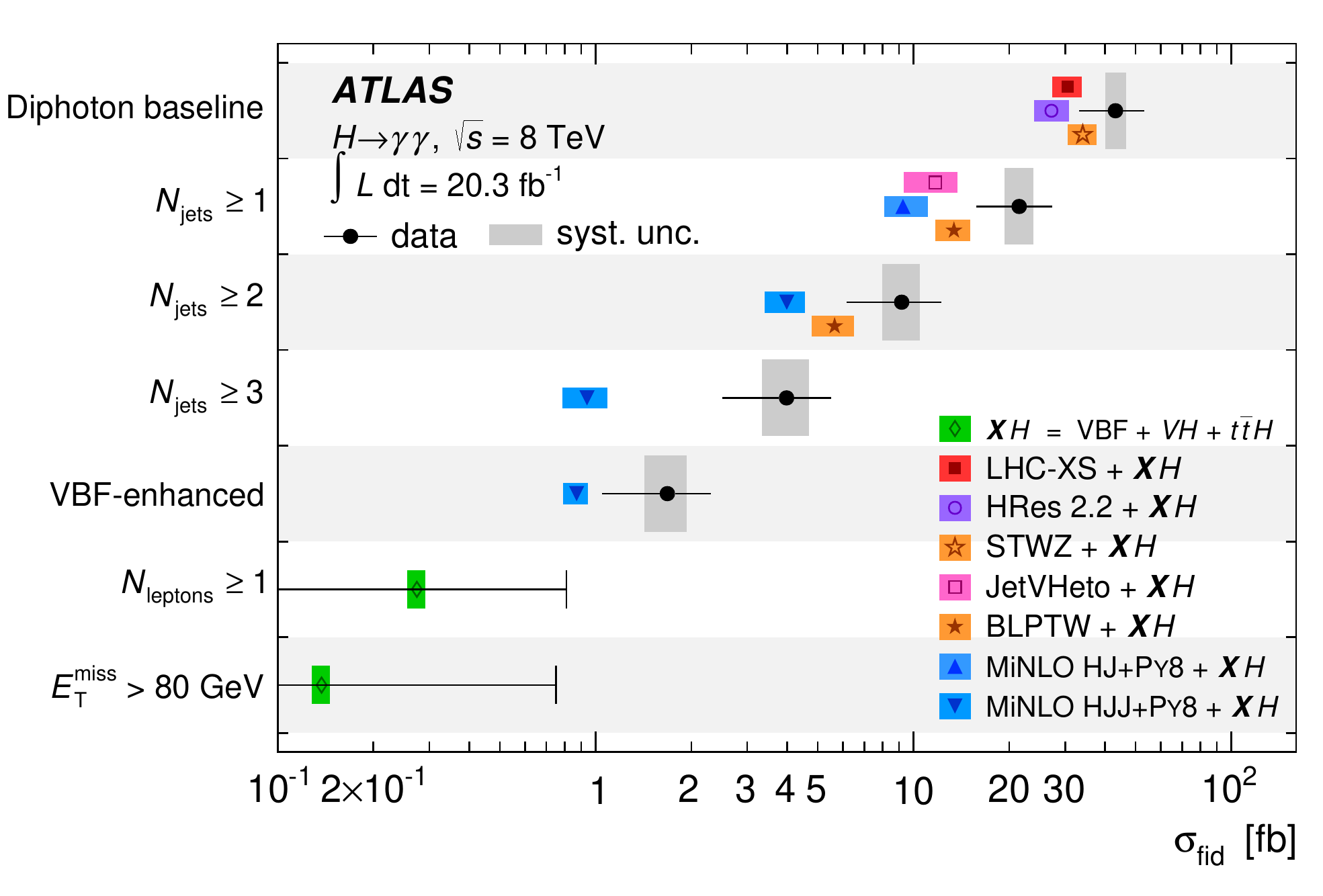}
    \caption[]{The measured cross sections and cross-section limits for $pp \rightarrow H\rightarrow \gamma\gamma$ in the seven fiducial regions defined in section~\ref{sec:evsel}. The intervals on the vertical axis each represent one of these fiducial regions. The data are shown as filled (black) circles. The error bar on each measured cross section represents the total uncertainty in the measurement, with the systematic uncertainty shown as dark grey rectangles. The error bar on each cross-section limit is shown at the 95\% confidence level. The data are compared to state-of-the-art theoretical predictions (see text for details). The width of each theoretical prediction represents the total uncertainty in that prediction. All regions include the SM prediction arising from VBF, $VH$ and $t\bar{t}H$, which are collectively labelled as $XH$. 
    } 
    \label{fig:fids}
  \end{center}
\end{figure}

The measured fiducial cross sections and cross-section limits are compared to a variety of theoretical predictions for SM Higgs boson production in Figure \ref{fig:fids}. The measured and predicted cross sections are also documented in table~\ref{tab:xs} and table \ref{tab:xsth}, respectively. The SM predictions are defined at the particle level and, in each fiducial region, are the sum of cross-section predictions for gluon fusion, VBF, $VH$ and $t\bar{t}H$, for $\mh = 125.4$~GeV, as discussed in section~\ref{sec:theory}.

\begin{table}[tb]
\centering
  \begin{tabular}{| l | c | }
  \hline
  Fiducial region & Measured cross section (fb) \\
  \hline
Baseline & $43.2 \pm   9.4 \, ({\rm stat.})  \,  {}^{+3.2}_{-2.9}  \, ({\rm syst.}) \pm     1.2 \, ({\rm lumi})$  \\        
$N_{\rm jets} \geq 1$ & $21.5 \pm    5.3 \, ({\rm stat.}) \,  {}^{+2.4}_{-2.2}  \,  ({\rm syst.}) \pm     0.6\, ({\rm lumi})$  \\
$N_{\rm jets} \geq 2$& $9.2 \pm    2.8 \, ({\rm stat.})    {}^{+1.3}_{-1.2} \, ({\rm syst.}) \pm     0.3 \, ({\rm lumi})$  \\
$N_{\rm jets} \geq 3$& $4.0 \pm    1.3 \, ({\rm stat.})  \, \pm 0.7 \, ({\rm syst.}) \pm     0.1\, ({\rm lumi})$  \\
VBF-enhanced  & $1.68 \pm    0.58\, ({\rm stat.})  {}^{+0.24}_{-0.25}  \, ({\rm syst.}) \pm     0.05 \, ({\rm lumi})$  \\
  \hline
$N_{\rm leptons} \geq 1$ & $< 0.80$   \\
$\met > 80$~GeV  & $< 0.74 $  \\
\hline
  \end{tabular}
  \caption{Measured cross sections in the baseline, $N_{\rm jets} \geq 1$, $N_{\rm jets} \geq 2$, $N_{\rm jets} \geq 3$ and VBF-enhanced fiducial regions, and cross-section limits at 95\% confidence level in the single-lepton and high-\met\ fiducial regions. The seven phase space regions are defined in section~\ref{sec:evsel}. 
  }
  \label{tab:xs}
\end{table}

\begin{table}[tb]
\centering
  \begin{tabular}{| l | c | c |}
  \hline
  Fiducial region & Theoretical prediction (fb) & Source \\
  \hline
Baseline & $30.5 \pm 3.3$   & LHC-XS~\cite{Heinemeyer:2013tqa} + $XH$ \\ 
 & $34.1 \, ^{+3.6}_{-3.5}$  & STWZ~\cite{Stewart:2013faa} + $XH$  \\      
 &  $27.2 \, ^{+3.6}_{-3.2}$ & \hres~\cite{Grazzini:2013mca} + $XH$ \\    
 \hline
$N_{\rm jets} \geq 1$ &  $13.8\pm 1.7$  & BLPTW~\cite{Boughezal:2013oha} + $XH$ \\
 &  $11.7\, ^{+2.0}_{-2.4}$ &  JetVHeto~\cite{Banfi:2012jm} + $XH$ \\
 & $9.3\, ^{+1.8}_{-1.2}$ & \minlohj\  + $XH$ \\
 \hline
$N_{\rm jets} \geq 2$ & $5.65 \pm 0.87$ &  BLPTW + $XH$ \\
 & $3.99 \, ^{+0.56}_{-0.59}$ &   \minlohjj\  + $XH$ \\
 \hline
$N_{\rm jets} \geq 3$ &  $0.94 \pm 0.15$ &  \minlohjj\  + $XH$\\
\hline
VBF-enhanced  &  $0.87 \pm 0.08$ & \minlohjj\  +  $XH$\\
\hline
$N_{\rm leptons} \geq 1$ & $0.27 \pm 0.02 $ &    $XH$\\
$\met > 80$~GeV  &  $0.14 \pm 0.01 $  &  $XH$\\
\hline
  \end{tabular}
  \caption{Theoretical predictions for the  cross sections in the baseline, $N_{\rm jets} \geq 1$, $N_{\rm jets} \geq 2$, $N_{\rm jets} \geq 3$, VBF-enhanced, single-lepton and high-\met\ fiducial regions. The uncertainties on the cross-section predictions are discussed in detail in Section~\ref{sec:theory} and include the effect of scale and PDF variation as well as the uncertainties on the $H\rightarrow\gamma\gamma$ branching ratio and non-perturbative modelling factors. The seven phase space regions are defined in section~\ref{sec:evsel}. The `$XH$' refers to the theoretical predictions for VBF, $VH$ and $t\bar{t}H$ derived using the \powhegpyt, and \pythia\ event generators discussed in section~\ref{sec:mc}.}
  \label{tab:xsth}
\end{table}

The cross section for  $pp \rightarrow H \rightarrow \gamma\gamma$ measured in the baseline fiducial region is
\begin{equation*} 
\sigma_{\rm fid} (pp\rightarrow H\rightarrow \gamma\gamma) =43.2 \pm   9.4 \, ({\rm stat.})  \,  {}^{+3.2}_{-2.9}  \, ({\rm syst.}) \pm     1.2 \, ({\rm lumi}) \, \, {\rm fb}. 
\end{equation*}
This can be compared with the Standard Model prediction for inclusive Higgs boson production of $30.5\pm 3.3$~fb, constructed using the LHC-XS prediction for the gluon fusion contribution.  
The ratio of the data to this theoretical prediction is $1.41\pm 0.36$, which is consistent with a dedicated measurement of the Higgs boson signal strength in the diphoton decay channel~\cite{higgscouplings}. The ratio of the data to the theoretical prediction obtained using STWZ or \hres\ for the gluon fusion contribution is $1.27 \pm 0.32$ and $1.59 \pm 0.42$, respectively.  
Although the measured cross section is larger than the range of theoretical predictions, the excess is not significant. 
The theoretical prediction obtained using \hres\ for the gluon fusion component is slightly smaller than the corresponding prediction based on LHC-XS,  because of missing electroweak and threshold resummation corrections (that enhance the gluon fusion contribution by a few percent~\cite{Heinemeyer:2013tqa}) and the use of different parton distribution functions (\ctten\ rather than {\sc MSTW2008nlo}). Conversely, the theoretical prediction obtained using STWZ for the gluon fusion component is slightly larger than the prediction based on LHC-XS, despite the missing electroweak corrections.\footnote{Recent theoretical predictions for Higgs boson production via gluon fusion at approximate-NNNLO accuracy in QCD give results that are similar to the STWZ prediction \cite{Ball:2013bra}.}

The measured cross section for events containing at least one jet is compared to three theoretical predictions. The theoretical predictions based on the BLPTW and JetVHeto calculations for the gluon fusion component of the cross section are in agreement with the data. For events containing at least two jets, the BLPTW-based prediction is in good agreement with the data. In both of these regions, the predictions obtained using \minlohj\ or \minlohjj\ for the gluon fusion component of the cross section give a slightly poorer description of the data, suggesting that the higher-order corrections included in the BLPTW and JetVHeto calculations are important. For events containing at least three jets in addition to the diphoton system, the prediction based on  \minlohjj\ is below the data by 2.1$\sigma$ significance. Finally, the measured cross section in the VBF-enhanced fiducial region is in satisfactory agreement with the theoretical prediction constructed from \minlohjj\ (gluon fusion) and \powheg\ (VBF). The VBF process makes up approximately 75\% of the cross section for a Standard Model Higgs boson in this region and the data to MC comparison is therefore sensitive to the modelling of the VBF process.

The 95\% confidence limits on the cross sections in the single-lepton and high-\met\ fiducial regions are 0.80~fb and 0.74~fb, respectively. These limits are $1\sigma$ below the corresponding expected limits of 1.23 fb and 1.06 fb, assuming the production of
Higgs bosons in association with leptons or \met\ follows the SM prediction, which is made up almost entirely from  $VH$ and $t\bar{t}H$ production. Although the limits are a factor of three to five larger than the SM prediction, they can be used to constrain models of Higgs boson production in association with dark matter or other exotic weakly interacting particles. 

\section{Differential cross sections}
\label{sec:results_diff}

\begin{figure}[tp]
  \begin{center}
    \subfigure[] {
      \includegraphics[width=0.46\textwidth]{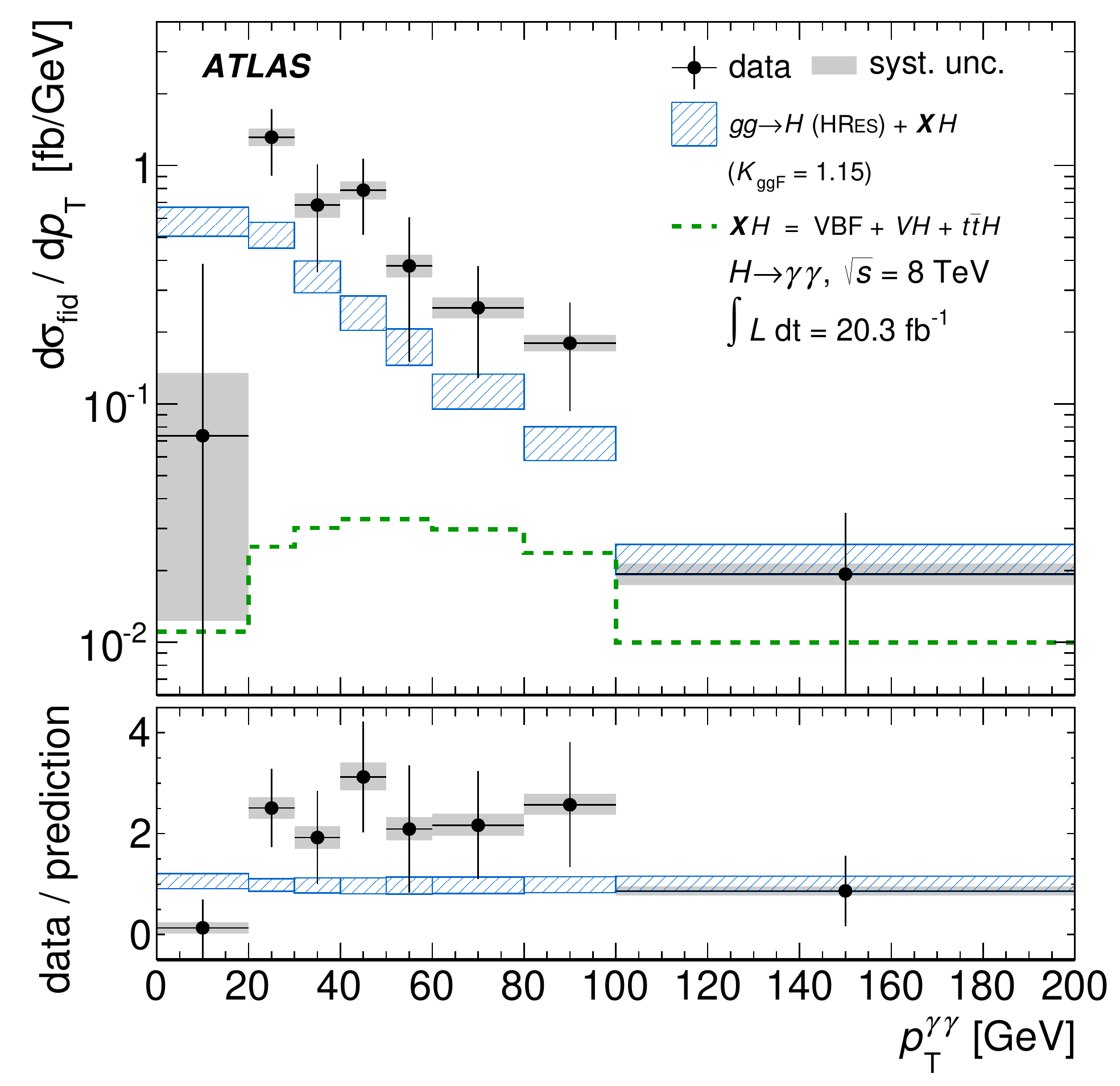} \quad
          }
    \subfigure[] {
      \includegraphics[width=0.46\textwidth]{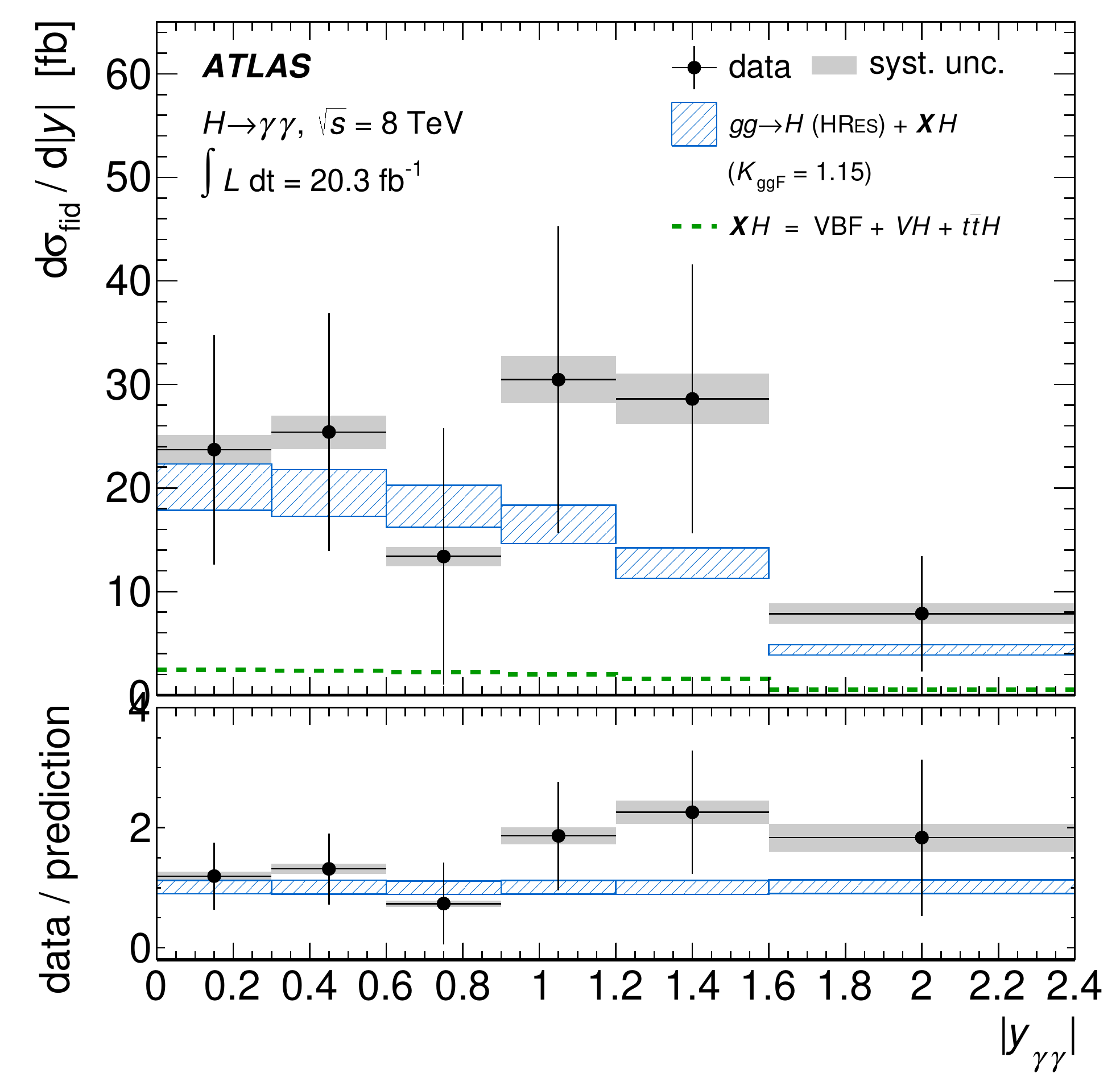}
         }
    \caption[]{The differential cross section for $pp \rightarrow H\rightarrow \gamma\gamma$ as a function of (a) the diphoton transverse momentum, \ptgg,  and (b) the absolute rapidity of the diphoton system, \ygg. The data are shown as filled (black) circles. The vertical error bar on each data point represents the total uncertainty in the measured cross section and the shaded (grey) band is the systematic component. The SM prediction, defined using the \hres\ prediction for gluon fusion and the default MC samples for the other production mechanisms, is presented as a hatched (blue) band, with the depth of the band reflecting the total theoretical uncertainty (see text for details). The small contribution from VBF, $VH$ and $t\bar{t}H$ is also shown separately as a dashed (green) line and denoted by $XH$. The \hres\ predictions are normalised to the total LHC-XS cross section~\cite{Heinemeyer:2013tqa} using a ${\rm K}$-factor of ${\rm K}_{\rm ggF}=1.15$.} 
    \label{fig:ptgg}
  \end{center}
\end{figure}

The differential cross sections, measured in the baseline fiducial volume defined by the kinematics of the two photons, are shown as a function of the diphoton transverse momentum and rapidity in figure \ref{fig:ptgg}. The data are compared to the SM prediction constructed from the \hres\ calculation for gluon fusion and the default MC samples for the other production mechanisms. The \hres\ calculation is normalised to the LHC-XS prediction using a ${\rm K}$-factor of ${\rm K}_{\rm ggF}=1.15$. The shapes of the distributions are satisfactorily described by the SM prediction, with an overall offset that is consistent with the cross-section measurement in the baseline fiducial region presented in the previous section. 

\begin{figure}[tp]
  \begin{center}
  \mbox{
    \subfigure[] {\includegraphics[width=0.46\textwidth]{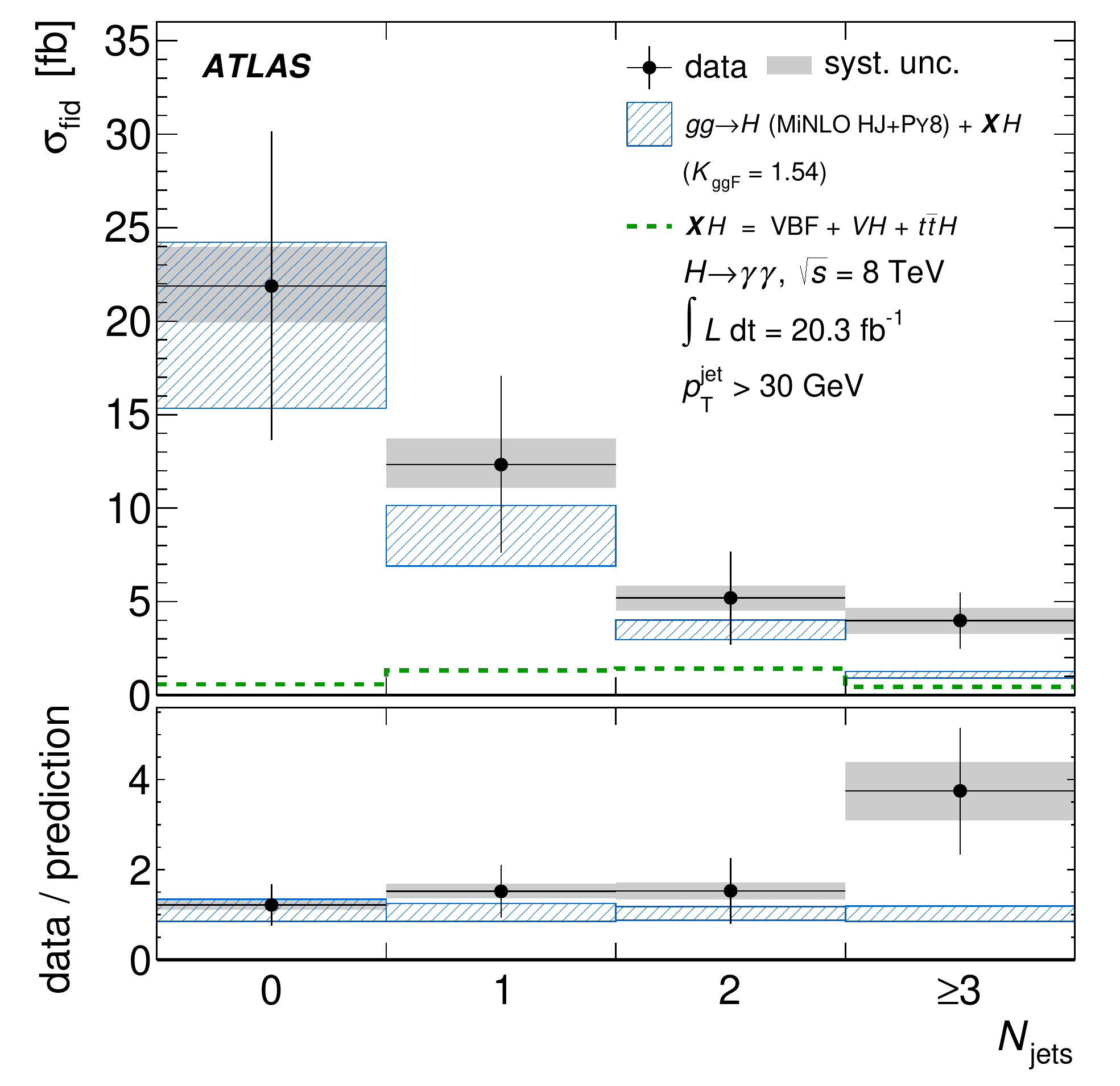}\quad}
    \subfigure[] {\includegraphics[width=0.46\textwidth]{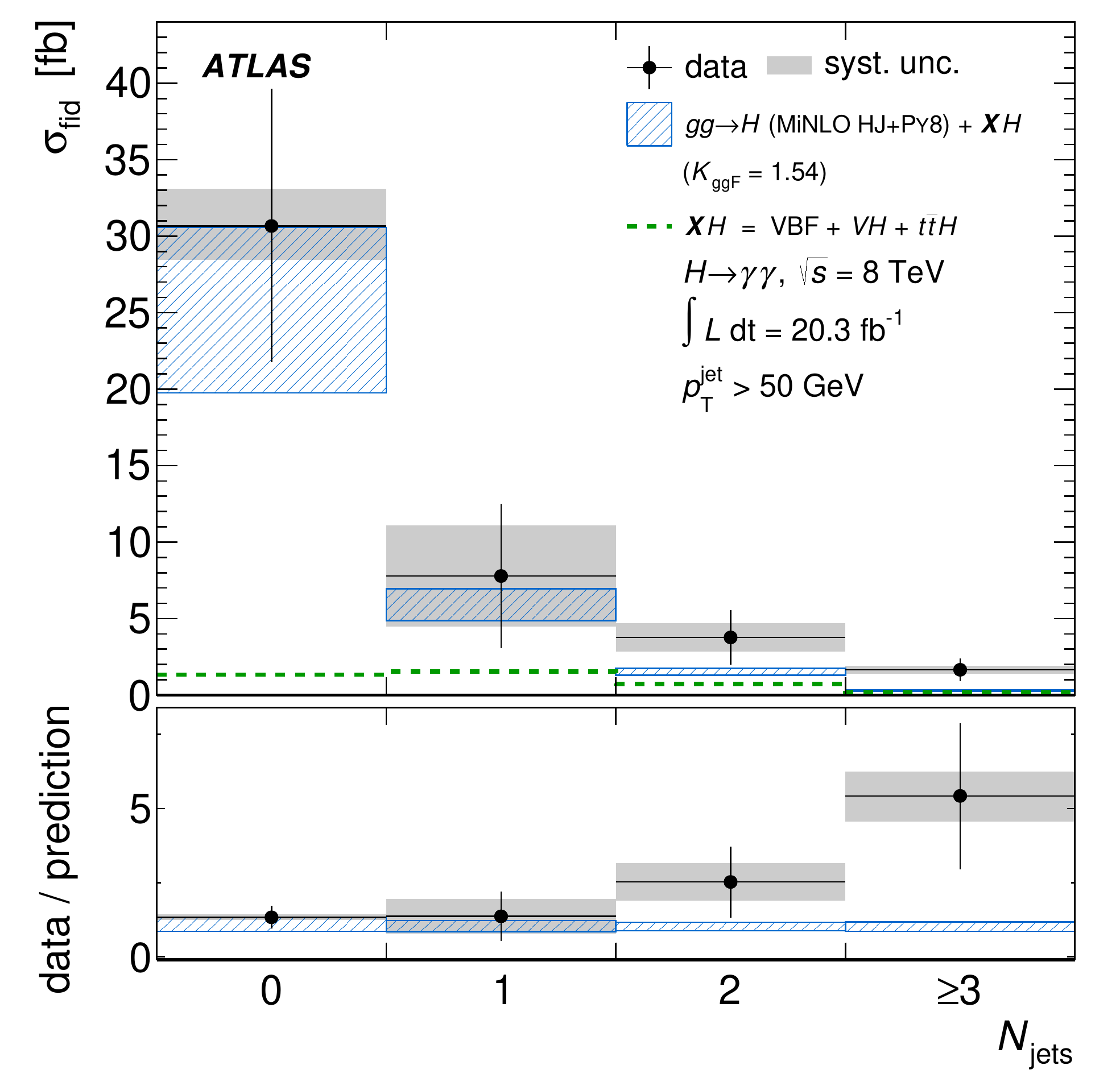}}
  }
  \caption[]{The differential cross section for $pp \rightarrow H\rightarrow \gamma\gamma$ as a function of (a) the jet multiplicity for $p_{\rm T}^{\rm jet} > 30$~GeV and (b) the jet multiplicity for $p_{\rm T}^{\rm jet} > 50$~GeV. The data and theoretical predictions are presented the same way as in figure~\ref{fig:ptgg}, although the SM prediction is now constructed using the \minlohj\ prediction for gluon fusion and the default MC samples for the other production mechanisms. The \minlohj\ prediction is normalised to the LHC-XS prediction using a ${\rm K}$-factor of ${\rm K}_{\rm ggF}=1.54$.
    }
    \label{fig:njet}
  \end{center}
\end{figure}

Figure~\ref{fig:njet} shows the differential cross section as a function of the jet multiplicity, which is calculated both for jets with $\pt > 30$~GeV and $\pt > 50$~GeV. The data are compared to the NLO+LL prediction provided by \minlohj\ for gluon fusion and the default MC samples for the other production mechanisms; the \minlohj\ prediction is normalised to the LHC-XS prediction using a ${\rm K}$-factor of ${\rm K}_{\rm ggF}=1.54$. The agreement between theory and data is satisfactory for both multiplicity distributions, with a non-significant excess of events in data at the highest jet multiplicities. The jet multiplicity distribution can be used to calculate the jet veto efficiency, which is defined as the fraction of the measured cross section that does not contain a jet with $\pt>30$~GeV. This variable directly tests the probability  of hard quark and gluon emission from inclusively produced Higgs boson events. The jet veto efficiency is measured to be 
$0.50 \, {}^{+0.10}_{-0.13} \, ({\rm stat.}) \pm 0.03 \, ({\rm syst.})$.  
This is approximately reproduced by the theoretical prediction from JetVHeto, which is $0.67 \pm 0.08$ for gluon fusion. The inclusion of all production mechanisms is expected to reduce the jet veto efficiency by approximately 0.06,  bringing the theoretical prediction into even better agreement with the data.

\begin{figure}[t]
  \begin{center}
  \mbox{
    \subfigure[]{\includegraphics[width=0.47\textwidth]{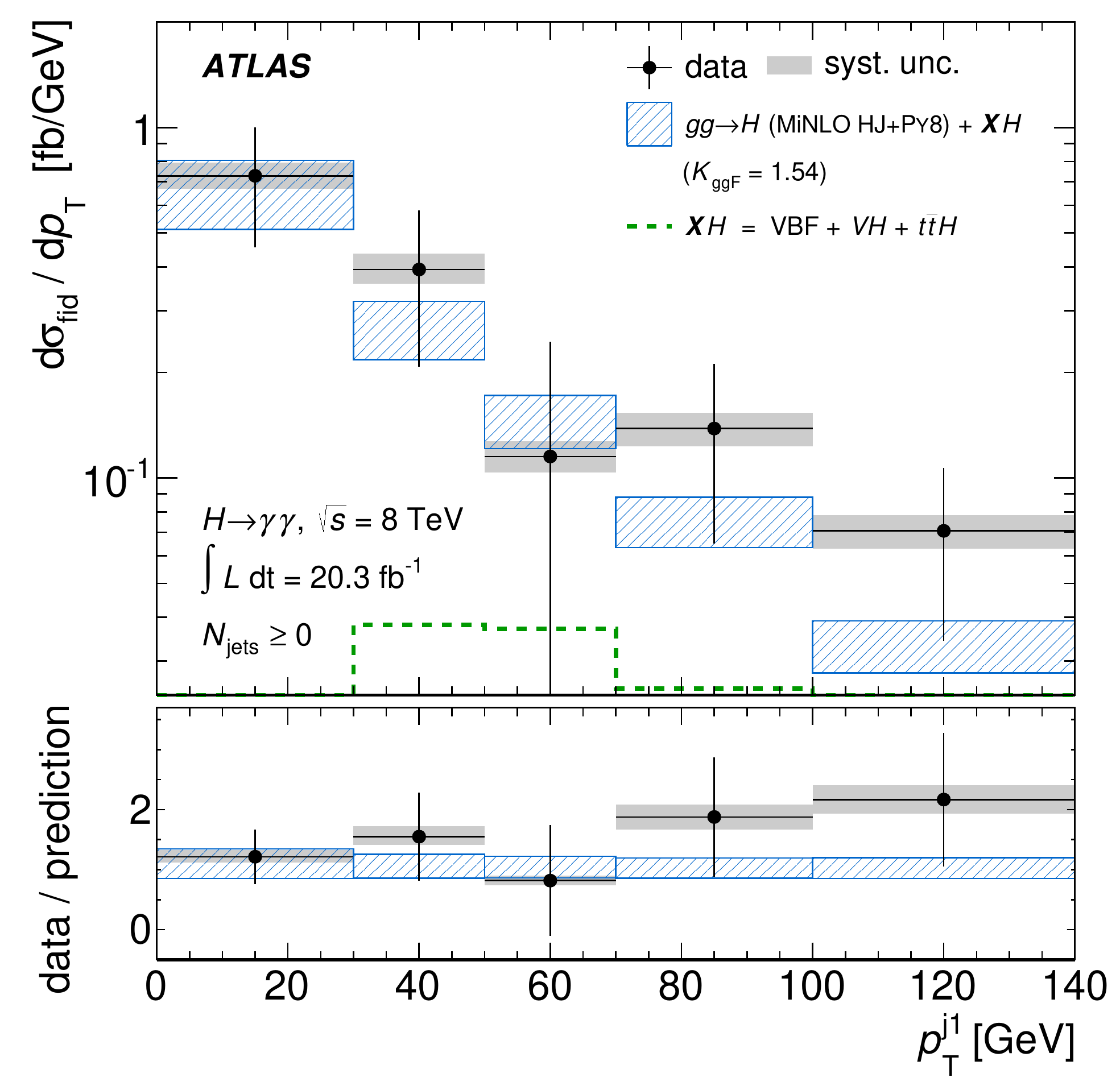}\quad}
    \subfigure[] {\includegraphics[width=0.47\textwidth]{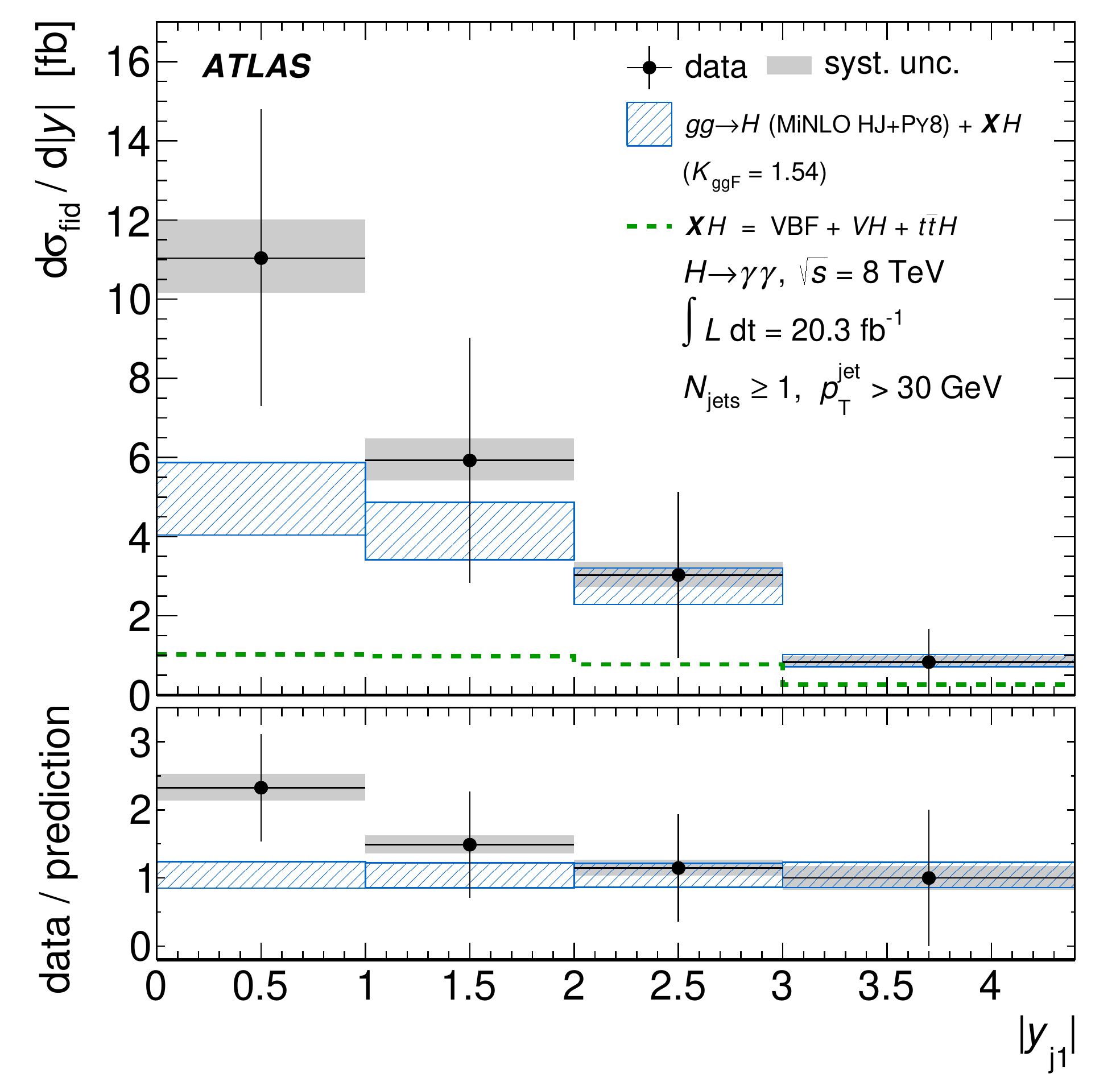}}
  }
  \vspace{0.5cm}
  \mbox{
      \subfigure[] {\includegraphics[width=0.47\textwidth]{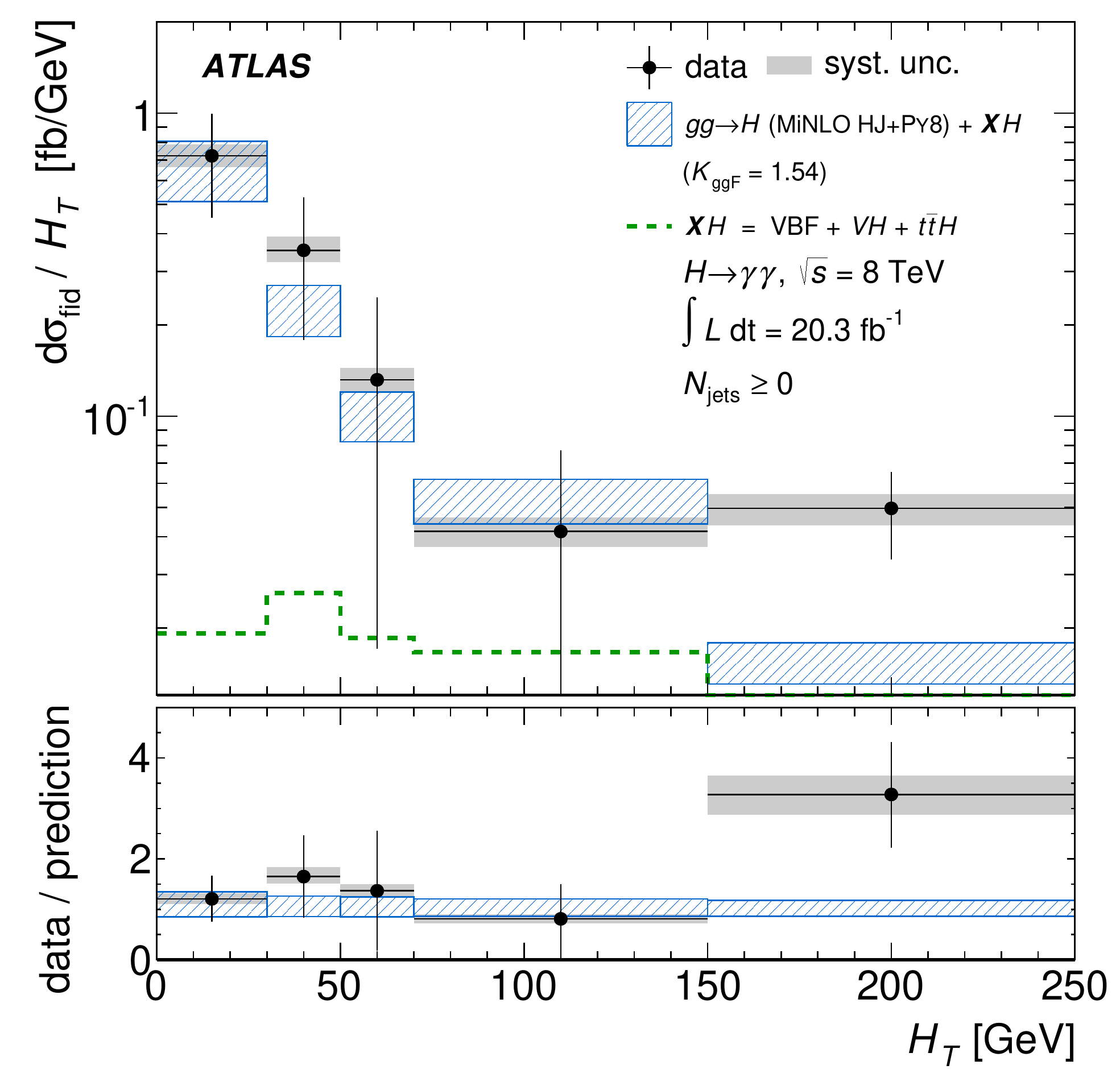}\quad}
    \subfigure[]{\includegraphics[width=0.47\textwidth]{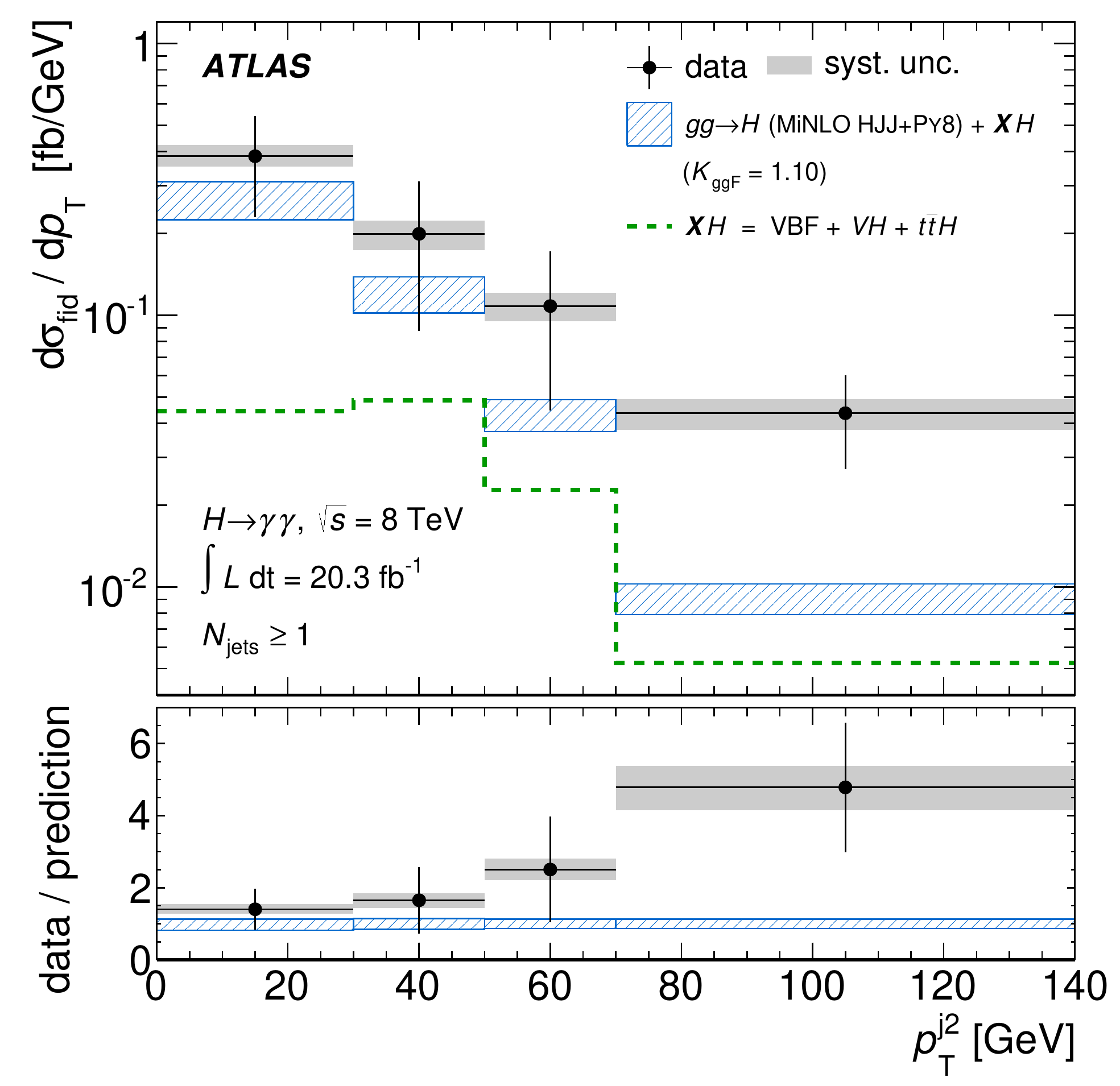}}

  }
  \caption[]{The differential cross section for $pp \rightarrow H\rightarrow \gamma\gamma$ as a function of (a) the leading jet transverse momentum, \ptjl,  (b) the leading jet absolute rapidity, \yjl, (c) the scalar sum of jet transverse momenta, $H_{\rm T}$, and (d) the subleading jet transverse momentum, \ptjsl.  The first bin in (a) and (c) represent 0-jet events that do not contain an additional jet with $\pt>30$~GeV. Similarly the first bin in (d) represents 1-jet events that do not contain an additional jet. The data and theoretical predictions are presented the same way as in figure~\ref{fig:ptgg}, although the SM prediction is now constructed using the \minlohj\ (or \minlohjj) prediction for gluon fusion and the default MC samples for the other production mechanisms. The \minlohj\ and \minlohjj\ predictions are normalised to the LHC-XS prediction using ${\rm K}$-factors of ${\rm K}_{\rm ggF}=1.54$ and ${\rm K}_{\rm ggF}=1.10$, respectively.
    }
    \label{fig:jetkin}
  \end{center}
\end{figure}

Figures \ref{fig:jetkin}(a) and \ref{fig:jetkin}(b) show the differential cross section as a function of the leading jet's transverse momentum and rapidity, respectively. Figure \ref{fig:jetkin}(c) shows the differential cross section as a function of $H_{\rm T}$. The shape of all these  distributions are in good agreement with the prediction provided by \minlohj\ for gluon fusion and the default MC samples for the other production mechanisms.  Figure \ref{fig:jetkin}(d) shows the differential cross section as a function of the subleading jet transverse momentum, the shape of which is satisfactorily described by the theoretical predictions provided by \minlohjj\ for gluon fusion and the default MC samples for the other production mechanisms. The \minlohjj\ prediction is normalised to the LHC-XS prediction using a ${\rm K}$-factor of ${\rm K}_{\rm ggF}=1.10$.

\begin{figure}[t]
  \begin{center}
    \subfigure[] {
      \includegraphics[width=0.47\textwidth]{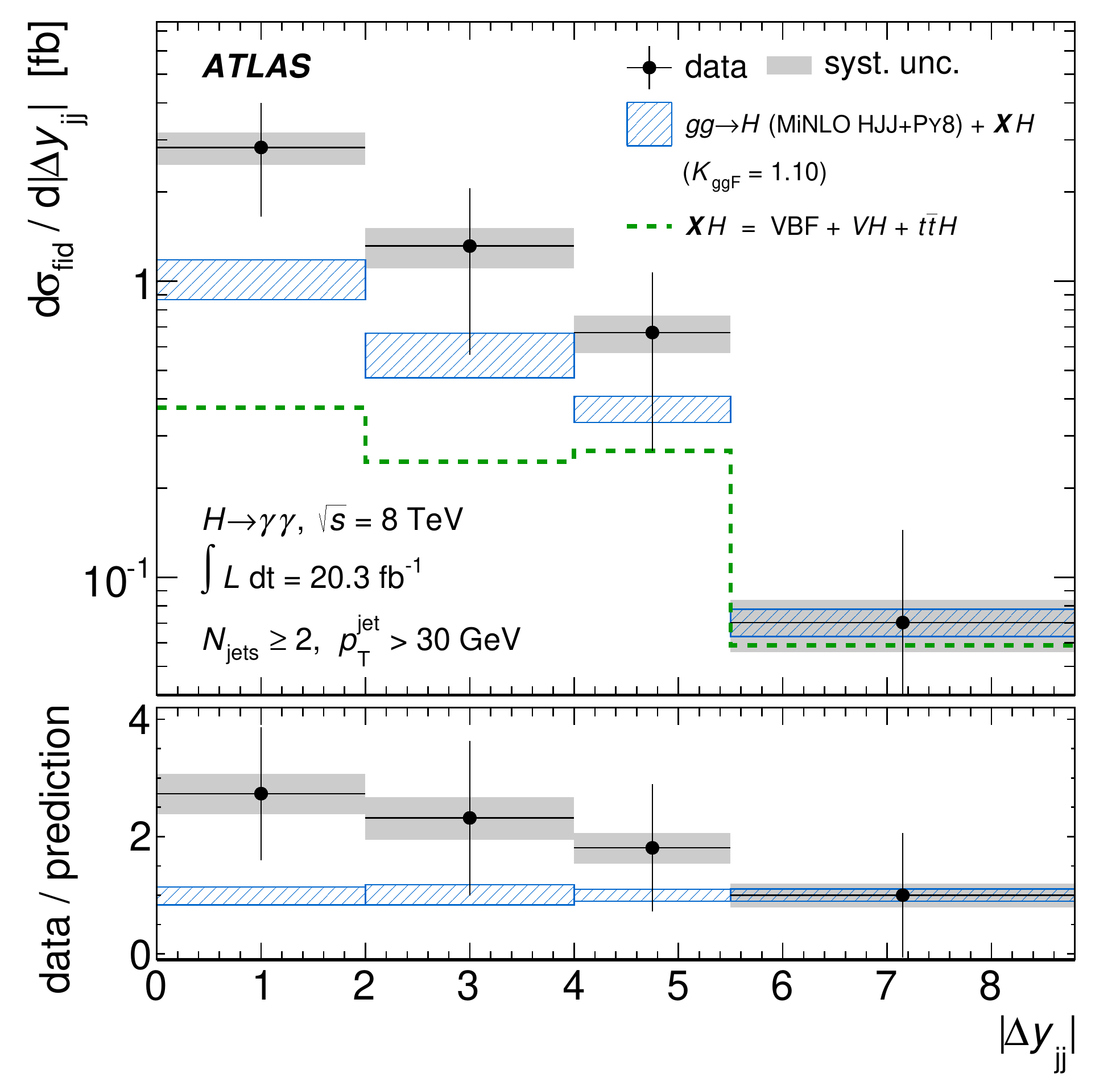}\quad
          }
    \subfigure[] {
      \includegraphics[width=0.47\textwidth]{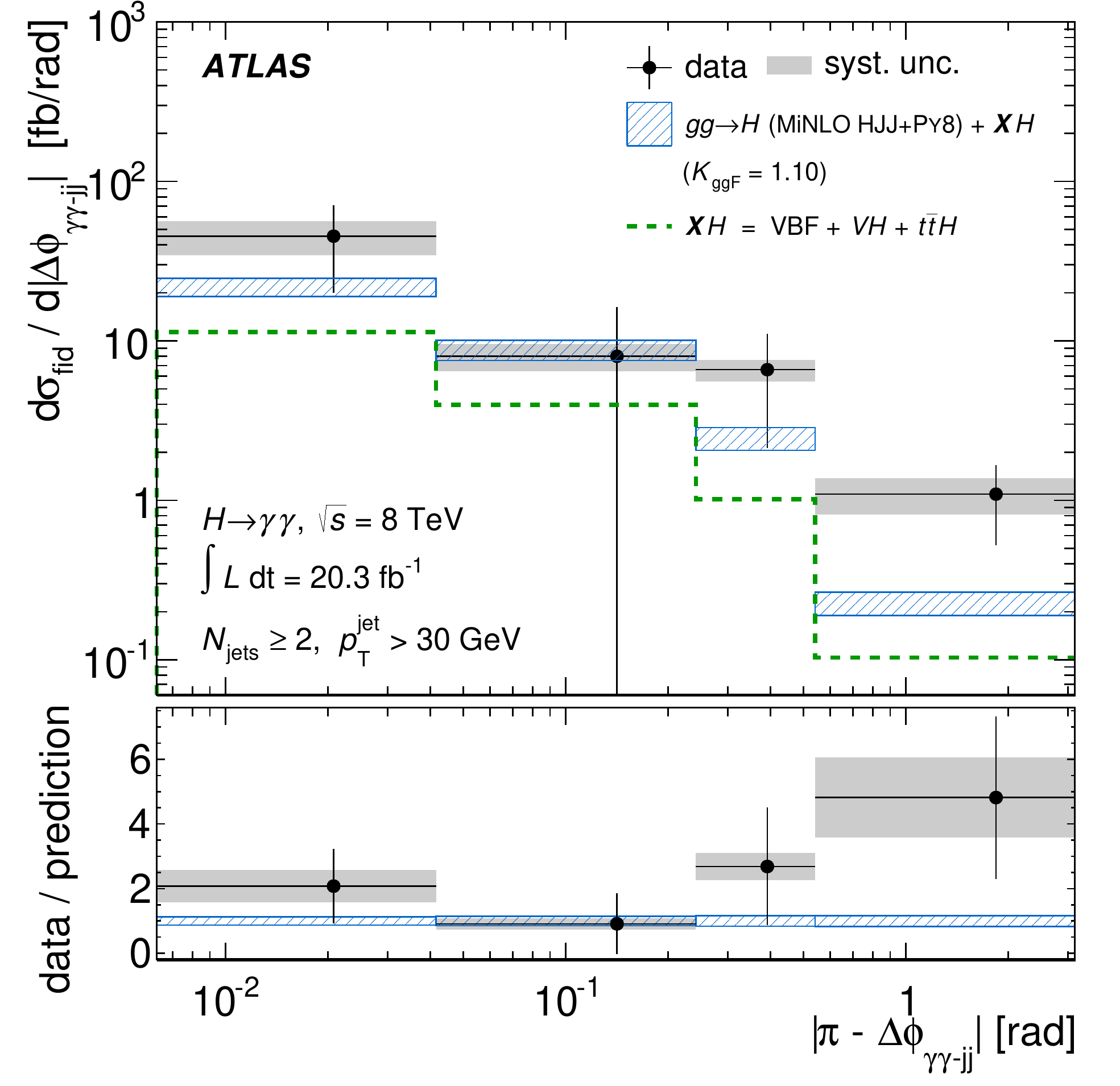}
         }
    \caption[]{The differential cross section for $pp \rightarrow H \rightarrow \gamma\gamma$ as a function of (a) the dijet rapidity separation, \deltayjj, and (b) the azimuthal angle between the dijet and diphoton systems presented as $|\pi - \Delta\phi_{\gamma\gamma,jj}|$. The data and theoretical predictions are presented the same way as in figure~\ref{fig:ptgg}, although the SM prediction is now defined using the \minlohjj\ prediction for gluon fusion and the default MC samples for the other production mechanisms. The \minlohjj\ prediction is normalised to the LHC-XS prediction using a ${\rm K}$-factor of ${\rm K}_{\rm ggF}=1.10$.} 
    \label{fig:mjj}
  \end{center}
\end{figure}

\begin{figure}[t]
  \begin{center}
    \subfigure[] {
      \includegraphics[width=0.47\textwidth]{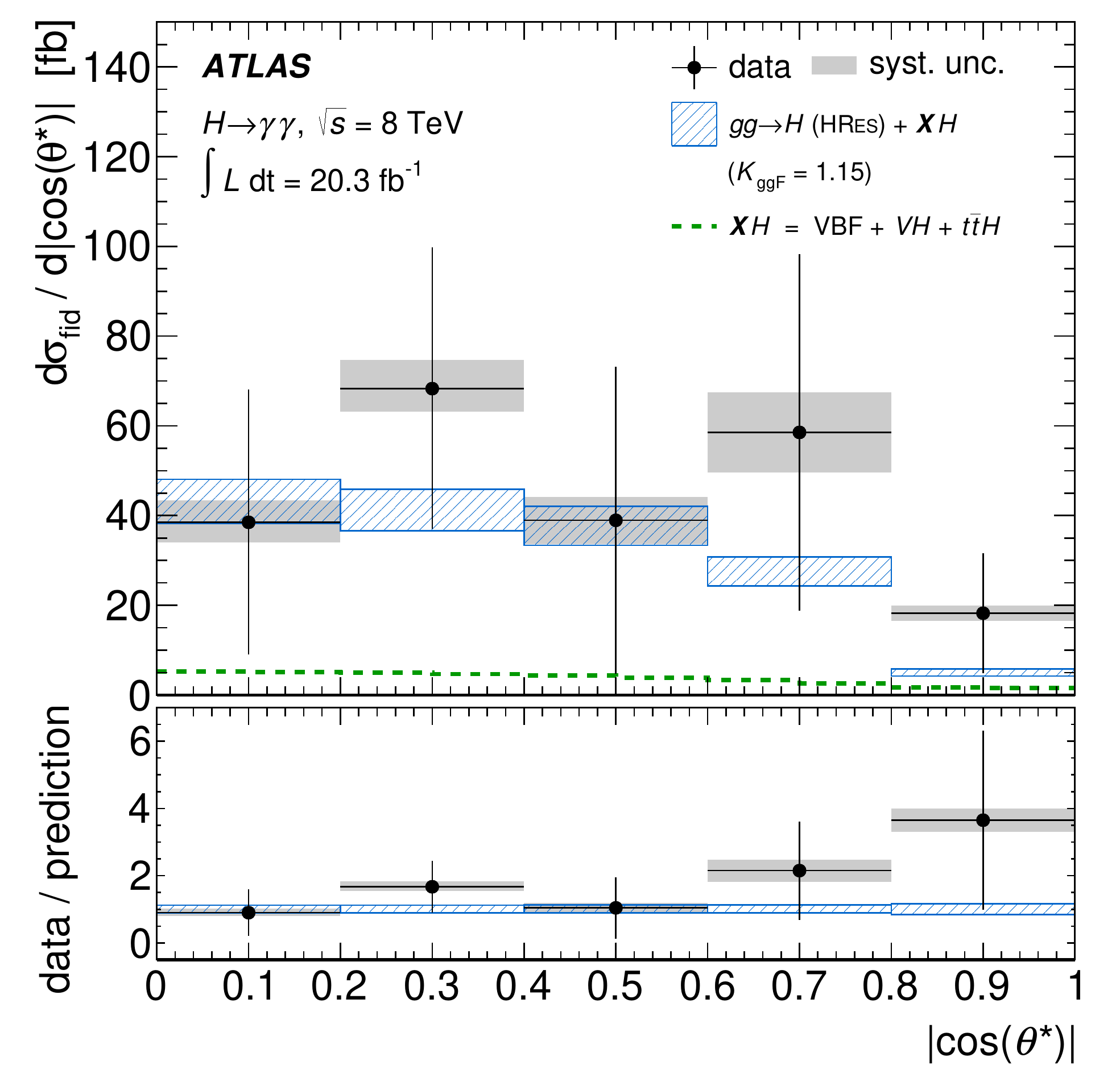}\quad
          }
    \subfigure[] {
      \includegraphics[width=0.47\textwidth]{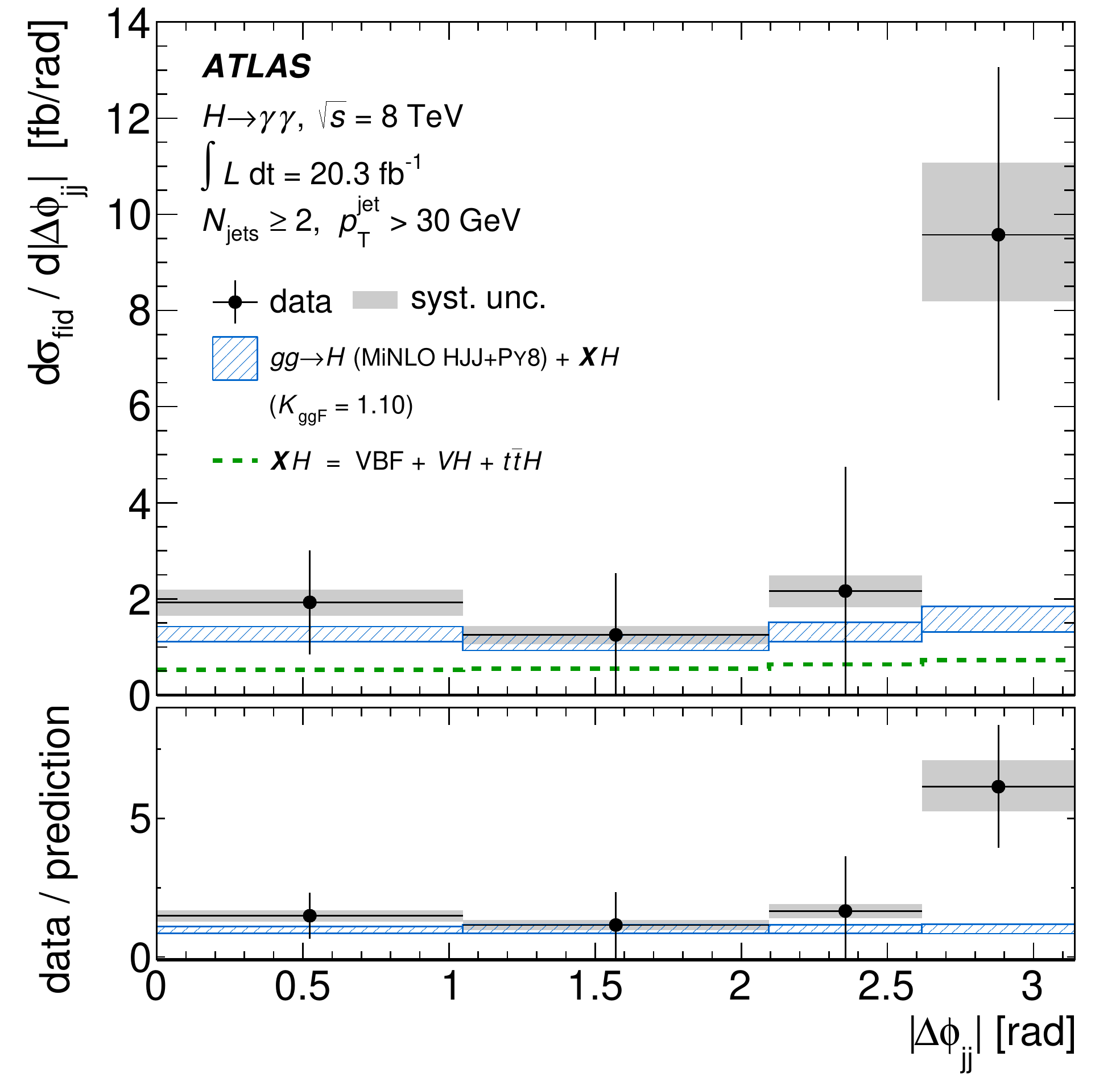} 
         }
    \caption[]{The differential cross section as a function of (a) the cosine of the photon decay angle in the Collins--Soper frame, \costhetastar, and (b) the azimuthal angle between the highest transverse momentum jets in events containing two or more jets, \dphijj.  The data and theoretical predictions are presented the same way as in figure~\ref{fig:ptgg}, although the SM prediction in (b) is now defined using the \minlohjj\ prediction for gluon fusion and the default MC samples for the other production mechanisms. The \hres\ and \minlohjj\ predictions are normalised to the LHC-XS prediction using ${\rm K}$-factors of ${\rm K}_{\rm ggF}=1.15$ and ${\rm K}_{\rm ggF}=1.10$, respectively.} 
    \label{fig:spinCP}
  \end{center}
\end{figure}

The differential cross sections as a function of the dijet rapidity separation, \deltayjj, and the azimuthal angle between the diphoton and dijet system, \dphiggjj, for events containing two or more jets, are shown in figure~\ref{fig:mjj}. These are standard variables used to discriminate between gluon fusion and vector-boson fusion production of the Higgs boson at the LHC~\cite{Aad:2013wqa}. The data are compared to the SM prediction provided by \minlohjj\ for gluon fusion and the default MC samples for the other production mechanisms. The shape of the SM prediction is in satisfactory agreement with the data.

The differential cross section as a function of the cosine of the photon decay angle in the Collins--Soper frame, \costhetastar, is shown in figure \ref{fig:spinCP}(a). This distribution is sensitive to the spin of the Higgs boson. The data are compatible with the results of earlier dedicated spin studies~\cite{Aad:2013xqa}, where the signal yields were extracted under the assumption of a particular spin hypothesis and not corrected for detector effects. The data are compared to the SM prediction defined using the \hres\ prediction for gluon fusion and the default MC samples for the other production mechanisms. The SM prediction is in good agreement with the data.

\begin{figure}[t]
  \begin{center}
  
    \subfigure[] {
      \includegraphics[width=0.47\textwidth]{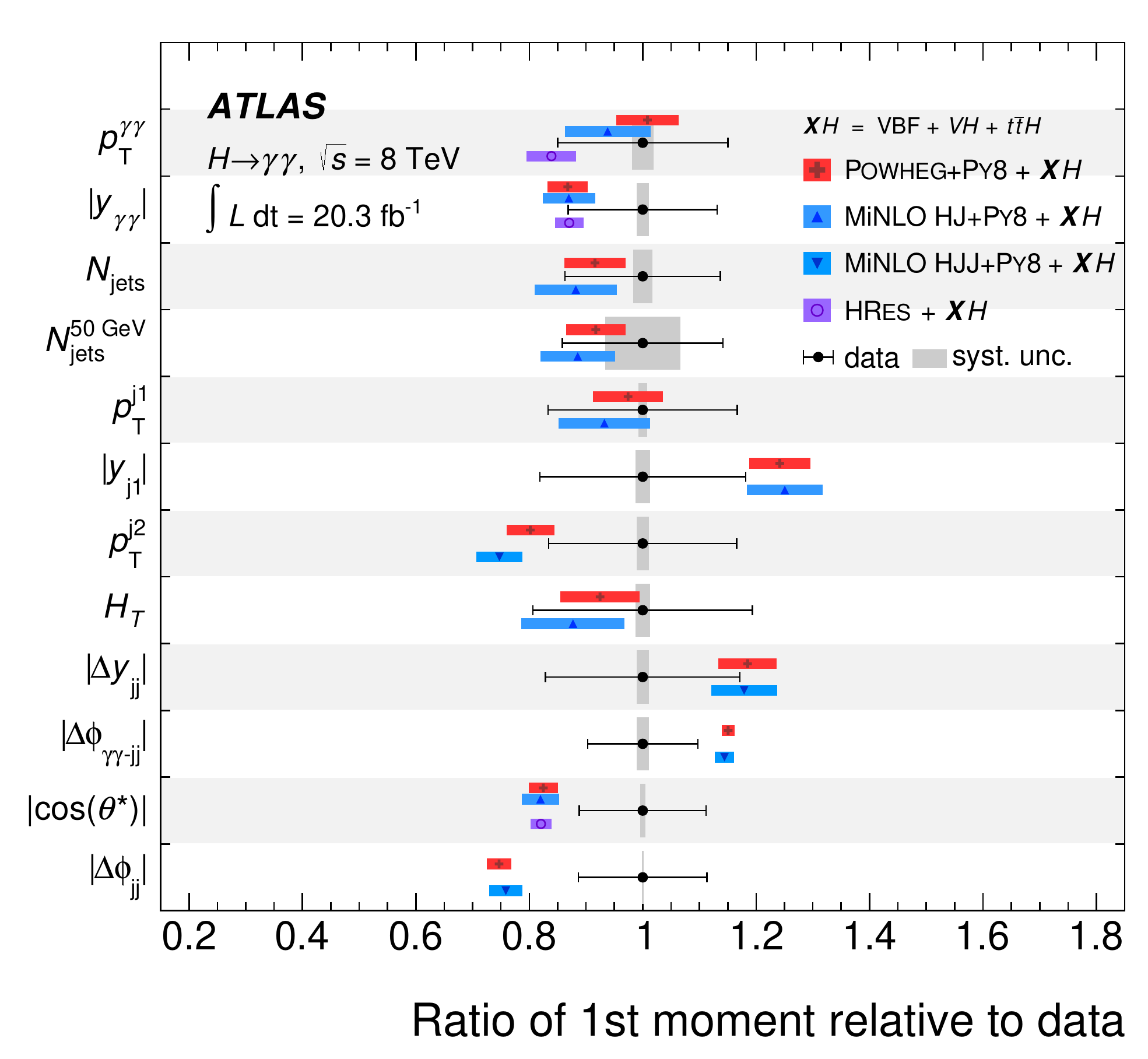}
          }
    \subfigure[] {
      \includegraphics[width=0.47\textwidth]{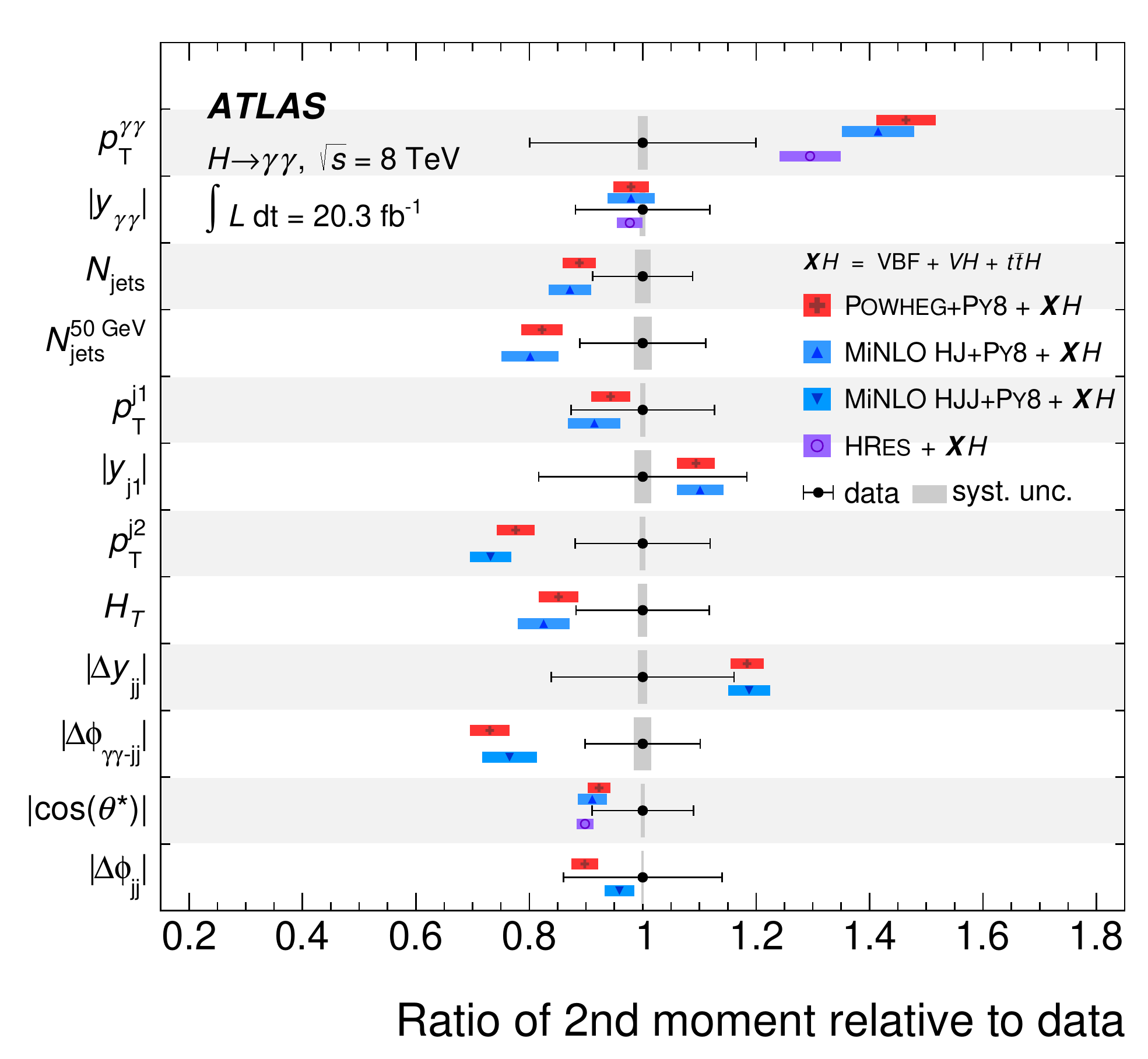}
         }
    \caption[]{(a) The ratio of the first moment (mean) of each differential distribution predicted by the theoretical models to that observed in the data.  (b) The ratio of the second moment (RMS) of each differential distribution predicted by the theoretical models to that observed in the data. The intervals on the vertical axes each represent one of the differential distributions. The band for each theoretical prediction represents the corresponding uncertainty in that prediction (see text for details). The error bar on the data represents the total uncertainty in the measurement, with the grey band representing the systematic-only uncertainty. } 
    \label{fig:summary}
  \end{center}
\end{figure}

The differential cross section as a function of the azimuthal angle between the jets in events containing two or more jets is shown in figure \ref{fig:spinCP}(b). The data are compared to the SM prediction defined using the \minlohjj\ prediction for gluon fusion and the default MC samples for the other production mechanisms. There is an upward deviation in data with respect to the SM prediction in the bin at $\dphijj  \sim \pi$, with an associated  significance of 2.3$\sigma$. This deviation remains present if the azimuthal angle between the jets is constructed using only central jets ($|y|<2.4$) with an increased JVF cut, which suggests that pileup is not responsible for the additional back-to-back jets. Similarly, the contribution of double parton scattering to $H+2 \, {\rm jet}$ production was estimated to be just 1.3\%, using the effective area parameter for double parton scattering measured in $W+2 \, {\rm jet}$ events at ATLAS~\cite{Aad:2013bjm}.

The azimuthal angle between the jets is sensitive to the charge conjugation and parity properties of the Higgs boson interactions. For example, in gluon fusion, a CP-even coupling has a dip at $\pi/2$ and peaks at $0$ and $\pi$, whereas a purely CP-odd coupling would present as a peak at $\pi/2$ and dips at $0$ and $\pi$ \cite{Klamke:2007cu,Andersen:2010zx, Dolan:2014upa}. For VBF, the SM prediction is approximately flat with a slight rise towards $\dphijj =\pi$ \cite{Plehn:2001nj}. Any additional anomalous CP-even or CP-odd contribution to the interaction between the Higgs boson and weak bosons would manifest itself as an additional oscillatory component, and any interference between the SM and anomalous couplings can produce distributions peaked at either $\dphijj =0$ or $\dphijj =\pi$ \cite{Plehn:2001nj}. The shape of the distribution is therefore sensitive to the relative contribution of gluon fusion and vector-boson fusion, as well as the tensor structure of the interactions between the Higgs boson and gluons or weak bosons. To further quantify the structure of the azimuthal angle between the two jets, an asymmetry is defined as
\begin{equation}
A_{\Delta\phi} = \frac{ \sigma(|\Delta\phi|<\frac{\pi}{3})  -  \sigma(\frac{\pi}{3} < |\Delta\phi|<\frac{2\pi}{3}) + \sigma(|\Delta\phi|>\frac{2\pi}{3}) }{\sigma(|\Delta\phi|<\frac{\pi}{3}) +  \sigma(\frac{\pi}{3} < |\Delta\phi|<\frac{2\pi}{3}) + \sigma(|\Delta\phi|>\frac{2\pi}{3})}
\end{equation}
which is motivated by a similar variable presented elsewhere~\cite{Andersen:2010zx}. The measured asymmetry in data is $A_{\Delta\phi}  = 0.72 \, {}^{+0.23}_{-0.29} \, ({\rm stat.}) {}^{+0.01}_{-0.02} \, ({\rm syst.})$. This can be compared to the Standard Model prediction of $A_{\Delta\phi}^{\rm SM}  = 0.43 \pm 0.02$, which is constructed from the \minlohjj\ prediction for gluon fusion and the standard VBF, $VH$ and $t\bar{t}H$ predictions using the event generators presented in section~\ref{sec:mc}. The uncertainty in this prediction includes scale and PDF uncertainties  for the gluon fusion and VBF components, plus an added uncertainty for gluon fusion which is derived from the envelope of predictions obtained from \minlohj, \minlohjj\ and \sherpa. The SM prediction is in agreement with the data. 

Figure \ref{fig:summary} shows the first and second moments of each differential distribution compared to a variety of theoretical predictions obtained from the MC event generators.  In general, the event generator predictions are in good agreement with the data. 
The increased jet activity and harder jet transverse momentum spectra suggest that there is more quark and gluon radiation in the data than in the theoretical predictions. 
However, the variables are correlated so this increase is not significant. The theoretical modelling is further explored for each of the differential distributions by performing a $\chi^2$ comparison with data in table \ref{tab:chi2table}. There is satisfactory agreement,  within statistical uncertainties, between theory and data for all $\chi^2$ tests.

The results presented in this section are published in HEPDATA \cite{hepdata}, with a complete breakdown of the uncertainties and their correlations, and a RIVET analysis routine is provided \cite{rivet}. The differential cross sections as a function of other variables have also been measured and are documented in appendix \ref{app:extra} and in HEPDATA. Each of these additional variables shows a high degree of correlation with the variables presented in this section. 

\begin{table}[tb]
\begin{center}
\small 
\begin{tabular}{ c | c c c c} 
\hline \hline 
 Variable & \powheg{}& \minlohj{}& \minlohjj{}& \hres{}\\ 
\hline 
\ptgg{} & 0.12& 0.10& 0.09& 0.12\\ 
\ygg & 0.81& 0.83& 0.83& 0.80\\ 
\costhetastar & 0.59& 0.57& 0.58& 0.56\\ 
$N_{\rm jets}$ & 0.42& 0.36& 0.30& - \\ 
$N_{\rm jets}^{\rm 50~GeV}$ & 0.33& 0.33& 0.30& - \\ 
\htj{} & 0.43& 0.39& 0.34& - \\ 
\ptjl{} & 0.84& 0.82& 0.79& - \\ 
\yjl & 0.64& 0.58& 0.51& - \\ 
\ptjsl{} & 0.34& 0.29& 0.23& - \\ 
\dphijj & 0.21& 0.28& 0.24& - \\ 
\deltayjj & 0.64& 0.58& 0.49& - \\ 
\dphiggjj & 0.45& 0.46& 0.42& - \\ 
\hline \hline 
\end{tabular} 
\caption{Probabilities from $\chi^2$ tests for the agreement between the differential cross section measurements and the theoretical predictions. Each prediction is normalised to the LHC-XS cross section before selection.\label{tab:chi2table}} 
\end{center} 
\end{table} 

\section{Summary and conclusion}
\label{sec:summary}

Measurements of  cross sections for Higgs boson production were presented in the diphoton decay channel for proton--proton collisions at a centre-of-mass energy of $\sqrt{s}=8$~TeV. The data were recorded by the ATLAS experiment at the CERN Large Hadron Collider and correspond to an integrated luminosity of 20.3~fb$^{-1}$. The data were corrected for detector inefficiency and resolution and are published in HEPDATA. The  $pp\rightarrow H \rightarrow \gamma\gamma$ cross section was measured to be 
$$43.2 \pm   9.4 \, ({\rm stat.})  \,  {}^{+3.2}_{-2.9}  \, ({\rm syst.}) \pm     1.2 \, ({\rm lumi}) \, \, {\rm fb},$$ 
for a Higgs boson of mass 125.4~GeV decaying to two isolated 
photons with transverse momentum greater than 35\% (25\%) of the diphoton invariant mass and have absolute pseudorapidity less than 2.37. Four additional fiducial cross sections and two cross-section limits were also presented. In addition, twelve differential cross sections were measured within the baseline fiducial volume defined by the kinematics of the two photons. Collectively, these measurements probe the Higgs boson kinematics, the jet activity produced in association with the Higgs boson, and the prevalence of vector-boson fusion, as well as the spin, charge conjugation and parity nature of the Higgs boson. In all cases, the data are in agreement with Standard Model expectations.

\section*{Acknowledgements}
We thank CERN for the very successful operation of the LHC, as well as the
support staff from our institutions without whom ATLAS could not be
operated efficiently.

We acknowledge the support of ANPCyT, Argentina; YerPhI, Armenia; ARC,
Australia; BMWF and FWF, Austria; ANAS, Azerbaijan; SSTC, Belarus; CNPq and FAPESP,
Brazil; NSERC, NRC and CFI, Canada; CERN; CONICYT, Chile; CAS, MOST and NSFC,
China; COLCIENCIAS, Colombia; MSMT CR, MPO CR and VSC CR, Czech Republic;
DNRF, DNSRC and Lundbeck Foundation, Denmark; EPLANET, ERC and NSRF, European Union;
IN2P3-CNRS, CEA-DSM/IRFU, France; GNSF, Georgia; BMBF, DFG, HGF, MPG and AvH
Foundation, Germany; GSRT and NSRF, Greece; ISF, MINERVA, GIF, I-CORE and Benoziyo Center,
Israel; INFN, Italy; MEXT and JSPS, Japan; CNRST, Morocco; FOM and NWO,
Netherlands; BRF and RCN, Norway; MNiSW and NCN, Poland; GRICES and FCT, Portugal; MNE/IFA, Romania; MES of Russia and ROSATOM, Russian Federation; JINR; MSTD,
Serbia; MSSR, Slovakia; ARRS and MIZ\v{S}, Slovenia; DST/NRF, South Africa;
MINECO, Spain; SRC and Wallenberg Foundation, Sweden; SER, SNSF and Cantons of
Bern and Geneva, Switzerland; NSC, Taiwan; TAEK, Turkey; STFC, the Royal
Society and Leverhulme Trust, United Kingdom; DOE and NSF, United States of
America.

The crucial computing support from all WLCG partners is acknowledged
gratefully, in particular from CERN and the ATLAS Tier-1 facilities at
TRIUMF (Canada), NDGF (Denmark, Norway, Sweden), CC-IN2P3 (France),
KIT/GridKA (Germany), INFN-CNAF (Italy), NL-T1 (Netherlands), PIC (Spain),
ASGC (Taiwan), RAL (UK) and BNL (USA) and in the Tier-2 facilities
worldwide.

\clearpage
\appendix
\section{Additional unfolded differential cross sections} \label{app:extra}

This appendix presents measurements of differential cross sections as
a function of eight additional variables that are compared with
theoretical predictions.

\begin{figure}[b]
  \begin{center}
    \subfigure[] {\includegraphics[width=0.47\textwidth]{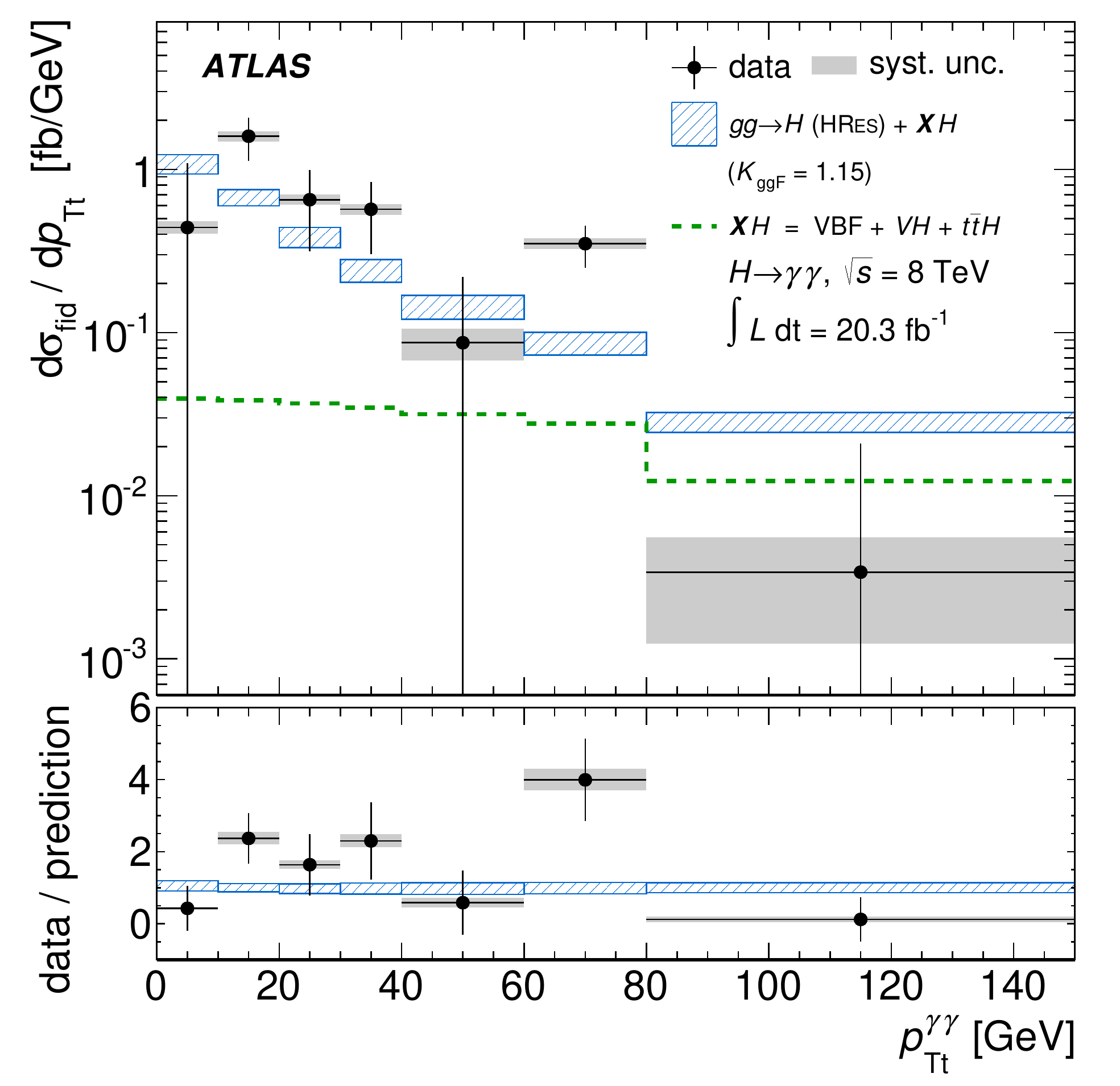}\quad}
    \subfigure[] { \includegraphics[width=0.47\textwidth]{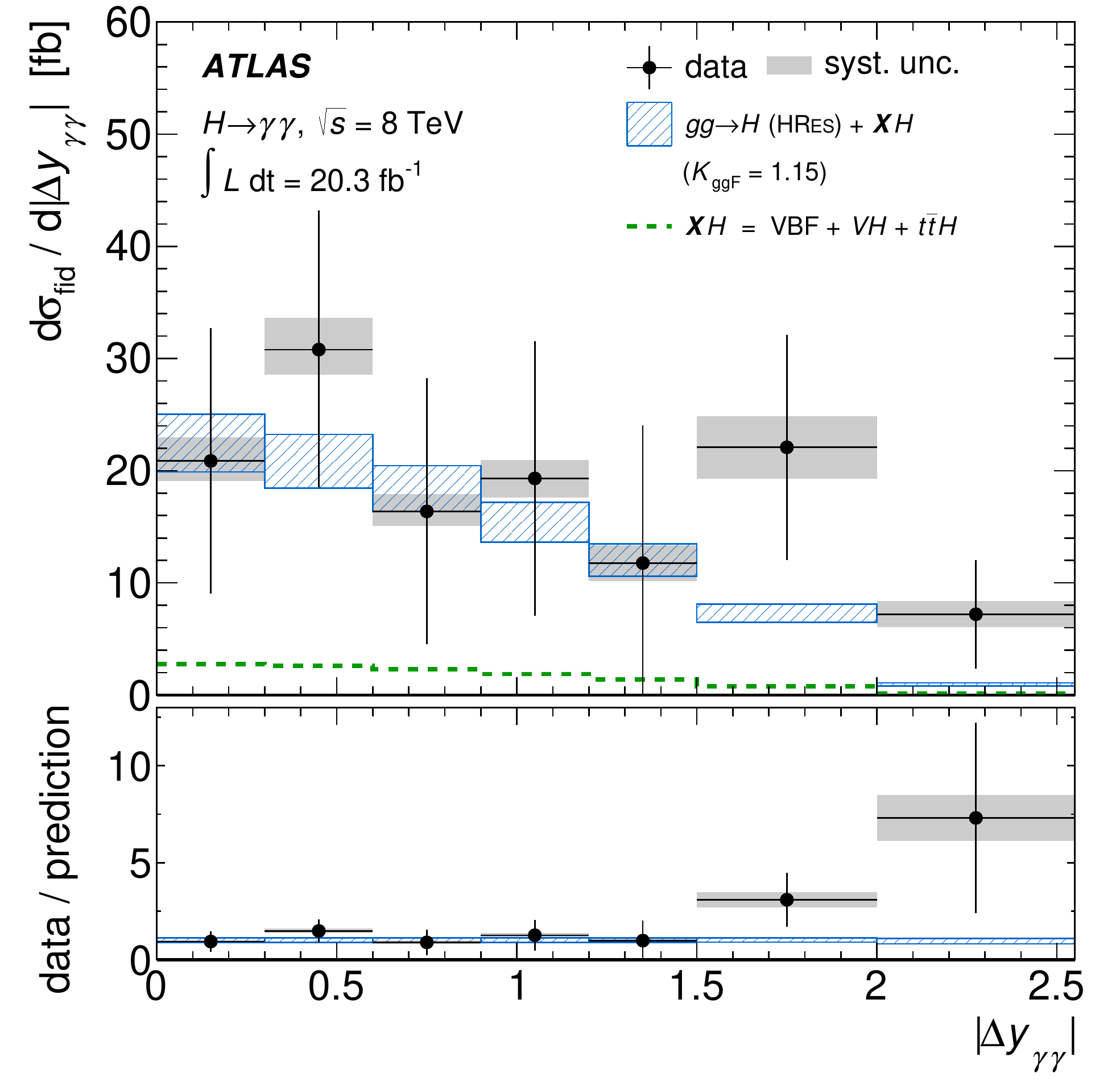}}
    \caption[]{The differential cross section for $pp \rightarrow
      H\rightarrow \gamma\gamma$ as a function of (a) \pttgg\ and (b)
      \deltaygg. The data are shown as filled (black) circles. The vertical
      error bar on each data point represents the total uncertainty in
      the measured cross section, and the shaded (grey) band is the
      systematic uncertainty component. The SM prediction, defined using the \hres\
      prediction for gluon fusion and the default MC samples for the
      other production mechanisms, is presented as a hatched (blue)
      band (see text for details). The small contribution from VBF,
      $VH$ and $t\bar{t}H$ is also shown separately as a dashed
      (green) line and denoted as $XH$. The \hres\ predictions are normalised to the total LHC-XS
      cross section~\cite{Heinemeyer:2013tqa} using a K-factor of
      ${\rm K}_{\rm ggF}=1.15$.} 
    \label{fig:extra1}
  \end{center}
\end{figure}

Figure~\ref{fig:extra1} shows the differential cross
section as a function of \pttgg{}, defined as the magnitude of the transverse
momentum of the diphoton system perpendicular to the diphoton thrust
axis~\cite{Aad:2013wqa}, as well as the rapidity separation between the two photons
\deltaygg{}. 

\begin{figure}[t]
  \begin{center}
    \mbox{
      \subfigure[] {\includegraphics[width=0.47\textwidth]{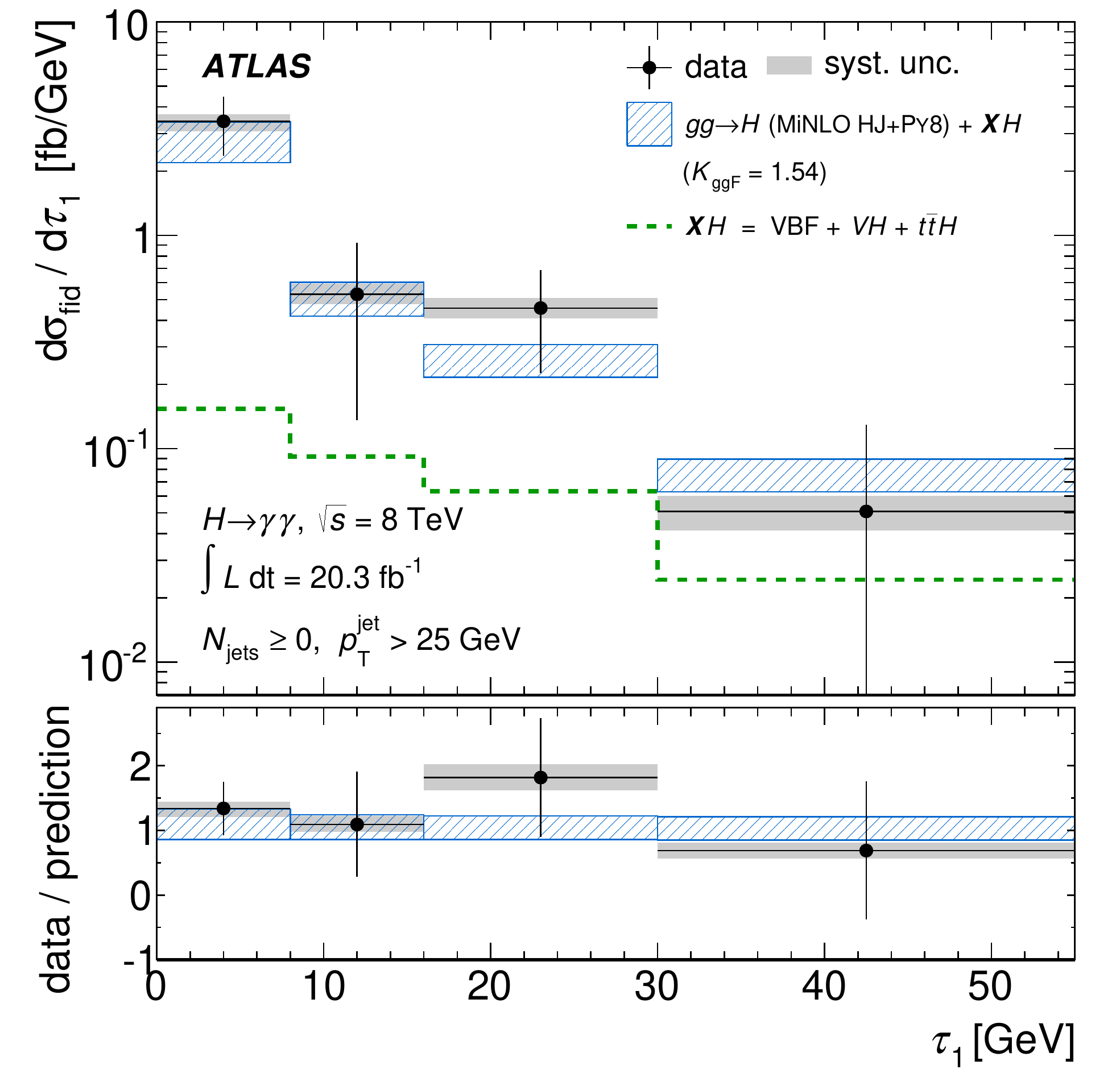}\quad}
      \subfigure[]{\includegraphics[width=0.47\textwidth]{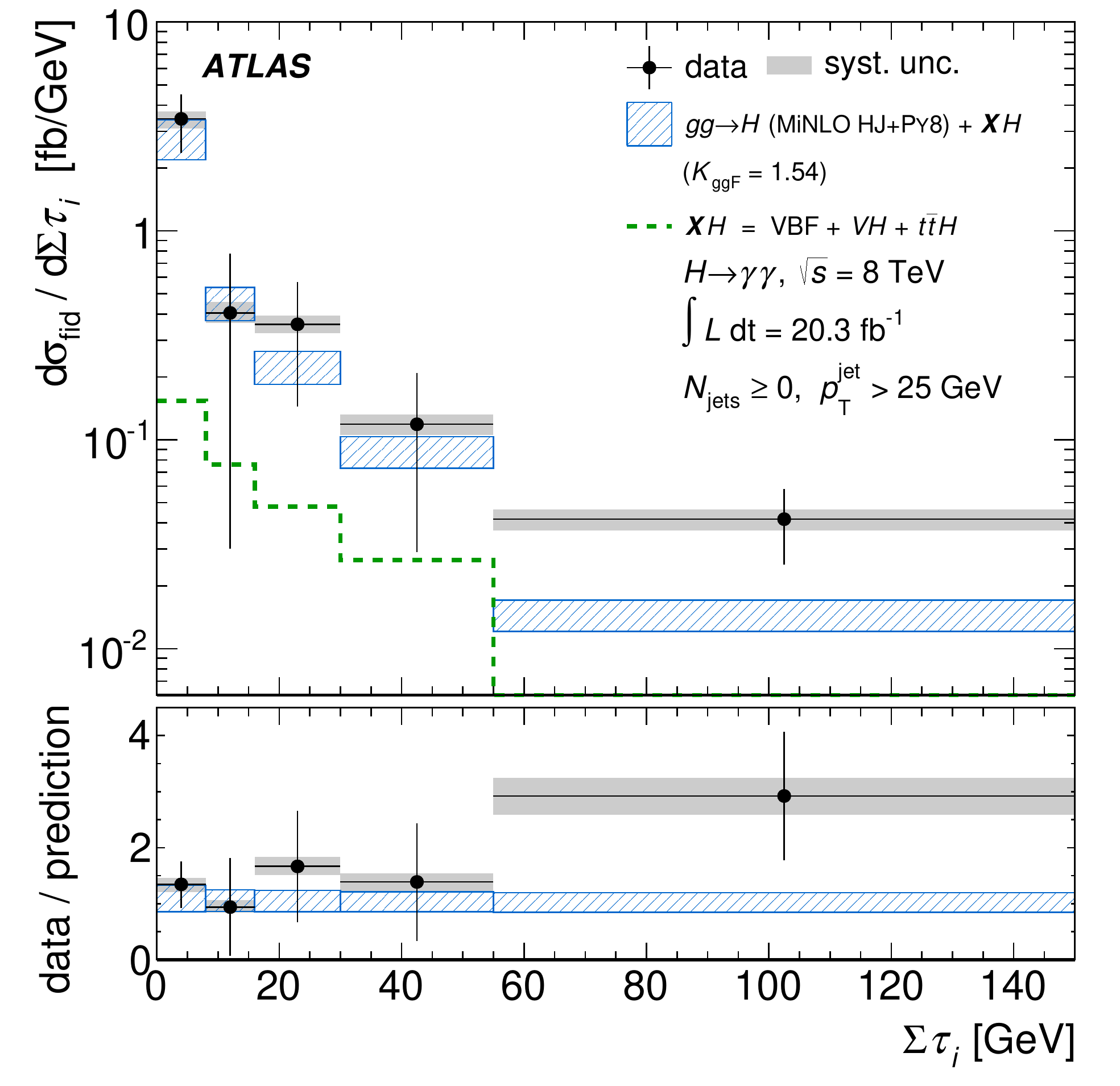}}
    }
    \caption[]{The differential cross section for $pp \rightarrow
      H\rightarrow \gamma\gamma$ as a function of (a) \taujet\ and (b) 
      \sumtaujet\ measured in the baseline fiducial region. 
      The data and theoretical predictions are presented the same way as in
      figure~\ref{fig:extra1}, although the SM prediction is now
      constructed using the \minlohj\ prediction for gluon fusion and the
      default MC samples for the other production mechanisms. 
      The first bin of these distributions contains the events for
      which no jet fullfils the $\tau>8$~GeV and $\pt > 25$~GeV requirements. The \minlohj\ prediction is normalised to the LHC-XS prediction using a ${\rm K}$-factor of ${\rm K}_{\rm ggF}=1.54$
      \label{fig:tau}
    }
  \end{center}
\end{figure}

Figure~\ref{fig:tau} presents measurements of 
the beam-thrust-like variables \taujet{} and \sumtaujet{}.
For a given jet, $\tau$ is defined by
\begin{eqnarray}
  \tau = \frac{m_{\mathrm{T}}}{2\cosh{y^*}}, & ~~~y^* =
  y - y_{\gamma\gamma}, & ~~~m_{\mathrm{T}} = \sqrt{p_{\mathrm{T}}^2 +
    m^2},
\end{eqnarray}
where $y$ is the jet rapidity and $m$ is the jet mass. The variable \taujet{} refers to the highest-$\tau{}$ jet, and \sumtaujet{} is the
scalar sum of $\tau{}$ for all jets with $\tau>8$~GeV, analogous to \ptjl{} and \htj{},
respectively. 
For large jet rapidities, $\tau$ corresponds to the small light-cone
component of the jet, $p^+_{\rm jet} = E_{\rm jet}-|p_{z,\rm jet}|$, while
the sum is closely analogous to the beam-thrust global event
shape~\cite{Stewart:2009yx} (both measured in the diphoton
rest frame). 

\begin{figure}[h]
  \begin{center}
    \mbox{
      \subfigure[] {\includegraphics[width=0.47\textwidth]{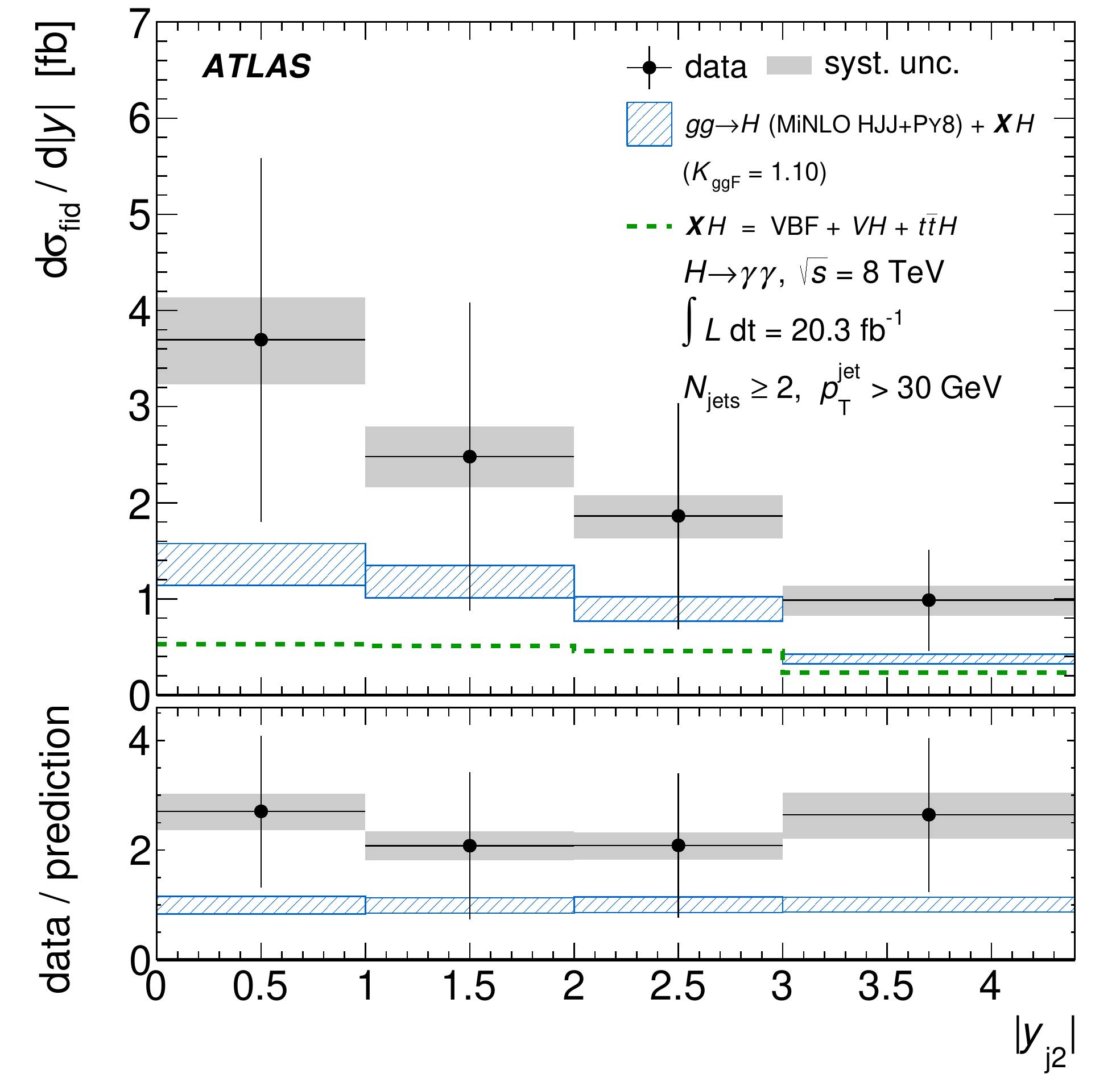}\quad}
      \subfigure[]{\includegraphics[width=0.47\textwidth]{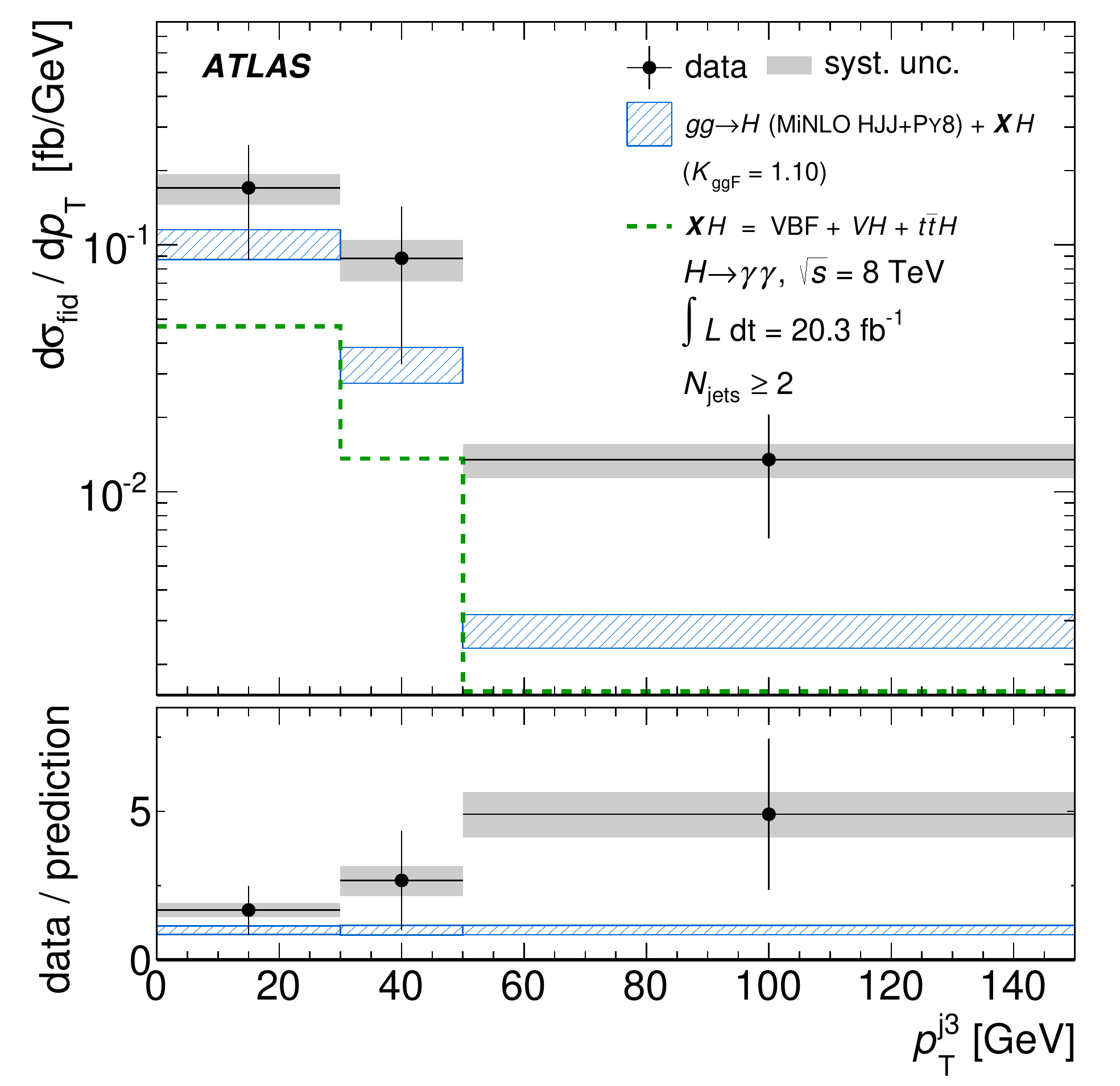}}
    }
    \mbox{
      \subfigure[]{\includegraphics[width=0.47\textwidth]{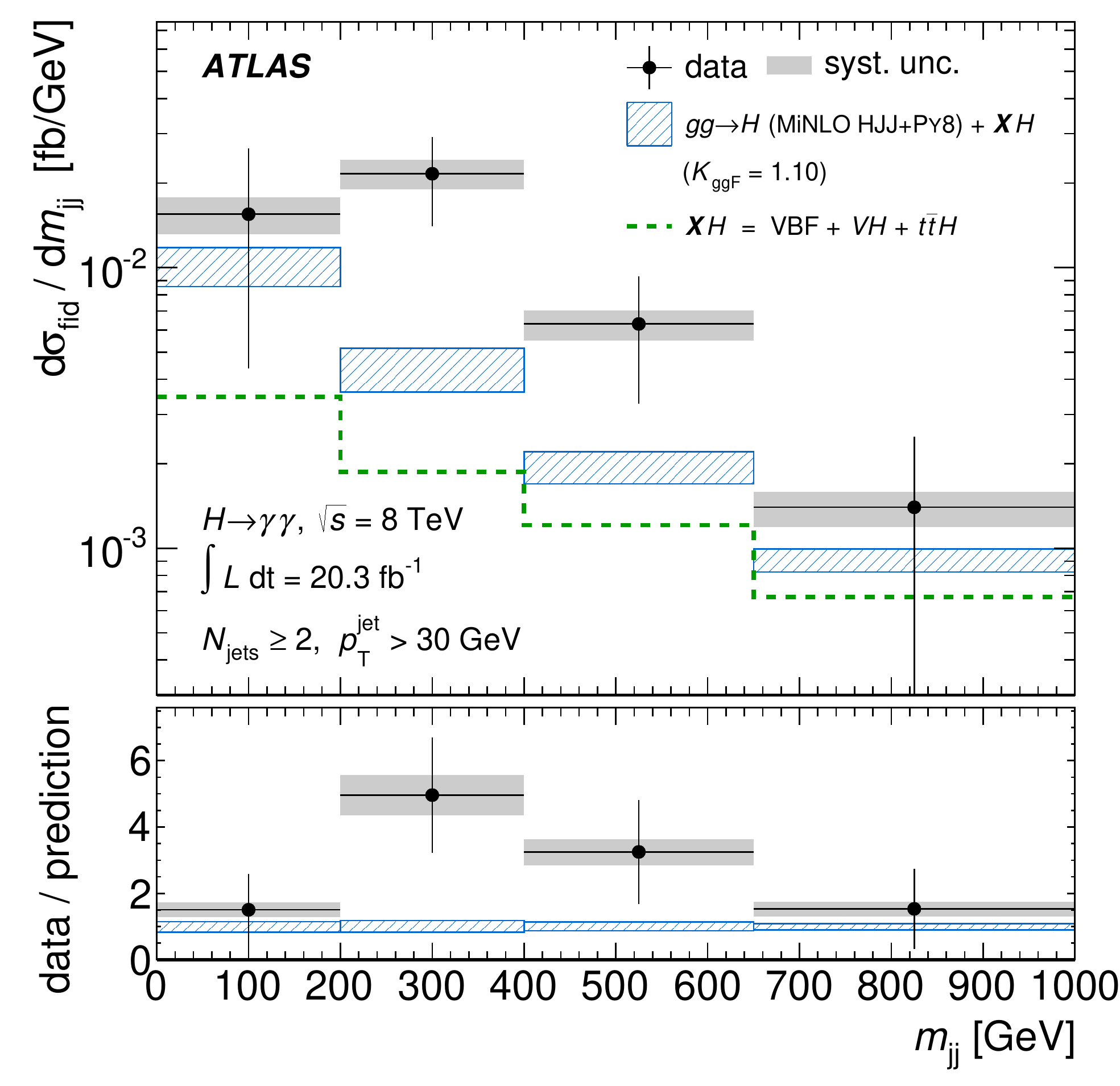}\quad}
      \subfigure[]{\includegraphics[width=0.47\textwidth]{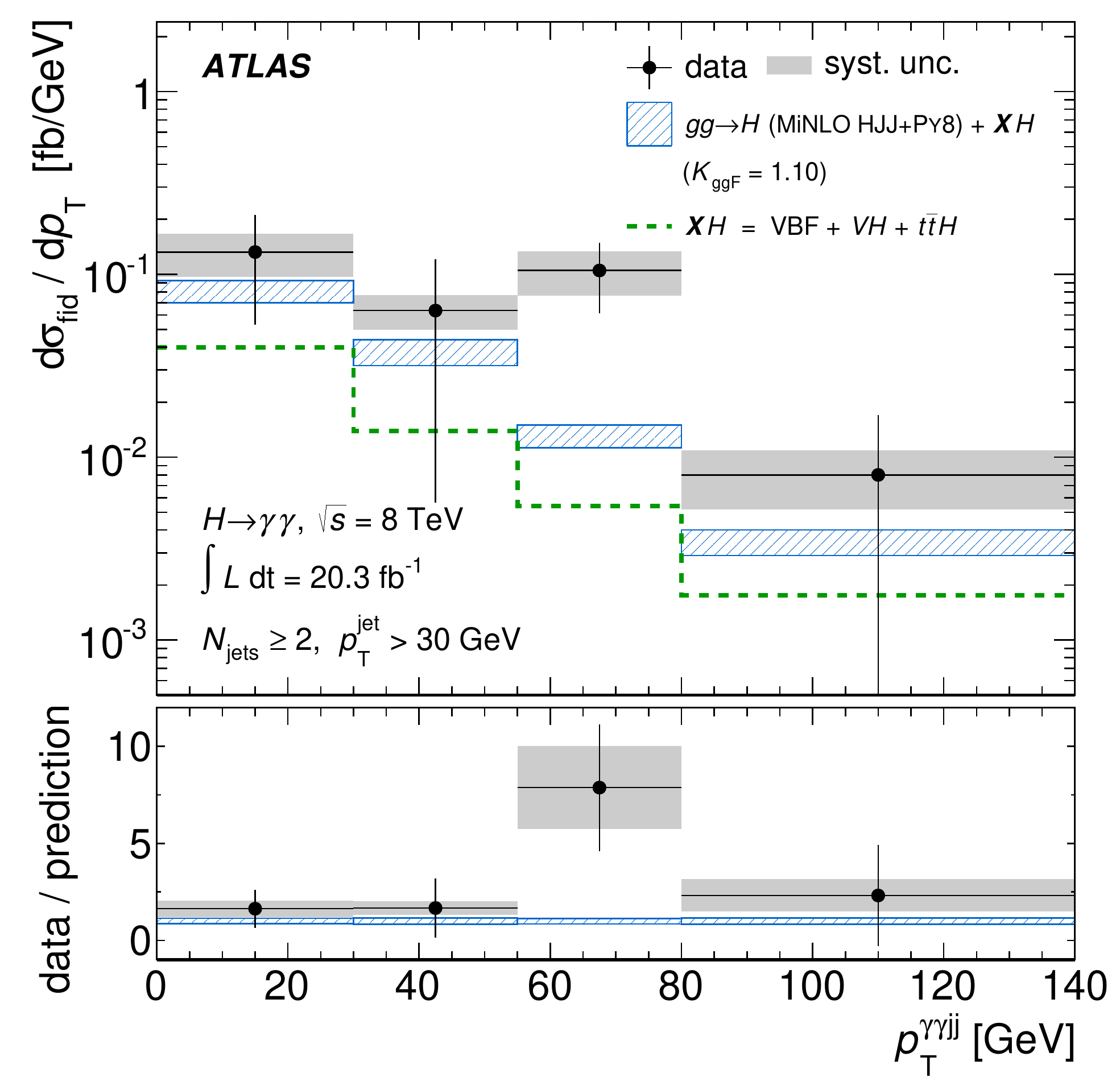}}
    }
    \caption[]{
      The differential cross section for $pp \rightarrow
      H\rightarrow \gamma\gamma$ as a function of 
      (a) the subleading jet rapidity \yjsl{}, (b) the third-leading
      jet transverse momentum \ptjssl{}, (c) the dijet invariant mass
      \mjj{}, and (d) the transverse momentum of the diphoton--dijet
      system \ptggjj{}.
      All variables are defined in the subset of the data containing two or more jets. 
      The first bin of the \ptjssl{} contains events with two jets
      with $\pt > 30$~GeV, but no third jet above this \pt{} threshold.
      The data and theoretical predictions are presented the same way as in
      figure~\ref{fig:extra1}, although the SM prediction is now
      constructed using the \minlohjj{} prediction for gluon fusion and the
      default MC samples for the other production mechanisms. 
      The \minlohjj\ prediction is normalised to the LHC-XS prediction
      using a ${\rm K}$-factor of ${\rm K}_{\rm ggF}=1.10$.
      \label{fig:extra2}
    }
  \end{center}
\end{figure}

Measurements of four additional variables are presented in figure~\ref{fig:extra2}:
the third-leading jet transverse momentum \ptjssl{}; the
sub-leading jet rapidity, \yjsl{}; the dijet invariant mass \mjj{}, and
the transverse momentum of the diphoton--dijet system \ptggjj{}.

\begin{figure}[h]
  \begin{center}
    \subfigure[]{\includegraphics[width=0.47\textwidth]{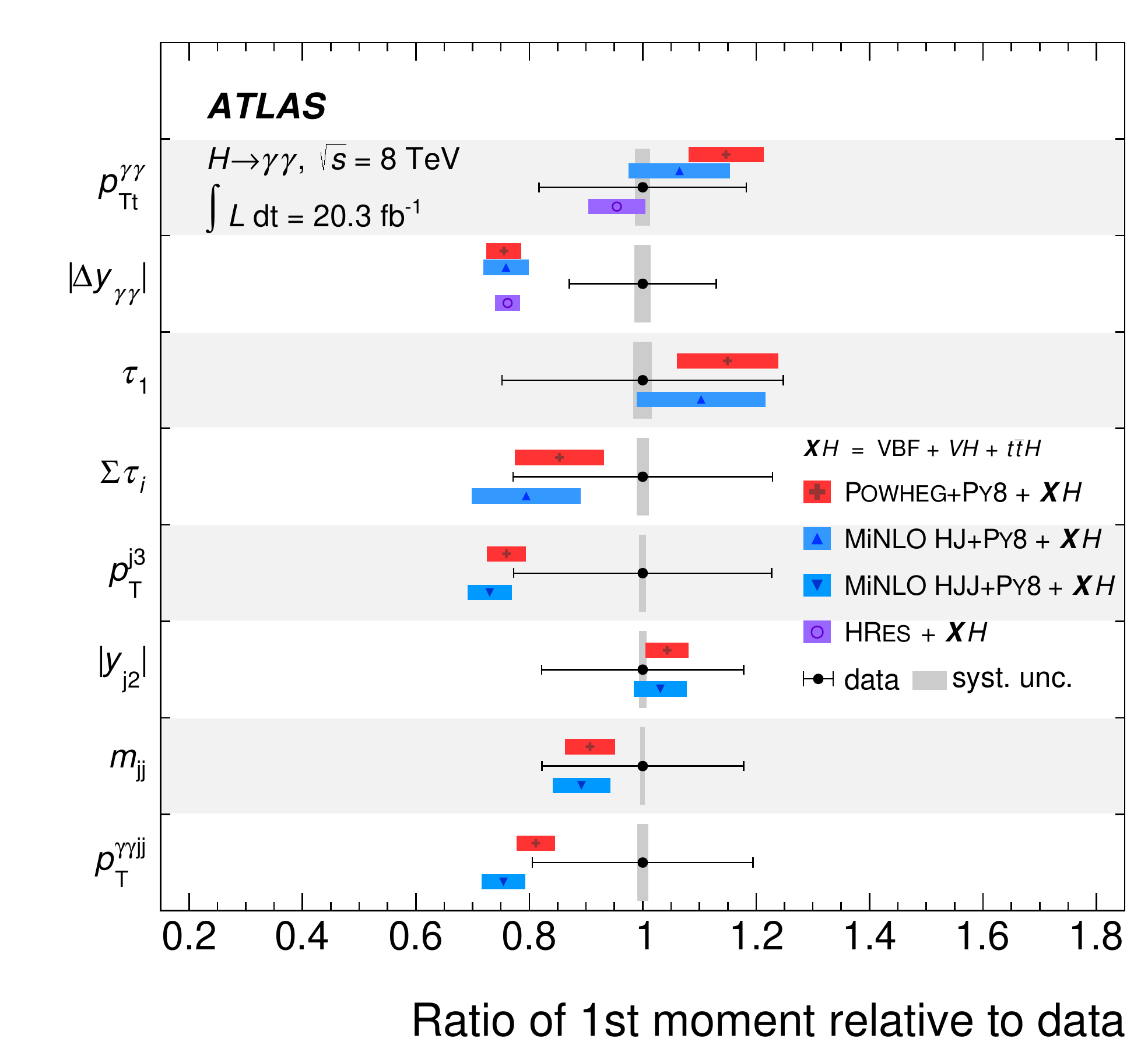}}\quad
    \subfigure[]{\includegraphics[width=0.47\textwidth]{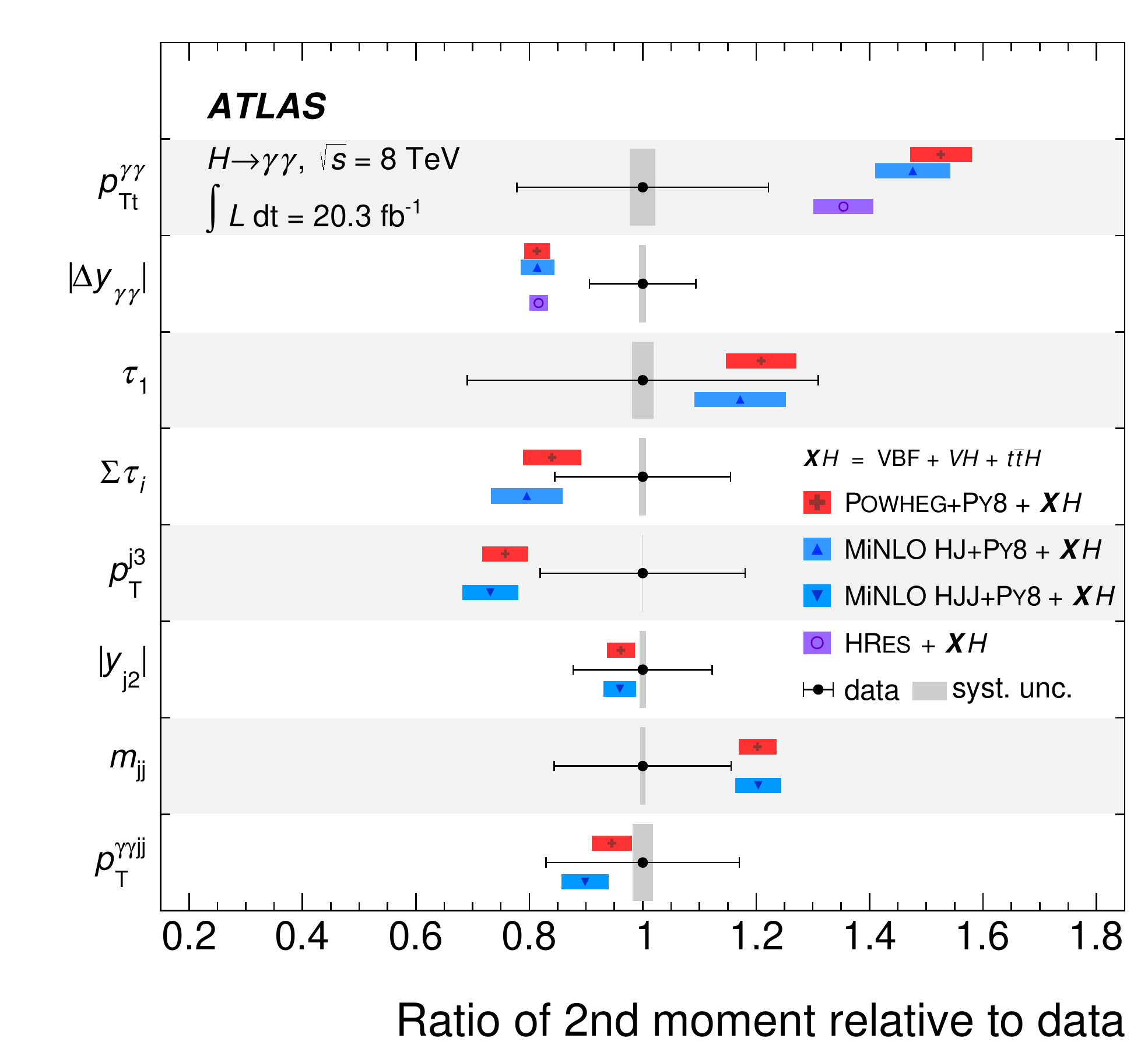}}
    \caption[]{
      (a) The ratio of the first moment (mean) of each differential
      distribution predicted by the theoretical models to that
      observed in the data. (b) The ratio of the second moment (RMS) 
      of each differential distribution predicted by the theoretical
      models to that observed in the data. The intervals on the vertical axes each represent one of the differential distributions. The band for each
      theoretical prediction represents the corresponding uncertainty (see text for details). The error bar on the data
      represents the total uncertainty in the measurement, with the
      grey band representing the systematic-only uncertainty.
      \label{fig:appvars_summary}
    }
  \end{center}
\end{figure}

Figure~\ref{fig:appvars_summary} shows the first and second moments of
each of the additional differential distributions. The data are compared to a variety of
theoretical predictions obtained from the MC event generators.  In
general, the event generator predictions are in good agreement with
the data, which is further quantified by a $\chi^2$ comparison
presented in table~\ref{tab:appvars_chi2table}.

\begin{table}[tb]
\begin{center}
\small 
\begin{tabular}{ c | c c c c} 
\hline \hline 
 Variable & \powheg{}& \minlohj{}& \minlohjj{}& \hres{}\\ 
\hline 
$p_{\mathrm{Tt}\gamma\gamma}$ & 0.02& 0.03& 0.03& 0.04\\ 
$\Delta y_{\gamma\gamma}$ & 0.73& 0.74& 0.74& 0.73\\ 
$\tau_{1}$ & 0.82& 0.86& 0.87& - \\ 
$\sum{\tau_{i}}$ & 0.68& 0.64& 0.58& - \\ 
\yjsl & 0.62& 0.56& 0.46& - \\ 
\mjj{} & 0.22& 0.18& 0.14& - \\ 
$p_{\mathrm{T}\gamma\gamma jj}$ & 0.34& 0.30& 0.25& - \\ 
$p_{\mathrm{T}}^{j3}$ & 0.45& 0.38& 0.32& - \\ 
\hline \hline 
\end{tabular} 
\caption{Probabilities from $\chi^2$ tests for the agreement between the differential cross section measurements and the theoretical predictions. Each prediction is normalised to the LHC-XS cross section before selection.\label{tab:appvars_chi2table}} 
\end{center} 
\end{table}

\clearpage
\section{Diphoton acceptance, photon isolation and non-perturbative correction factors for parton-level gluon fusion calculations} \label{app:corr}

The diphoton acceptance factors that are applied to parton-level calculations of Higgs production via gluon fusion in order to correctly account for the diphoton selection criteria applied to the Higgs decay products are shown in 
table~\ref{tab:fid_acceptance} for the fiducial and differential cross sections presented in this paper. 
Complementary isolation efficiency and non-perturbative correction factors that account for the efficiency of the photon isolation criterion and the impact of hadronisation and underlying event activity are presented in tables~\ref{tab:isoeff} and \ref{tab:NPcorr}, respectively. The isolation efficiency is determined as the fraction of selected diphoton events (i.e. within the kinematic acceptance) that also satisfy the isolation criteria, and is determined using samples without hadronisation and underlying events. The non-perturbative correction factors are defined as the ratio of cross sections produced with and without hadronisation and underlying event. The default non-perturbative correction factor is taken to be the centre of the envelope of correction factors obtained from multiple event generators and/or event generator tunes, with the uncertainty taken to be half of the envelope. The variations in non-perturbative correction factors were obtained using AU2 (\pythia\ \cite{ATLASUE2}), UE-EE-4-LO (\herwig++ \cite{Gieseke:2003hm,Bellm:2013lba}) and AUET2B-LO, AUET2B-CTEQ6L1, AMBT2B-LO and AMBT-CTEQ6L1 (\pythiasix, \cite{ATLASUE}). The prediction for combined $WH$, $ZH$ and $t\bar{t}H$ production is given in table \ref{tab:XH}.

\begin{sidewaystable}[htpb]
\caption{Diphoton kinematic acceptances in percent for gluon fusion for each fiducial region/variable bin studied in this paper, defined as the probability to fulfil the diphoton kinematic criteria: $p_{\rm T}/m_{\gamma\gamma}<0.35$ (0.25) for the leading (subleading) photon and $|\eta_{\gamma}|<2.37$. The factors are evaluated using the \powheg{} event generator with MPI modelling and hadronisation turned off. Consistent results for the diphoton variables are obtained by \hres{} 2.2. Uncertainties are taken from PDF variations. QCD scale varaitions have a negligible impact on these factors. The range of each bin is given in Table~\ref{tab:binning}.\label{tab:fid_acceptance}} 
\begin{center}
\small
\begin{tabular}{l| c  c  c  c  c  c  c  c } 
\hline \hline 
 Bin    & 1 & 2 & 3 & 4 & 5 & 6 & 7 & 8\\ \hline \hline 
Baseline                        &$   62.6\pm    0.9$\\ 
$\geq 1$ jet                     &$   61.1\pm    0.7$\\ 
$\geq 2$ jet                     &$   61.9\pm    0.6$\\ 
$\geq 3$ jet                     &$   63.8\pm    0.6$\\ 
VBF-enhanced                     &$   55.0\pm    0.5$\\ 
$p_{\mathrm{T}}^{j3}$            &$   61.2\pm    0.6$ &$   62.2\pm    0.6$ &$   68.9\pm    0.6$\\ 
$N_{\rm jets}$                   &$   63.6\pm    1.0$ &$   60.8\pm    0.7$ &$   61.2\pm    0.6$ &$   63.8\pm    0.6$\\ 
$N_{\rm jets}^{\rm 50~GeV}$      &$   62.9\pm    0.9$ &$   60.7\pm    0.6$ &$   64.3\pm    0.6$ &$   67.8\pm    0.6$\\ 
\yjl                             &$   61.5\pm    0.6$ &$   61.5\pm    0.7$ &$   60.2\pm    0.7$ &$   60.1\pm    0.7$\\ 
\yjsl                            &$   61.1\pm    0.5$ &$   61.6\pm    0.6$ &$   63.1\pm    0.6$ &$   63.4\pm    0.6$\\ 
\ptjsl{}                         &$   60.8\pm    0.7$ &$   60.1\pm    0.6$ &$   62.9\pm    0.6$ &$   67.5\pm    0.6$\\ 
\mjj{}                           &$   61.0\pm    0.6$ &$   62.7\pm    0.6$ &$   64.2\pm    0.6$ &$   62.6\pm    0.6$\\ 
\deltayjj                        &$   62.1\pm    0.6$ &$   62.0\pm    0.6$ &$   60.1\pm    0.6$ &$   60.9\pm    0.5$\\ 
\dphijj                          &$   54.4\pm    0.5$ &$   79.6\pm    0.8$ &$   66.7\pm    0.7$ &$   42.5\pm    0.4$\\ 
$p_{\mathrm{T}\gamma\gamma jj}$  &$   60.3\pm    0.6$ &$   62.3\pm    0.6$ &$   63.3\pm    0.6$ &$   67.4\pm    0.6$\\ 
\dphiggjj                        &$   57.5\pm    0.6$ &$   65.7\pm    0.7$ &$   61.5\pm    0.6$ &$   60.8\pm    0.5$\\ 
$\tau_{1}$                       &$   59.5\pm    1.0$ &$   61.5\pm    0.7$ &$   59.7\pm    0.6$ &$   61.8\pm    0.6$\\ 
$\sum{\tau_{i}}$                 &$   63.0\pm    1.0$ &$   61.3\pm    0.7$ &$   59.8\pm    0.6$ &$   60.5\pm    0.5$ &$   67.3\pm    0.6$\\ 
\ptjl{}                          &$   63.6\pm    1.0$ &$   60.7\pm    0.8$ &$   59.8\pm    0.7$ &$   59.6\pm    0.6$ &$   60.5\pm    0.5$\\ 
\htj{}                           &$   63.6\pm    1.0$ &$   60.9\pm    0.8$ &$   60.1\pm    0.7$ &$   59.3\pm    0.6$ &$   63.8\pm    0.6$\\ 
\ygg                             &$   73.9\pm    0.7$ &$   74.0\pm    0.7$ &$   74.2\pm    0.6$ &$   73.9\pm    0.6$ &$   69.9\pm    0.6$ &$   38.8\pm    0.5$\\ 
\ptgg{}                          &$   62.1\pm    1.1$ &$   64.9\pm    1.0$ &$   64.1\pm    0.9$ &$   62.8\pm    0.8$ &$   61.7\pm    0.7$ &$   60.1\pm    0.7$ &$   60.2\pm    0.7$ &$   59.9\pm    0.5$\\ 
\hline \hline 
\end{tabular} 
\end{center} 
\end{sidewaystable}

\begin{sidewaystable}[htpb]
\caption{Isolation efficiencies in percent for gluon fusion $H\to\gamma\gamma$ for each fiducial region/variable bin measured in this analysis. The isolation efficiency is defined as the probability for both photons to fulfil the isolation criteria ($E_{\rm T}^{\rm iso}<14$~GeV as described in the text) for events that pass the diphoton kinematic criteria. Uncertainties are assigned in the same way as for the non-perturbative correction factors: by varying the fragmentation and underlying event modelling. These factors can be multiplied by the kinematic acceptance factors (see table~\ref{tab:fid_acceptance}) to extrapolate an inclusive gluon fusion Higgs prediction to the fiducial volume used in this analysis. The range of each bin is given in Table~\ref{tab:binning}.\label{tab:isoeff}} 
\begin{center}
\small
\begin{tabular}{l| c  c  c  c  c  c  c  c  c  c } 
\hline \hline 
 Bin    & 1 & 2 & 3 & 4 & 5 & 6 & 7 & 8 & 9 & 10\\ \hline \hline 
Baseline                        &$   98.1\pm    0.5$\\ 
$\geq 1$ jet                     &$   97.5\pm    0.6$\\ 
$\geq 2$ jet                     &$   96.4\pm    1.0$\\ 
$\geq 3$ jet                     &$   95.1\pm    1.9$\\ 
VBF-enhanced                     &$   97.0\pm    3.0$\\ 
$p_{\mathrm{T}}^{j3}$            &$   96.7\pm    1.1$ &$   95.3\pm    2.1$ &$   93.8\pm    3.8$\\ 
$N_{\rm jets}$                   &$   98.5\pm    0.6$ &$   97.9\pm    0.7$ &$   96.7\pm    1.1$ &$   94.9\pm    1.9$\\ 
$N_{\rm jets}^{\rm 50~GeV}$      &$   98.4\pm    0.6$ &$   97.3\pm    0.7$ &$   95.7\pm    1.7$ &$   94.6\pm    3.7$\\ 
\yjl                             &$   97.1\pm    0.7$ &$   97.4\pm    0.8$ &$   97.7\pm    1.2$ &$   97.9\pm    1.8$\\ 
\yjsl                            &$   95.9\pm    1.5$ &$   96.1\pm    1.6$ &$   96.5\pm    2.1$ &$   97.1\pm    3.2$\\ 
\ptjsl{}                         &$   97.9\pm    0.7$ &$   96.7\pm    1.2$ &$   96.0\pm    2.0$ &$   95.1\pm    2.6$\\ 
\mjj{}                           &$   96.5\pm    1.2$ &$   95.8\pm    1.8$ &$   95.7\pm    3.2$ &$   95.6\pm    4.8$\\ 
\deltayjj                        &$   96.0\pm    1.2$ &$   96.4\pm    1.7$ &$   97.1\pm    3.6$ &$   97.7\pm    7.8$\\ 
\dphijj                          &$   96.1\pm    1.6$ &$   96.3\pm    1.6$ &$   96.5\pm    2.2$ &$   95.5\pm    2.4$\\ 
$p_{\mathrm{T}\gamma\gamma jj}$  &$   97.4\pm    1.5$ &$   96.1\pm    1.8$ &$   94.6\pm    3.0$ &$   92.4\pm    3.8$\\ 
\dphiggjj                        &$   93.9\pm    3.2$ &$   95.9\pm    2.0$ &$   96.7\pm    1.4$ &$   96.9\pm    2.2$\\ 
$\tau_{1}$                       &$   98.1\pm    0.6$ &$   97.9\pm    0.9$ &$   97.3\pm    0.8$ &$   96.9\pm    1.3$\\ 
$\sum{\tau_{i}}$                 &$   98.4\pm    0.6$ &$   98.1\pm    1.0$ &$   97.5\pm    0.9$ &$   97.0\pm    1.2$ &$   95.8\pm    1.4$\\ 
\ptjl{}                          &$   98.5\pm    0.6$ &$   98.1\pm    0.8$ &$   97.6\pm    1.0$ &$   97.3\pm    1.2$ &$   96.8\pm    1.4$\\ 
\htj{}                           &$   98.5\pm    0.6$ &$   98.2\pm    0.9$ &$   97.9\pm    1.1$ &$   97.3\pm    0.9$ &$   96.4\pm    1.4$\\ 
\ygg                             &$   98.0\pm    0.7$ &$   98.1\pm    0.7$ &$   98.0\pm    0.7$ &$   98.2\pm    0.7$ &$   98.2\pm    0.7$ &$   98.0\pm    0.9$\\ 
$\Delta y_{\gamma\gamma}$        &$   97.8\pm    0.6$ &$   97.8\pm    0.7$ &$   98.0\pm    0.7$ &$   98.2\pm    0.7$ &$   98.5\pm    0.8$ &$   98.9\pm    0.9$ &$   99.2\pm    2.3$\\ 
$p_{\mathrm{Tt}\gamma\gamma}$    &$   98.2\pm    0.5$ &$   98.2\pm    0.6$ &$   98.4\pm    0.9$ &$   98.3\pm    1.0$ &$   98.3\pm    0.9$ &$   98.0\pm    1.1$ &$   97.4\pm    1.0$\\ 
\ptgg{}                          &$   99.5\pm    0.5$ &$   98.4\pm    0.9$ &$   97.6\pm    1.1$ &$   97.1\pm    1.1$ &$   96.7\pm    1.2$ &$   96.5\pm    1.0$ &$   96.6\pm    1.3$ &$   97.4\pm    1.0$\\ 
\costhetastar                    &$   97.8\pm    0.7$ &$   97.8\pm    0.8$ &$   97.9\pm    0.8$ &$   98.0\pm    0.8$ &$   98.2\pm    0.8$ &$   98.4\pm    0.9$ &$   98.7\pm    0.9$ &$   98.9\pm    1.1$ &$   97.9\pm    2.0$ &$   97.9\pm    2.3$\\ 
\hline \hline 
\end{tabular} 
\end{center} 
\end{sidewaystable} 

\begin{sidewaystable}[htpb]
\caption{Non-perturbative correction factors in percent accounting for the impact of hadronisation and the underlying event activity for all measured variables and fiducial regions. Uncertainties are evaluated by deriving these factors using different generators and tunes as described in the text. The range of each bin is given in Table~\ref{tab:binning}.\label{tab:NPcorr}} 
\begin{center}
\small
\begin{tabular}{l| c  c  c  c  c  c  c  c  c  c } 
\hline \hline 
 Bin    & 1 & 2 & 3 & 4 & 5 & 6 & 7 & 8 & 9 & 10\\ \hline \hline 
Baseline                        &$ 99.9\pm   0.6$\\ 
$\geq 1$ jet                     &$ 99.2\pm   2.2$\\ 
$\geq 2$ jet                     &$ 98.0\pm   4.5$\\ 
$\geq 3$ jet                     &$ 96.2\pm   5.9$\\ 
VBF-enhanced                     &$ 98.8\pm   5.9$\\ 
$p_{\mathrm{T}}^{j3}$            &$ 98.8\pm   4.0$ &$ 96.9\pm   7.1$ &$ 95.5\pm   3.3$\\ 
$N_{\rm jets}$                   &$100.9\pm   2.1$ &$ 99.7\pm   2.8$ &$ 98.8\pm   4.0$ &$ 96.3\pm   5.9$\\ 
$N_{\rm jets}^{\rm 50~GeV}$      &$100.6\pm   1.2$ &$ 99.4\pm   2.5$ &$ 98.4\pm   4.1$ &$ 95.3\pm   3.3$\\ 
\yjl                             &$ 99.7\pm   2.1$ &$ 99.4\pm   2.1$ &$ 99.2\pm   2.3$ &$ 98.2\pm   3.5$\\ 
\yjsl                            &$ 97.7\pm   4.0$ &$ 98.2\pm   4.3$ &$ 98.6\pm   5.1$ &$ 98.1\pm   6.2$\\ 
\ptjsl{}                         &$ 99.7\pm   2.8$ &$ 98.3\pm   5.0$ &$ 98.0\pm   4.6$ &$ 97.8\pm   3.6$\\ 
\mjj{}                           &$ 97.1\pm   4.2$ &$ 99.0\pm   4.5$ &$ 99.2\pm   4.9$ &$ 99.0\pm   5.8$\\ 
\deltayjj                        &$ 97.5\pm   3.7$ &$ 99.0\pm   5.3$ &$ 98.7\pm   5.9$ &$ 98.8\pm   7.0$\\ 
\dphijj                          &$ 96.6\pm   4.8$ &$ 98.6\pm   4.2$ &$ 98.1\pm   4.4$ &$100.2\pm   5.4$\\ 
$p_{\mathrm{T}\gamma\gamma jj}$  &$ 97.0\pm   6.5$ &$ 99.5\pm   3.0$ &$ 98.8\pm   4.5$ &$ 97.9\pm   5.3$\\ 
\dphiggjj                        &$101.6\pm   5.4$ &$ 98.3\pm   4.3$ &$ 98.6\pm   4.8$ &$ 95.2\pm   4.2$\\ 
$\tau_{1}$                       &$100.5\pm   1.7$ &$ 99.0\pm   2.6$ &$100.4\pm   2.3$ &$100.2\pm   2.0$\\ 
$\sum{\tau_{i}}$                 &$100.5\pm   1.7$ &$ 99.4\pm   2.4$ &$100.6\pm   2.3$ &$ 99.5\pm   2.6$ &$ 99.1\pm   2.5$\\ 
\ptjl{}                          &$100.9\pm   2.1$ &$ 99.5\pm   2.2$ &$ 99.2\pm   2.4$ &$ 99.1\pm   2.3$ &$ 99.0\pm   2.4$\\ 
\htj{}                           &$100.9\pm   2.1$ &$ 99.5\pm   2.6$ &$ 99.4\pm   3.2$ &$ 99.0\pm   2.6$ &$ 98.5\pm   3.6$\\ 
\ygg                             &$100.0\pm   0.6$ &$100.1\pm   0.6$ &$ 99.9\pm   0.4$ &$100.1\pm   0.6$ &$100.0\pm   0.6$ &$100.3\pm   0.6$\\ 
$\Delta y_{\gamma\gamma}$        &$100.1\pm   0.6$ &$100.0\pm   0.7$ &$100.0\pm   0.6$ &$100.2\pm   0.8$ &$100.2\pm   0.6$ &$100.2\pm   0.5$ &$100.5\pm   1.5$\\ 
$p_{\mathrm{Tt}\gamma\gamma}$    &$ 99.2\pm   1.3$ &$101.0\pm   1.1$ &$100.4\pm   0.7$ &$100.1\pm   0.7$ &$100.2\pm   0.7$ &$100.0\pm   0.7$ &$ 99.9\pm   0.7$\\ 
\ptgg{}                          &$ 99.5\pm   0.9$ &$101.0\pm   1.0$ &$100.4\pm   0.9$ &$100.2\pm   0.7$ &$100.2\pm   1.0$ &$ 99.8\pm   0.7$ &$100.0\pm   0.8$ &$100.0\pm   0.6$\\ 
\costhetastar                    &$100.2\pm   0.7$ &$100.1\pm   0.6$ &$100.2\pm   0.7$ &$ 99.9\pm   0.7$ &$100.2\pm   0.6$ &$100.0\pm   0.5$ &$100.2\pm   0.7$ &$100.0\pm   0.6$ &$100.0\pm   1.0$ &$100.2\pm   1.1$\\ 
\hline \hline 
\end{tabular} 
\end{center} 
\end{sidewaystable} 

\begin{sidewaystable}[htpb]
\caption{Fiducial cross sections (fb) for combined $WH$, $ZH$ and $t\bar{t}H$ production in each variable bin and fiducial region. The range of each bin is given in Table~\ref{tab:binning}. The uncertainties on the cross-section predictions are discussed in detail in Section~\ref{sec:theory} and include the effect of scale and PDF variation as well as the uncertainties on the $H\rightarrow\gamma\gamma$ branching ratio. \label{tab:XH}} 
\begin{center}
\small
\begin{tabular}{l| c  c  c  c  c  c  c  c } 
\hline \hline 
 Bin    & 1 & 2 & 3 & 4 & 5 & 6 & 7 & 8\\ \hline \hline 
Baseline                        & $   3.75 \pm    0.21$\\ 
$\geq 1$ jet                     & $   3.17 \pm    0.18$\\ 
$\geq 2$ jet                     & $   1.84 \pm    0.10$\\ 
$\geq 3$ jet                     & $   0.43 \pm    0.03$\\ 
VBF-enhanced                     & $   0.68 \pm    0.04$\\ 
$N_{\ell} \geq 1$                & $   0.27 \pm    0.02$\\ 
$E_{\rm T}^{\rm miss} > 80 {\rm GeV}$ & $   0.14 \pm    0.01$\\ 
\ptgg{}                          & $   0.22 \pm    0.01$ & $   0.25 \pm    0.01$ & $   0.30 \pm    0.02$ & $   0.33 \pm    0.02$ & $   0.33 \pm    0.02$ & $   0.59 \pm    0.03$ & $   0.47 \pm    0.03$ & $   1.00 \pm    0.06$\\ 
\ygg                             & $   0.73 \pm    0.04$ & $   0.71 \pm    0.04$ & $   0.66 \pm    0.04$ & $   0.59 \pm    0.03$ & $   0.62 \pm    0.03$ & $   0.43 \pm    0.02$\\ 
\costhetastar                    & $   1.04 \pm    0.06$ & $   0.96 \pm    0.05$ & $   0.82 \pm    0.05$ & $   0.60 \pm    0.03$ & $   0.33 \pm    0.02$\\ 
$N_{\rm jets}$                   & $   0.58 \pm    0.03$ & $   1.33 \pm    0.07$ & $   1.41 \pm    0.08$ & $   0.43 \pm    0.03$\\ 
$N_{\rm jets}^{\rm 50~GeV}$      & $   1.34 \pm    0.07$ & $   1.54 \pm    0.08$ & $   0.71 \pm    0.04$ & $   0.16 \pm    0.01$\\ 
\htj{}                           & $   0.58 \pm    0.03$ & $   0.52 \pm    0.03$ & $   0.37 \pm    0.02$ & $   1.33 \pm    0.07$ & $   0.63 \pm    0.04$\\ 
\ptjl{}                          & $   0.58 \pm    0.03$ & $   0.76 \pm    0.04$ & $   0.74 \pm    0.04$ & $   0.75 \pm    0.04$ & $   0.49 \pm    0.03$\\ 
\yjl                             & $   1.03 \pm    0.06$ & $   0.98 \pm    0.05$ & $   0.77 \pm    0.04$ & $   0.38 \pm    0.02$\\ 
\ptjsl{}                         & $   1.33 \pm    0.07$ & $   0.97 \pm    0.05$ & $   0.46 \pm    0.03$ & $   0.37 \pm    0.02$\\ 
\yjsl                            & $   0.53 \pm    0.03$ & $   0.51 \pm    0.03$ & $   0.46 \pm    0.02$ & $   0.33 \pm    0.02$\\ 
$p_{\mathrm{T}}^{j3}$            & $   1.40 \pm    0.08$ & $   0.27 \pm    0.02$ & $   0.15 \pm    0.01$\\ 
\mjj{}                           & $   0.69 \pm    0.04$ & $   0.37 \pm    0.02$ & $   0.30 \pm    0.02$ & $   0.23 \pm    0.01$\\ 
\deltayjj                        & $   0.75 \pm    0.05$ & $   0.49 \pm    0.03$ & $   0.40 \pm    0.02$ & $   0.19 \pm    0.01$\\ 
\dphijj                          & $   0.55 \pm    0.03$ & $   0.57 \pm    0.03$ & $   0.33 \pm    0.02$ & $   0.38 \pm    0.02$\\ 
$p_{\mathrm{T}\gamma\gamma jj}$  & $   1.20 \pm    0.07$ & $   0.35 \pm    0.02$ & $  0.136 \pm   0.008$ & $  0.106 \pm   0.008$\\ 
\dphiggjj                        & $   0.27 \pm    0.02$ & $   0.30 \pm    0.02$ & $   0.79 \pm    0.04$ & $   0.47 \pm    0.03$\\ 
$p_{\mathrm{Tt}\gamma\gamma}$    & $   0.39 \pm    0.02$ & $   0.38 \pm    0.02$ & $   0.37 \pm    0.02$ & $   0.35 \pm    0.02$ & $   0.63 \pm    0.04$ & $   0.55 \pm    0.03$ & $   0.86 \pm    0.05$\\ 
$\Delta y_{\gamma\gamma}$        & $   0.83 \pm    0.05$ & $   0.78 \pm    0.04$ & $   0.69 \pm    0.04$ & $   0.56 \pm    0.03$ & $   0.42 \pm    0.02$ & $   0.38 \pm    0.02$ & $  0.077 \pm   0.004$\\ 
$\tau_{1}$                       & $   1.23 \pm    0.07$ & $   0.73 \pm    0.04$ & $   0.88 \pm    0.05$ & $   0.61 \pm    0.04$\\ 
$\sum{\tau_{i}}$                 & $   1.23 \pm    0.07$ & $   0.61 \pm    0.03$ & $   0.67 \pm    0.04$ & $   0.66 \pm    0.04$ & $   0.50 \pm    0.03$\\ 
\ptjl{}, $N_{\rm jets} = 1$      & $   0.52 \pm    0.03$ & $   0.73 \pm    0.04$\\ 
\hline \hline 
\end{tabular} 
\end{center} 
\end{sidewaystable}

\begin{sidewaystable}[htpb]
\caption{Bin ranges for each of the studied variables.\label{tab:binning}} 
\begin{center}
\small
\begin{tabular}{l| r  r  r  r  r  r  r  r  r  r } 
\hline \hline 
 Bin    & 1~~ & 2~~ & 3~~ & 4~~ & 5~~ & 6~~ & 7~~ & 8~~ & 9~~ & 10~~\\ \hline \hline 
$p_{\mathrm{T}}^{j3}$            & 0--30 & 30--50 & 50--150\\ 
$N_{\rm jets}$, $N_{\rm jets}^{\rm 50~GeV}$      & 0 & 1 & 2 & $\geq 3$\\ 
\yjl                             & 0.0--1.0 & 1.0--2.0 & 2.0--3.0 & 3.0--4.4\\ 
\yjsl                            & 0.0--1.0 & 1.0--2.0 & 2.0--3.0 & 3.0--4.4\\ 
\ptjsl{}                         & 0--30 & 30--50 & 50--70 & 70--140\\ 
\mjj{}                           & 0--200 & 200--400 & 400--650 & 650--1000\\ 
\deltayjj                        & 0.0--2.0 & 2.0--4.0 & 4.0--5.5 & 5.5--8.8\\ 
\dphijj                          & 0--$\pi$/3 & $\pi$/3--2$\pi$/3 & 2$\pi$/3--5$\pi$/6 & 5$\pi$/6--$\pi$\\ 
$p_{\mathrm{T}\gamma\gamma jj}$  & 0--30 & 30--55 & 55--80 & 80--140\\ 
\dphiggjj                        & 0.0--2.6 & 2.6--2.9 & 2.9--3.1 & 3.1--$\pi$\\ 
$\tau_{1}$                       & 0--8 & 8--16 & 16--30 & 30--55\\ 
$\sum{\tau_{i}}$                 & 0--8 & 8--16 & 16--30 & 30--55 & 55--150\\ 
\ptjl{}                          & 0--30 & 30--50 & 50--70 & 70--100 & 100--140\\ 
\htj{}                           & 0--30 & 30--50 & 50--70 & 70--150 & 150--250\\ 
\ygg                             & 0.0--0.3 & 0.3--0.6 & 0.6--0.9 & 0.9--1.2 & 1.2--1.6 & 1.6--2.4\\ 
$\Delta y_{\gamma\gamma}$        & 0.0--0.3 & 0.3--0.6 & 0.6--0.9 & 0.9--1.2 & 1.2--1.5 & 1.5--2.0 & 2.0--2.5\\ 
$p_{\mathrm{Tt}\gamma\gamma}$    & 0--10 & 10--20 & 20--30 & 30--40 & 40--60 & 60--80 & 80--150\\ 
\ptgg{}                          & 0--20 & 20--30 & 30--40 & 40--50 & 50--60 & 60--80 & 80--100 & 100--200\\ 
\costhetastar                    & 0.0--0.1 & 0.1--0.2 & 0.2--0.3 & 0.3--0.4 & 0.4--0.5 & 0.5--0.6 & 0.6--0.7 & 0.7--0.8 & 0.8--0.9 & 0.9--1.0\\ 
\hline \hline 
\end{tabular} 
\end{center} 
\end{sidewaystable}

\clearpage
\bibliographystyle{JHEP}
\bibliography{higgspaper}

\onecolumn
\clearpage
\begin{flushleft}
{\Large The ATLAS Collaboration}

\bigskip

G.~Aad$^{\rm 84}$,
B.~Abbott$^{\rm 112}$,
J.~Abdallah$^{\rm 152}$,
S.~Abdel~Khalek$^{\rm 116}$,
O.~Abdinov$^{\rm 11}$,
R.~Aben$^{\rm 106}$,
B.~Abi$^{\rm 113}$,
M.~Abolins$^{\rm 89}$,
O.S.~AbouZeid$^{\rm 159}$,
H.~Abramowicz$^{\rm 154}$
% ...
\bigskip
\\
% ...
$^{11}$ Institute of Physics, Azerbaijan Academy of Sciences, Baku, Azerbaijan\\
% ...
$^{84}$ CPPM, Aix-Marseille Universit{\'e} and CNRS/IN2P3, Marseille, France\\
% ...
$^{89}$ Department of Physics and Astronomy, Michigan State University, East Lansing MI, United States of America\\
% ...
$^{106}$ Nikhef National Institute for Subatomic Physics and University of Amsterdam, Amsterdam, Netherlands\\
% ...
$^{112}$ Homer L. Dodge Department of Physics and Astronomy, University of Oklahoma, Norman OK, United States of America\\
$^{113}$ Department of Physics, Oklahoma State University, Stillwater OK, United States of America\\
% ...
$^{116}$ LAL, Universit{\'e} Paris-Sud and CNRS/IN2P3, Orsay, France\\
% ...
$^{152}$ Institute of Physics, Academia Sinica, Taipei, Taiwan\\
% ...
$^{154}$ Raymond and Beverly Sackler School of Physics and Astronomy, Tel Aviv University, Tel Aviv, Israel\\
% ...
$^{159}$ Department of Physics, University of Toronto, Toronto ON, Canada\\
% ...
\end{flushleft}

\end{document}